\documentclass{msuphddissertation}

\usepackage{amsmath,amssymb,amsthm,paralist,color,bm,braket,graphicx,subfig,bbold,multirow,url,array}
\usepackage[square,numbers,sort&compress]{natbib}

\newcolumntype{C}[1]{>{\centering\let\newline\\\arraybackslash\hspace{0pt}}m{#1}}

\author{ Joshua Vredevoogd } 
\title{ Relativistic Viscous Hydrodynamics for High Energy Heavy Ion Collisions } 
\field{ Physics } 

\DeclareMathSizes{12}{12}{10}{8}

\begin{document}

\maketitlepage 
\begin{abstract}

It has been over a decade since the first experimental data from gold nuclei collisions 
at the Relativistic Heavy Ion Collider suggested hydrodynamic behavior. 
While early ideal hydrodynamical models were surprisingly accurate in
their predictions, they ignored that the large longitudinal velocity gradient meant that
even small shear viscosity would produce large corrections to the transverse dynamics. In addition, 
much less was known about the equation of state predicted by lattice calculations of quantum chromodynamics,
which predicts a soft region as the degrees of freedom change from quarks to hadrons but no
first-order phase transition. Furthermore, the effects of late, dilute stage rescattering 
were handled within the hydrodynamic framework 
to temperatures where local kinetic equilibrium is difficult to justify.
This dissertation presents a three-dimensional viscous hydrodynamics code with a realistic equation of state 
coupled consistently to a hadron resonance gas calculation. The code presented here is capable 
of making significant comparisons to experimental data as part of an effort to learn 
about the structure of experimental constraints on the microscopic interactions of dense, hot quark matter.

\end{abstract}




\begin{dedication} 
To my lovely wife and consistent source of inspiration, Sara.

\end{dedication}




\TOC


\LOF


\newpage
\setcounter{page}{0}\pagenumbering{arabic}
\begin{doublespace}

\chapter{Heavy Ion Collisions}


It has been understood for over half a century that the protons and neutrons that make up
atomic nuclei have substructure.  The symmetry structure of the many new particles that
were being discovered in the 1950's and 1960's led to an elegant explanation in terms of constituent
particles that Gell-Mann coined as quarks.  The quark model successfully predicted the 
Omega baryon, so named for being the only undiscovered combination of the three 
quarks (up, down, and strange) that made up all known particles at the time.  Since 
that time, heavier quarks have been theorized and discovered. 
Notably, the top quark was discovered by the Tevatron collider at Fermilab in collisions
of protons with anti-protons at extremely high relative momentum.  Such 
high momentum collisions, of single hadrons or of nuclei,
form the experimental basis of our knowledge about the interactions of quarks 
and the corresponding force carriers, gluons.  The accepted theory for explaining
these interactions is called Quantum ChromoDynamics (QCD), which has been studied 
in great detail and strenuously tested over the last half century.  

Among the unique features of QCD is that each quark is confined
by the strong force and can
never be completely separated from at least one or two partner quarks.  
Confinement is borne out experimentally: despite
the extreme amount of energy available to separate quarks in heavy ion collisions, 
collider experiments only observe mesons and baryons, 
made up of two and three constituent quarks respectively,
and never a bare quark.  In QCD, this is explained by the
introduction of color charge.  Isolating objects with non-zero
color charge is forbidden, and since quarks carry a single color charge,
they can never be observed in isolation.
In the opposite regime, when quarks are extremely close, they 
interact only weakly, a result known as asymptotic freedom.  This is exactly
the opposite of Quantum Electrodynamics where the bare charge is infinite
and renormalization is required to reproduce couplings observable
at large distances as terms are added to the perturbation series.  

Since the strong force increases dramatically as quarks become
separated, there is the possibility of creating a new and interesting
state of matter when the average distance between quarks becomes small.
Creating this new state of matter, often called the Quark-Gluon Plasma (QGP),
in which the quarks are freed from hadrons
and form a plasma was one of the goals of the heavy ion collisions 
at the Relativistic Heavy Ion Collider (RHIC),
and the associated STAR (Solenoid Tracker At RHIC) 
and PHENIX (Pioneering High Energy Nuclear Interaction eXperiment) experiments \citep{Bass:1999zq}. 
While such a phase transition had been expected,
clean experimental signatures that the transition has occurred have been
somewhat difficult to isolate and, for instance, there is at present no evidence for 
behavior associated with a first-order phase transition.
At this time, this is true even in the data from new RHIC initiative 
searching for the critical point in the QCD phase diagram at larger
baryon chemical potential
\citep{Adamczyk:2012ku}.  
Still, significant experimental evidence suggests that the matter created at RHIC 
interacts strongly and collectively, notably 
away-side jet quenching \citep{Adcox:2001jp,Adams:2003im,Adams:2005ph,Adler:2002tq}
and flow observables \citep{Ackermann:2000tr,Adler:2003kt,Abelev:2008ae,Afanasiev:2009wq}.  

\begin{figure}
\centerline{\includegraphics[width=0.7 \textwidth]{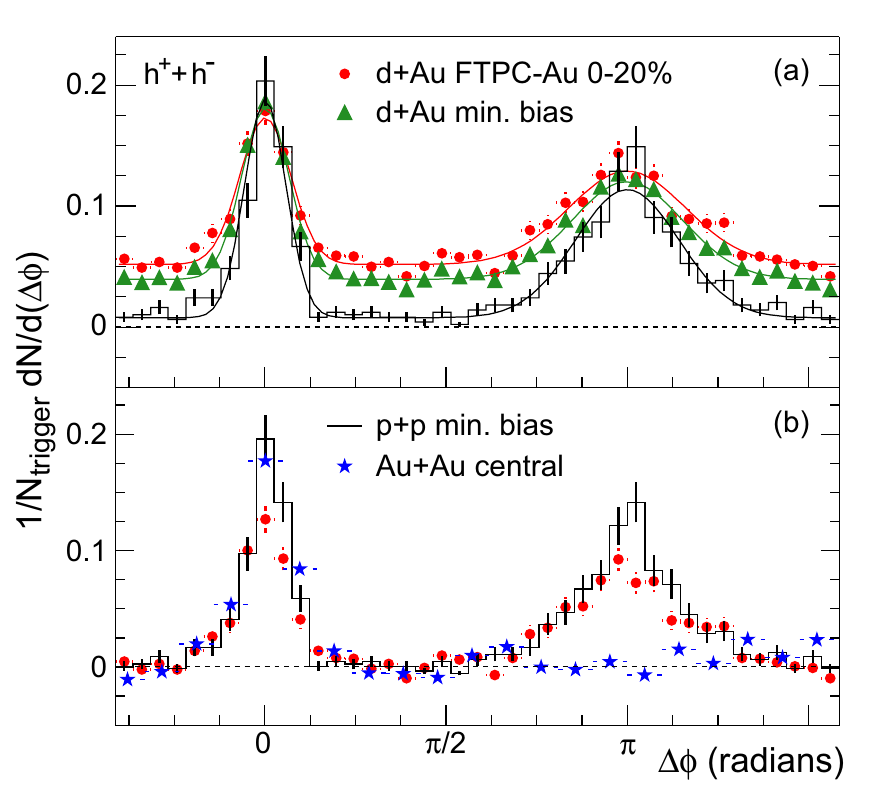}}
\caption{ 
Away side suppression of high transverse momentum hadrons 
in Au+Au collisions as compared to the 
same data for d+Au and p+p at constant energy per nucleon,
$\sqrt{s_{\text{NN}}} = 200$ GeV, as measured by the STAR experiment \citep{Adams:2003im}.
For interpretation of the references to color in this and all 
other figures, the reader is referred to the electronic version of this dissertation.
}
\label{fig:hadronSuppression}
\end{figure}

Jets are formed in hadronic collisions at high relative momentum when there is 
a hard scattering of a pair of quarks or gluons.  If the pair scatter away from the 
hadronic matter, the energy due to their separation grows rapidly.  
The separation energy becomes much larger than the energy required to generate additional
quarks and so each quark or gluon begins to form quark-antiquark pairs until they are able to form
a collection of colorless combinations.  This process results in a group of high momentum hadrons
at small relative momentum known as a jet.  In proton collisions, jets are almost always found in
back-to-back pairs with similar total energy due to momentum conservation.
This is not the case in heavy ion collisions, as Figure \ref{fig:hadronSuppression} demonstrates
for Au+Au collisions at a center of mass energy per nucleon pair ($\sqrt{s_{\text{NN}}}$) of 200 GeV.
The figure shows the distribution of charged particles whenever there is a very high transverse
momentum particle present shown as a function of the angular difference in the transverse plane
compared to that particle.  To assist in interpreting this plot, we define the 
coordinate system of heavy ion physics.
The transverse direction is radially outward from the interaction
point in the plane orthogonal to the motion of the colliding nuclei, where the direction along the 
the motion of the colliding nuclei is referred to as longitudinal.  We will frequently refer to 
particles by their transverse momentum, $p_T$, and their longitudinal rapidity, $y$.
The final coordinate is the azimuthal angle in the transverse plane, $\phi$, 
meaning that the coordinate system in particle momentum space is cylindrical.

In Figure \ref{fig:hadronSuppression}, the highest transverse momentum particle is 
defined to be at $\Delta \phi = 0$ and only events with such a high transverse
momentum particle are shown.  Since events are selected on high transverse
momentum, this particle is called the trigger particle and the associated jet 
is called the trigger jet.
We expect a cluster of particles distributed around $\Delta \phi = 0$
associated with the near-side jet, and another peak around $\Delta \phi = \pi$ corresponding
to the away-side jet.  This is strictly due to momentum conservation in the transverse plane.
The particles in the trigger jet in the peak around zero are only slightly altered depending on 
the system size, but the away-side jet that appears as a peak at $\Delta \phi = \pi$ disappears for
central Au+Au collisions, those that produce the largest total number of low momentum particles.  
This implies that the matter created in a heavy ion collisions is opaque 
to high momentum particles, though the theory of momentum
transfer from high momentum quarks to a thermal QCD medium is 
not well understood  \citep{Armesto:2011ht}.

\begin{figure}
\centerline{\includegraphics[width=0.75 \textwidth]{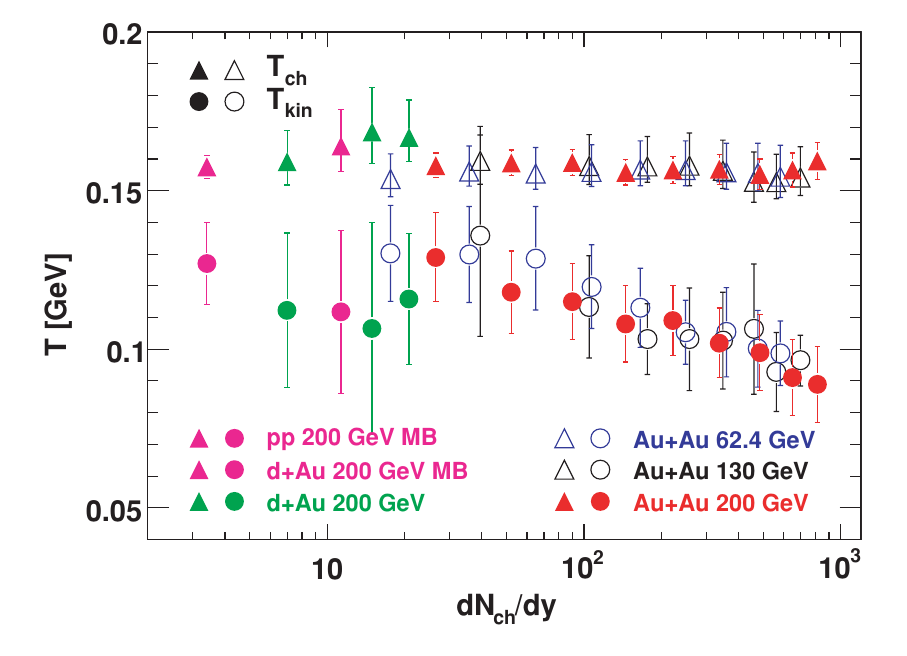}}
\caption{ 
The temperature extracted from blast wave fits to low mass particle
spectra from the STAR experiment for various systems
at various beam energies per nucleon pair \citep{Abelev:2008ab}.  
}
\label{fig:spectraTemp}
\end{figure}

The matter created in high energy heavy ion collisions
also appears to interact long enough to cool.  Figure 
\ref{fig:spectraTemp} shows the chemical and kinetic temperatures
indicated by blast wave fits to the transverse momentum dependence
of the particle spectra as a function of the charged particle multiplicity.  
In its simplest form, the blast wave model assumes that all particles were emitted from a 
sphere of uniform density characterized by five parameters:
\begin{itemize}
\item A single kinetic temperature ($T_{\text{kin}}$),
\item a single chemical temperature ($T_{\text{ch}}$),
\item a linear velocity profile with unknown average velocity ($\langle \beta \rangle$),
\item a spherical radius (R), and 
\item a baryon chemical potential ($\mu_B$).
\end{itemize}
These five parameters are sufficient to
predict the probability of observing a particular particle at a particular
momentum, and therefore, they can all be determined by a fit
to the spectrum low mass particles (pion, kaon, and proton).
While this model is overly simplistic and does not include
important physics, the parameters are a useful way of discussing
trends in heavy ion collisions.

The chemical temperature extracted via the blast wave fit
does not change as the number of charged particles 
or the beam energy increases and is consistent with the ratios observed in 
high energy proton-proton collisions.  This trend does not carry over to the kinetic 
temperature indicated by the momentum distribution, which decreases as particle 
production increases.  This softening of the spectra should not be associated with resonance
production, since pions below 500 MeV/c were excluded, but instead should be related to 
the increasing importance of rescattering at increasingly temperatures.
In peripheral collisions, which are shown in the figure
as those with smaller multiplicities for the same beam energy, the spectra show 
no characteristic difference from proton-proton collisions in either the chemistry
or the kinetics.  This suggests that multiple scattering plays essentially the same 
role in these collisions.  However, the effective kinetic temperature decreases
smoothly for all beam energies as the hot region grows and the number of 
soft collisions, and therefore the system size, increase.  This suggests 
that rescattering remains important for a considerable period after 
hadrons are created.  A hydrodynamic model of heavy ion collisions
must address the temperature range between kinetic and chemical freezeout.
Modelers have taken two strategies -- one 
can continue using a hydrodynamic model that takes into account the 
motion of the different particle species relative to one another; or one
can couple the calculation to a gas model.  In this work, we chose the latter.

\begin{figure}
\centerline{\includegraphics[width=0.75 \textwidth]{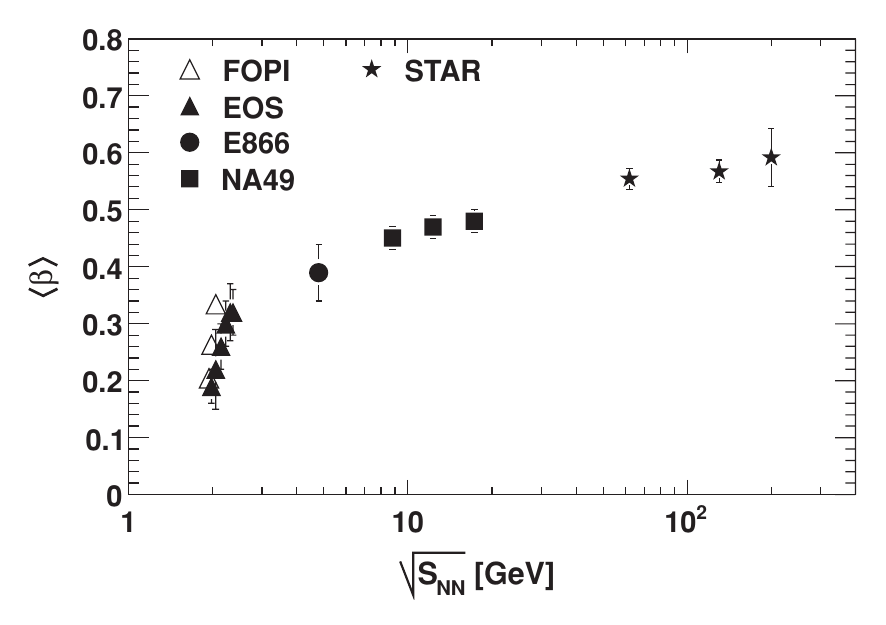}}
\caption{ 
Average particle velocity extracted from the particle spectra of heavy ion collisions
of various nuclei via a blast wave fit as a function of the beam energy per nucleon pair \citep{Abelev:2008ab}.  
Low momentum pions ($p_T < 0.5$ GeV/c) are removed from the fit due to 
resonance background.  Strong collective velocity is a strong indicator of the
onset of collective behavior for collisions at center of mass energy per nucleon pair greater
than a few GeV.
}
\label{fig:avVelocity}
\end{figure}

The transverse velocity indicated by the blast-wave fit  is further evidence 
of collective behavior in heavy ion collisions.  Figure \ref{fig:avVelocity} shows 
the average velocity indicated by the blast-wave fit as a function
of beam energy per nucleon pair for heavy ion collisions.  A rapid increase in the 
average collective velocity is observed in the region around $\sqrt{s_{NN}} = 2$ GeV and a 
continued steady increase toward the RHIC experiments.  As a baseline,
the STAR experiment finds that for proton collisions at $\sqrt{s_{NN}} = 200$ GeV, 
the apparent average collective velocity is $<\beta>_{pp} = 0.24 \pm 0.08$.
The collective velocity in proton-proton collisions
is not zero, as one might anticipate from a system expected to be too
small to develop collective flow, but it is considerably smaller than
 $<\beta>_{AA} = 0.59 \pm 0.05$ as observed in heavy ion collisions \citep{Abelev:2008ab}.
This suggests that around a few GeV in beam energy
the nature of heavy collisions changes, and collective response becomes
critical to explaining the matter's behavior.

\begin{figure}
\centerline{\includegraphics[width=0.75 \textwidth]{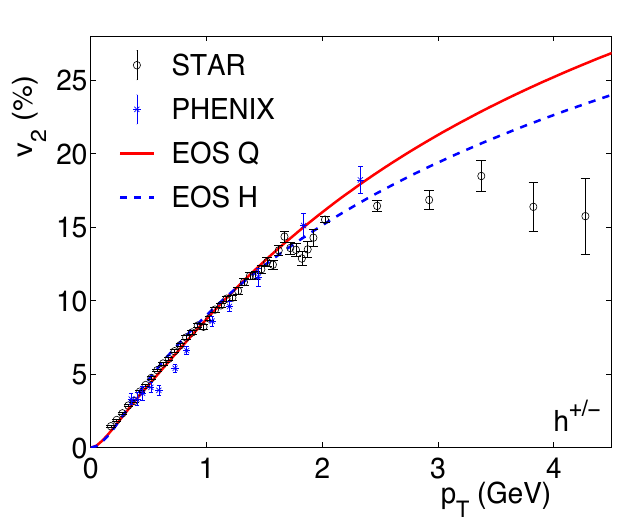}}
\caption{ 
Momentum dependence of the anisotropic flow observed for all charged particles 
by the STAR and PHENIX experiments as compared to the output 
of early ideal hydrodynamic simulations \citep{kolbThesis} for mid-central collisions at 
$\sqrt{s_{NN}} =$ 200 GeV.  The simulations show excellent agreement
with the experimental data up to $p_T = 2$ GeV, where hard processes are increasingly
important.
}
\label{fig:idealHydroV2}
\end{figure}

In addition to the onset of significant outward flow, the data taken at RHIC
show azimuthal anisotropy of flow as shown in Figure \ref{fig:idealHydroV2},
which shows agreement between ideal hydrodynamic models and
experimental data up to $p_T = 2$ GeV.  This anisotropic (or elliptic) flow is 
generated by the anisotropy in the initial state from the finite impact 
parameter.  The elliptical interaction region means larger pressure
gradients along the minor axis leading to more rapid hydrodynamic
expansion.  This additional flow produces additional particles at 
moderate momenta aligned with the minor axis across a large
range of rapidities and therefore the second Fourier coefficient,
\begin{equation}  \label{eq:ellipticFlow}
v_2 = < \cos{ 2\phi_{\text{RP}}} >,
\end{equation}
increases where $\phi_{\text{RP}}$ is the azimuthal angle relative 
to the impact parameter.  The agreement between ideal hydrodynamic
models was considered strong evidence for the formation of the 
Quark-Gluon Plasma  \citep{Kolb:2003dz,Adams:2005dq}.  
At the time, there were some caveats required. 
Ideal hydrodynamics overestimated the duration of the collisions
as seen in the longitudinal source size \citep{Heinz:2002un}, but systematic improvements
to hydrodynamic calculations have resolved these discrepancies \citep{Pratt:2008qv}.
Elliptic flow is a very important observable in heavy ion collisions and 
will be a focus of Chapters Five and Six.

\begin{figure}
\centerline{\includegraphics[width=0.8 \textwidth]{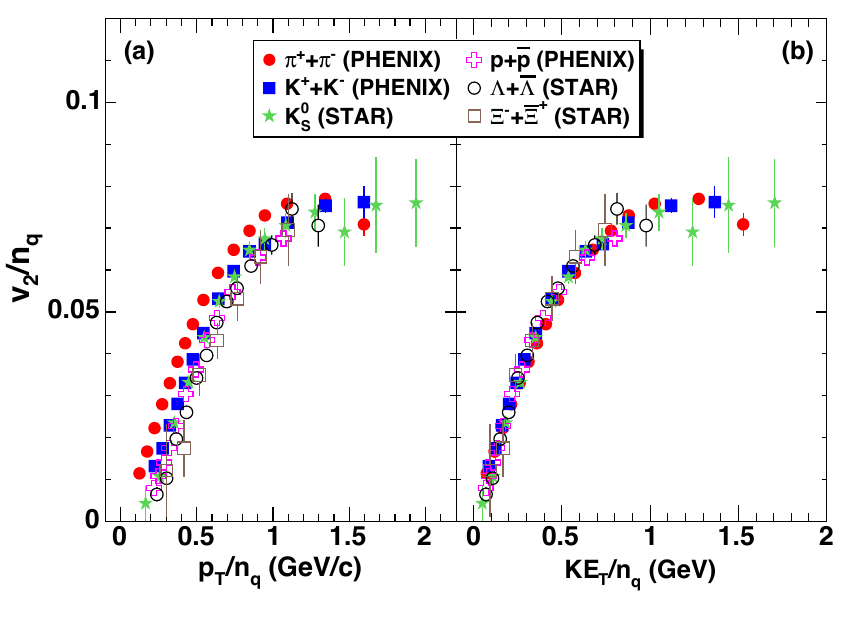}}
\caption{ 
The anisotropic flow as a function scaled by the number of quarks as a function
of the kinetic energy scaled by the number of quarks \citep{Adare:2006ti}.  
This result suggests that elliptic flow is generated in the deconfined phase
and is transferred to hadrons through quark recombination.
}
\label{fig:nqScaling}
\end{figure}

The truly novel feature of the Quark-Gluon Plasma is that quarks and gluons
move freely and not in the bound particle states of a hadronic gas.
While Figure \ref{fig:idealHydroV2} demonstrates that the matter 
created in a heavy ion collision exhibits
hydrodynamic behavior, Figure \ref{fig:nqScaling} further shows that 
the anisotropic flow scales with the number of constituent quarks in
the baryons and mesons observed in the final state.
Scaling is not evident as a function of the transverse momentum but 
instead the transverse kinetic energy, $\text{KE}_T = \sqrt{m^2 + p_T^2} - m$,
which corresponds to the kinetic energy available after creating
the particle itself.  If the anisotropic flow were generated in the 
liquid phase, translated to the hadronic matter by recombination,
and subsequent rescatterings in the gas phase did
not spoil the symmetry, the quantity would be expected to be 
the same for all particle species.  The observation that this 
occurs in heavy ion collisions suggests that flow is
developed prior to recombination.

\begin{figure}
\centerline{\includegraphics[width=0.8 \textwidth]{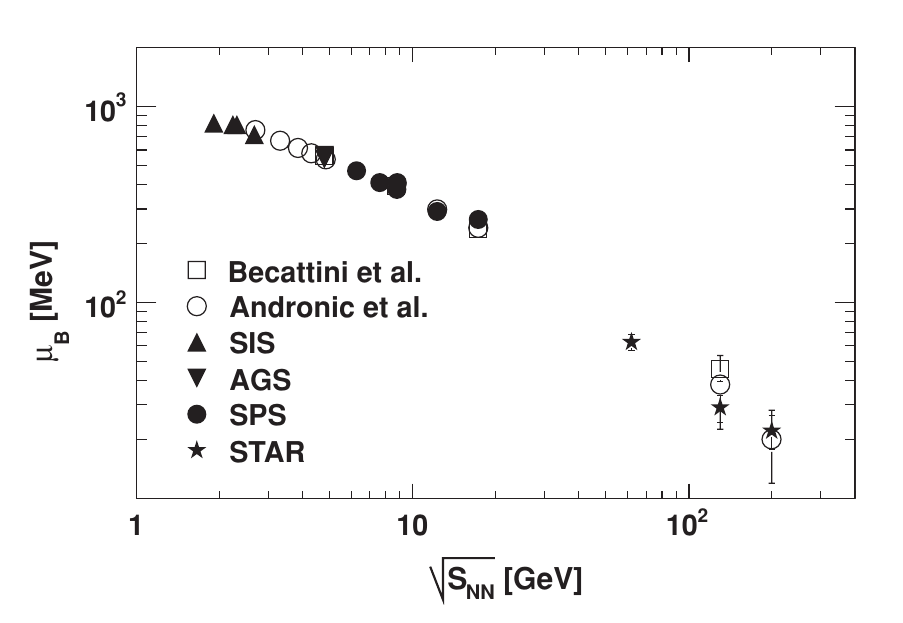}}
\caption{ 
Baryon chemical potential indicated by particle ratios at midrapidity
as measured by the STAR experiment at higher energies and 
other experiments at lower beam energies \citep{Abelev:2008ab}.  
At beam energy of a few GeV, 
even the matter at midrapidity maintains the same chemical
nature as the original nucleus.  This is in contrast to the situation
at $\sqrt{s_{NN}} = $ 200 GeV where the apparent chemical potential
is reduced by almost two order of magnitude from that of stable
nuclear matter.  
}
\label{fig:ExpChemPot}
\end{figure} 

Experimental data, shown in Figure \ref{fig:ExpChemPot},
demonstrate that the observed chemical potential in midrapidity
particles is very small compared to the temperature
at the highest RHIC energies.  This means that there
is essentially no preference in the particle ratios between baryons and 
mesons other than that attributable to their differing masses.  
Therefore, we chose a hydrodynamic description of the evolution
of the quark matter that ignores the effects of non-zero baryon number.
In addition to reducing computational costs slightly, there is the theoretical
issue that lattice calculations are not yet able to make reliable predictions 
for the equation of state of quark matter at finite chemical potential.  
Theoretical progress on this matter is ongoing \citep{Schmidt:2006us,Huovinen:2011xc}, 
but in light of the significant uncertainty
we choose instead to limit our experimental observables 
to those at midrapidity to avoid these issues.  

At the onset of this project, the state of the art in hydrodynamical modeling
were viscous codes with trivial longitudinal expansion based on boost invariance
\citep{Song:2007ux,Luzum:2008cw,Dusling:2007gi}.
Boost invariance is based on the observation that, in the center of momentum frame,
small boosts should not influence observables
since the original nucleons are at much larger rapidity than the created
particles \citep{Bjorken:1982qr}.  
If this symmetry is observed exactly,
then the system will have no variation in the coordinate
\begin{equation}
\eta = \frac{1}{2} \log \left( \frac{t+z}{t-z} \right),
\end{equation}
where $t$ is time and $z$ is longitudinal distance relative to the symmetry axis,
which we refer to as spatial rapidity. This is due to the correspondence of 
spatial rapidity to momentum rapidity,
\begin{equation}
y = \frac{1}{2} \log \left( \frac{E+p_z}{E-p_z} \right),
\end{equation}
where $p_z$ is the longitudinal momentum.
This correspondence is clearest when considering the product of a particle's momentum
with the collective velocity, as will arise in the phase space density,
\begin{equation}
p^\mu u_\mu = \sqrt{m^2 + p_{\perp}} \cosh{(y-\eta)},
\end{equation}
meaning that there will be a penalty for emission at a different rapidity that
scales as $\exp^{-\cosh{(y-\eta)}}$.
If one defines the proper time in a frame with constant $\eta$ as a replacement
for the usual lab frame time coordinate, which is defined $\tau^2 = t^2 - z^2$,
one maintains an orthogonal set of coordinates with metric 
$g_{\mu \nu} = \text{diag} \{1, -1, -1, -\tau^2 \}$.
Furthermore, the hydrodynamic equations of motion respect boost invariance,
meaning that a system born into boost invariance maintains this symmetry
at all times even in the presence of shear or bulk viscosity.
In the limit of infinite beam energy, the symmetry is exact.

\begin{figure}
\centerline{\includegraphics[width=0.7 \textwidth]{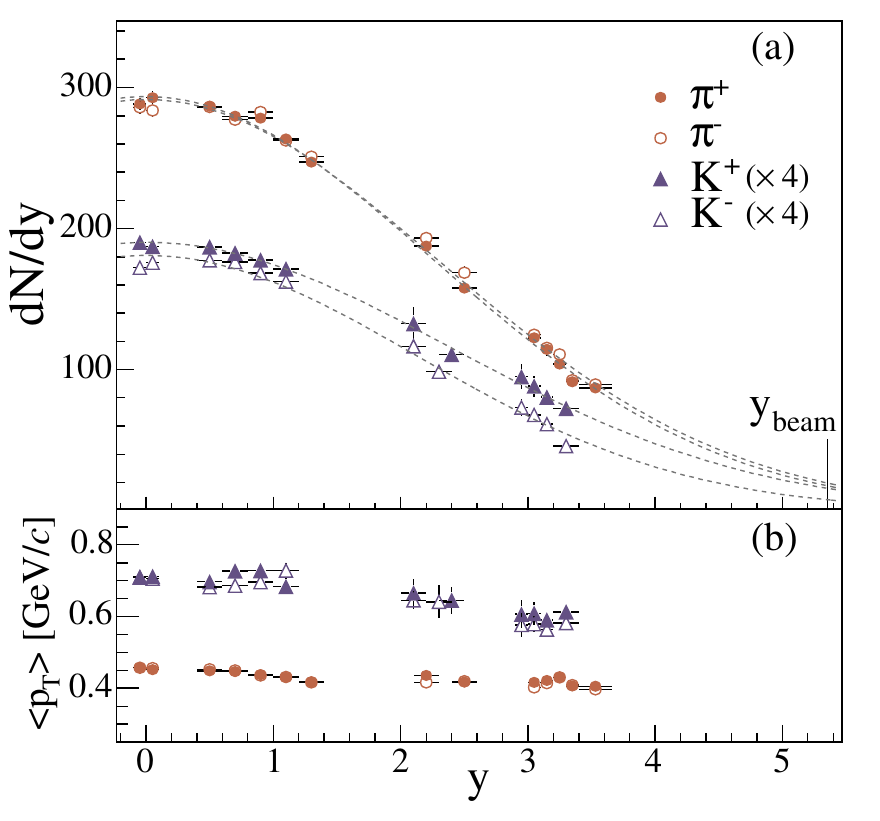}}
\caption{ 
Pion (circles) and kaon (triangles) spectrum observed by the BRAHMS experiment \citep{Bearden:2004yx}
as a function of their longitudinal rapidity including Gaussian fits (dotted lines).  
Shows that pion source is does not have an especially large width 
and boost invariance is a questionable assumption even for describing behavior at midrapidity.
}
\label{fig:longPions}
\end{figure} 

As noted earlier, for collisions at RHIC, boost invariance has proved a useful approximation
for calculating midrapidity observables such as elliptic flow and particle spectra 
\citep{Song:2007ux,Luzum:2008cw,Dusling:2007gi}.  
That said, the distribution of pions is still well described by a Gaussian in 
longitudinal rapidity. 
This was also the case at lower beam energies, though 
width increases to about 2.3 units of rapidity by $\sqrt{s_{\text{NN}}} = 200$ GeV,
as shown in Figure \ref{fig:longPions}.  
While this means that boost invariance is increasingly valid, 
the range of validity was unknown as no three-dimensional
simulation that included viscosity had been completed.  In light of this, one goal of this 
project was to explore the validity of the boost invariant assumption 
in viscous hydrodynamic models regarding the prediction of midrapidity observables.

Taken in aggregate, the experimental evidence from relativistic heavy ion collisions
points to the formation of a new liquid phase.  This liquid remains in thermal contact
long enough to develop significant radial and elliptic flow.  However, the rapid 
longitudinal expansion means that even if the shear viscosity in the liquid phase is 
small, the viscous corrections to the transverse expansion will be important. 
For these reasons, we choose to model the high temperature regions with
relativistic viscous hydrodynamics. 
Since larger systems cool further following the formation of hadrons, the 
lower temperature regions will be modeled as a hadronic gas.
With this two-component model, we will investigate the sensitivity of
our predictions at midrapidity to the addition of non-trivial longitudinal
dynamics and uncertainties in the initial condition to hydrodynamics.

\chapter{Theory of Fluid Dynamics} 

The experimental results discussed in Chapter 1 suggest
that the matter created in heavy ion collisions can be 
described by hydrodynamics.  However, 
thermal contact is maintained to lower temperatures than could be described 
by hydrodynamics.  Because of this, our model will couple an hadronic
gas at low temperatures to viscous hydrodynamics at higher temperatures.
The focus of this chapter will be on hydrodynamic theory underlying 
the higher temperature model.

In this chapter, we pursue one of the possible methods of 
developing fluid dynamics. We consider
the macroscopic, collective behavior of systems of free moving 
particles interacting with one another -- a highly interactive limit of a gas.
This will be of particular use when coupling the hydrodynamic phase to 
the gaseous phase where several results from this derivation will be 
of direct use.
While this is a useful way of deriving the 
equations of motion of hydrodynamics, the resulting equations 
are more general than a derivation from kinetic theory might imply.  
Most notably, the quark-gluon plasma is expected to be strongly-interacting,
while kinetic theory is weakly-interacting.
The equations of motion apply to systems where entropy is conserved locally,
collective motion is sufficient to describe the dynamics of interest, and the acceleration
of that collective velocity comes from the absence or presence of material nearby.
The equations are useful only when one can express an equation of state to close the
set of equations.  For the case of ideal hydrodynamics with no charges, this is to
provide the pressure as a function of energy density.

\section{Kinetic Theory and Fluids}

In the case of relativistic heavy ion collisions, the underlying microscopic theory at high 
temperature should be Quantum ChromoDynamics
for a bulk system.  The present state of the theory has not produced a dynamical 
theory evolving microscopic degrees of freedom that would be needed
to describe the observations from heavy ion collisions, 
but it has produced an equation of state for hot quark matter 
from lattice models (see Chapter 3) \citep{Borsanyi:2010cj,Bazavov:2009zn}.

The general structure of hydrodynamic theory can be derived from breaking 
the system into a system of fluid elements, each of which contains many colliding particles
\citep{huangStatMech, Romatschke:2009im, Betz:2008me, Huovinen:2008te}.
If the particles collide relatively infrequently, the system is dominated
by particle diffusion; whereas fluid dynamics considers a system for 
which particles collide on a shorter time scale than the system's
shape changes.  The frequent collisions will tend to move
the distribution of particle momenta toward the equilibrium distribution 
as collisions randomize each particles' momentum.
In the case that quantum corrections to this distribution are not important,
this momentum distribution is the Maxwell-Boltzmann distribution, given by
\begin{equation}  \label{eq:MBDist}
f (p_\mu u^\mu) \propto \frac{d^6 N}{d^3 r d^3 p} \propto  \exp ( - p_\mu u^\mu / T ),
\end{equation}
where $f$ is the phase space distribution, $p$ is the particle momentum, 
$u$ is the fluid velocity, and $T$ is the temperature.

The phase space distribution leads to the number density current
and to the stress energy tensor.  For the number density, one integrates
over all momentum modes, which we will denote by
\begin{equation} \label{eq:momIntegral}
\int d \omega = \int \frac{ d^3 p}{(2\pi)^3 p_0},
\end{equation}
where $p_0$ is the relativistic energy.  Using this notation, one can then define
the number density current to be
\begin{equation} \label{eq:ndCurrent}
n^\mu =  \int d\omega ~ p^\mu f(p \cdot u),  
\end{equation}
and the stress energy tensor to be 
\begin{equation} \label{eq:boltzSE}
T^{\mu \nu} = \int d\omega ~ p^\mu p^\nu  f( p \cdot u). 
\end{equation}

For a collisionless system, the phase space density is conserved in the 
frame that moves with the particle's momentum, $p^\mu$.  If the particles are allowed 
to collide, these can be included as a source term to the equation called the collision
integral.  This is summarized by
\begin{equation} \label{eq:MB}
p^\mu \partial_\mu f (u^\mu p_\mu) = \it{C},
\end{equation}
where $\it{C}$ describes the effects of collisions. 
This equation is often referred to as the Boltzmann equation, though it 
has many names depending upon the addition of other effects.
It is generally considered the starting point for kinetic theory and is 
fundamental to the description of gases.

As described above, hydrodynamics can be seen as a limit of 
kinetic theory where collisions do not allow the system to significantly
deviate from the equilibrium distribution.  If one then integrates the Boltzmann
equation over momentum, one obtains
\begin{equation}  \label{eq:ktConsNum}
\int d\omega ~ p^\mu \partial_\mu f = \partial_\mu n^\mu = 0 = \int d\omega ~ \it{C},
\end{equation}
where the expression is zero assuming that collisions are local and locally conserve
particle number in the aggregate.
If one takes the first moment of Boltzmann equation with respect to momentum,
one obtains
\begin{equation} \label{eq:ktConsSE}
\int d\omega ~ p^\nu p^\mu \partial_\mu f = \partial_\mu T^{\mu \nu} = 0 = \int d\omega ~ p^\nu \it{C},
\end{equation}
which is zero if the collisions also conserve momentum and energy.
Higher moments of the Boltzmann equation also vanish exactly if the system never
deviates from local equilibrium.  
This defines the stress energy tensor ($T^{\mu \nu}$) from the perspective 
of kinetic theory which will serve as our connection to hydrodynamic theory, most
notably when coupling the hydrodynamic module to the gaseous module.

In order to calculate the connection with hydrodynamics, we consider a small volume of 
particles distributed according to the equilibrium distribution with no collective motion.
If we calculate $T^{tt}$ from kinetic theory as in Eq. \ref{eq:boltzSE}, we obtain
\begin{equation} \label{eq:ktTtt}
T^{tt}(\vec{r}) = \int \frac{d^3p}{p_0} p_0^2 f (\vec{p},\vec{r}) = \epsilon,
\end{equation}
where $\epsilon$ is the energy density.
Since this frame is defined by having zero net momentum, all off-diagonal elements
of the tensor are zero at equilibrium in this frame due to symmetry.  
Spatial elements along the diagonal, however, are proportional to the
average of a momentum squared, which is non-zero and given by
\begin{equation} \label{eq:ktTxx}
T^{xx}(\vec{r}) = \int \frac{d^3p}{p_0} p_x^2 f ( \vec{p}, \vec{r}) = P,
\end{equation}
where $P$ is the pressure.
The identification of this quantity as the pressure can be understood by noting that
the linear momentum transferred to the boundary of the fluid element and the
frequency of collisions with the boundary are both proportional to the momentum \citep{huangStatMech}.
This will be confirmed later when we investigate the conservation of 
stress-energy in the following section.

One can also note directly from Eq. \ref{eq:ktTtt} and Eq. \ref{eq:ktTxx} that the
equation of state for a massless gas is entirely determined.  From relativistic 
kinematics one knows that
\begin{equation}
p_0^2 - \vec{p} \cdot \vec{p} = m^2 ,
\end{equation}
where $m$ is the mass of the particle.  But then if one examines the trace of $T^{\mu \nu}$,
\begin{equation} \label{eq:masslessTTrace}
T_\mu^\mu = T^{tt} - T^{xx} - T^{yy} - T^{zz} = \int \frac{d^3p}{p_0} (p_0^2 - p_x^2 - p_y^2 - p_z^2) f (\vec{p},\vec{r}) = 0,
\end{equation}
where the final step uses the fact that the particles are massless.
Changing from the kinetic theory description to the language of hydrodynamics,
Equation \ref{eq:masslessTTrace}
\begin{equation}
\epsilon = 3 P,
\end{equation}
which suffices to provide an equation of state for ideal hydrodynamics.
Note also that if the mass is non-zero, then the trace of the stress-energy tensor
is positive and the pressure at constant energy density is reduced.
This is relevant for temperatures larger than the pion mass where more
mesonic and baryonic states become relevant and the equation of state 
begins to soften (Chapter 3).

\section{Ideal Hydrodynamics}

The equations of relativistic hydrodynamics can be determined by adding
the requisite relativistic structure to the stress-energy tensor.  The only available
non-scalar quantities are the velocity of the frame in which the fluid element has no collective 
velocity and the metric tensor.  No derivatives are permitted in the stress energy tensor
since their inclusion would require the introduction an additional microscopic length scale 
over which we assume that the stress energy tensor will not vary.  Easing of this assumption
will come in the form of viscous corrections to be discussed in later sections.
From our earlier symmetry consideration, the stress energy
tensor has no off-diagonal structure and trace reads $T^{\mu \nu} =  \text{diag}\{ \epsilon, P, P, P  \} $
in the frame with no collective velocity.  This is sufficient to determine the relativistic 
structure \citep{Romatschke:2009im}, which for ideal hydrodynamics is given by
\begin{equation}  \label{eq:SE}
T^{\mu \nu} = u^\mu u^\nu (\epsilon + P)  -  g^{\mu \nu} P,
\end{equation}
where $u^\mu$ is the collective velocity and $g^{\mu \nu}$ is the metric tensor.
We take the metric to be $T^{\mu \nu} = \text{diag}\{1, -1, -1, -1 \}$ in Minkowski space, and
$u^\mu$ transforms as a vector and has 
unit measure, $u_\mu u^\mu = 1$.  Implicitly, we have chosen the Landau frame 
where the collective velocity moves with the energy density, 
and not the Eckhart frame where the collective velocity moves 
with particle number.  This is due in part to interest in systems with zero
net charge where particle number, and therefore the Eckhart frame, would not be well defined.

The equations of motion of ideal hydrodynamics are then just the conservation of the 
stress energy tensor as from Eq. \ref{eq:ktConsSE}.  The portion of temporal component 
along the collective velocity is given by:
\begin{eqnarray}
0 &=& u_\nu \partial_\mu T^{\mu \nu} = u_\nu \partial_\mu \left[ u^\mu u^\nu (\epsilon + P) - g^{\mu \nu} P \right], \nonumber \\
0 &=& D \epsilon +  \left(\epsilon + P \right) \partial_\mu u^\mu,
\end{eqnarray}
where $D = u^\mu \partial_\mu$ is the comoving time derivative.  To obtain the familiar
non-relativistic limit, assume that $v^2$ is small and that $\gamma \approx 1$.  This yields
the familiar expression:
\begin{equation} \label{eq:idealCont}
\partial_t \epsilon + \vec{v} \cdot \vec{\nabla} \epsilon = - (\epsilon + P) \vec{\nabla} \cdot \vec{v}.
\end{equation}
The divergence of the collective velocity is often called the expansion rate and 
can be related to the change in volume of the fluid element by  $dV/dt = V \vec{\nabla} \cdot \vec{v}$.
Equation \ref{eq:idealCont} can then recast as $dE = - P dV$ which is the fundamental
thermodynamic relation for a system with constant entropy.  For this reason,
the first equation of ideal hydrodynamics is often framed in terms of the local conservation of entropy.
Furthermore, it underscores the more general applicability of hydrodynamics to
statistical systems at or near the maximal entropy state.

The Euler equation is related to the conservation of the spatial elements of the 
stress-energy tensor.   To extract this independently of the first equation we use the projector
$\Delta^{\mu \nu} = g^{\mu \nu} - u^\mu u^\nu$ which selects out the spatial indices
in the frame of the matter.  Applying this to the equation of stress-energy conservation yields
\begin{eqnarray}
0 &=& \Delta^\alpha_\nu \partial_\mu T^{\mu \nu} = \partial_\mu T^{\mu \alpha} - u^\alpha \left( D\epsilon + (\epsilon + P ) \partial_\mu u^\mu \right), \nonumber \\
0 &=& (\epsilon + P) D u^\alpha + u^\alpha DP - g^{\alpha \mu} \partial_\mu P.
\end{eqnarray}
When taking the same limit as before, one obtains
\begin{equation} \label{eq:idealEuler}
(\epsilon + P ) \partial_t \vec{v} + (\epsilon +P) \left[\vec{v} \cdot \vec{\nabla} \right] \vec{v} = - \vec{\nabla} P - \vec{v} \partial_t P,
\end{equation}
which reduces to $(\epsilon + P) \partial_t \vec{v} = - \vec{\nabla} P$ in the frame of the matter.
This should be interpreted to mean that acceleration of the frame of the collective velocity 
is due to the pressure gradient observed in that frame. 

\section{Viscous Hydrodynamics}

\subsection{Navier-Stokes Hydrodynamics}

The equations of ideal hydrodynamics are derived from the assumption that local 
equilibrium is exact, or equivalently, assuming that the mean free path 
of the microscopic particles is identically zero.  If this is the case, 
a uniform density fluid with a plane of static fluid adjacent to a plane of
fluid in motion would be a stable condition; the two planes would never affect one 
another because only the divergence of the collective velocity, 
rather than a mixed partial derivative like $\partial_y v^x$, 
enters into the equations of motion.  That is, pressure is a force
orthogonal to the surface of the fluid element in the rest frame; shear viscosity
allows force along the surface.

Experience with liquids indicates that this is not a physical conclusion; if one pulls a sheet 
of material through a fluid, nearby fluid begins to move with the sheet.  Friction in
the fluid allows linear momentum to be transferred between adjacent fluid elements
and their collective velocities would tend to equalize.  Navier-Stokes
hydrodynamics introduces a viscous correction to the spatial elements of the 
stress-energy tensor from such a frictional force \citep{huangStatMech}.  Physically, this should
be separated into a traceless contribution and an effect on the trace, which correspond
to shear ($\eta$) and bulk ($\zeta$) viscosity respectively.
The viscous components are proportional to the velocity gradient with a 
corresponding transport coefficient that characterizes the amount of internal friction in 
the fluid. In the traceless case, this the shear viscosity.  
These results can be summarized as
\begin{equation}
\delta T^{ij} \propto \eta \frac{\partial v^i}{\partial x^j}, ~~ \delta T^{ij} \propto \zeta \delta^{ij} \vec{\nabla} \cdot \vec{v}.
\end{equation}

These terms are collected into a correction to the stress energy tensor
 --  $\pi^{\mu \nu}$ -- and referred to as the shear tensor.  
 The shear tensor is mostly clear written in the fluid frame where it takes the form
 \begin{equation} \label{eq:NSfom}
 \pi^{ij} = -\eta \left(\partial_i v^j + \partial_j v^i - \frac{2}{3} \delta^{ij} \vec{\nabla} \cdot \vec{v}\right),
 \end{equation}
where $\delta^{ij} = 1$ if $i=j$ and is otherwise zero, and the latin indices are meant to
indicate that this applies only to spatial coordinates.
In the frame of the matter, the ideal part of the stress energy tensor is only the outward 
pressure, $\delta^{ij}P$, and Eq. \ref{eq:NSfom} is added as a correction to the equations
of motion (Eqs. \ref{eq:idealCont} and \ref{eq:idealEuler}).  
These additions result in the Navier-Stokes equations of viscous hydrodynamics.
Note that adding these corrections relativistically is non-trivial as one needs to correct for both
the frame motion in both the derivatives and the velocities.  This will be addressed fully when
discussing the Israel-Stewart equations of motion.

\subsection{The Maxwell-Cattaneo Equation}

Solving the Navier-Stokes equations is notoriously difficult as the differential equations are 
parabolic, for which solutions tend to be unstable, and famously unsolved in the general case.  
In addition to being difficult to solve, the relativistic Navier-Stokes equations are 
not necessarily causal as the system responds instantaneously to changes in the velocity gradients.

While it is not clear that either of these difficulties are important for heavy ion physics,
both can be addressed within the Maxwell-Cattaneo framework
\citep{Romatschke:2009im}.  Essentially this framework
 assumes that instead of defining the change to the effective pressure to be exactly
proportional to the velocity gradients, it instead relaxes toward the Navier-Stokes value.
This means that the shear tensor can no longer be computed directly from the other 
dynamical variables or their derivatives.  Therefore, the shear tensor itself is 
promoted to a dynamical variable and must be tabulated separately.
In addition, the shear tensor now must have its own equations of motion.
The form of these equations of motion in the frame of the matter are taken 
to be of the form
\begin{equation} \label{eq:MaxCatt}
\partial_t \left(  \frac{\pi^{ij}}{\sigma_\eta} \right) = \frac{ - \left( \pi^{ij} - \pi^{ij}_{(NS)} \right)}{\sigma_\eta \tau_\pi},
\end{equation}
where $\pi^{ij}_{(NS)}$ is the Navier-Stokes shear tensor given by Eq. \ref{eq:NSfom}, 
$\tau_\pi$ is a new transport coefficient that characterizes the time scale 
on which the system approaches Navier-Stokes viscosity, and $\sigma_\eta$ 
is a scaling factor required to guarantee the Second Law of Thermodynamics
\citep{Pratt:2007gj} (see the following subsection).

Historically, this equation informed the search for an extension of hydrodynamic theory that would 
would be hyperbolic and causal in the general case.  
The equations of motion for the shear tensor can be derived in this form 
from a gradient expansion in kinetic theory or from general entropy production considerations
as will be shown in the next two subsections.
That said, the new transport coefficients, $\tau_\pi$ and $\sigma_\eta$ 
will not be independent of one another complicating the correspondence of the Israel-Stewart theory to the 
Maxwell-Cattaneo equation and its interpretation as a relaxation equation.

\subsection{ Israel-Stewart Hydrodynamics and Entropy} 

Since the system is away from equilibrium, the entropy should be reduced from its 
equilibrium value.  One expects the entropy should increase quadratically as the shear tensor
increases simply from symmetry at equilibrium so the lowest order correction to the entropy current
would be
\begin{equation}  \label{eq:ViscS}
s^\alpha = u^\alpha \left( s_{\text{eq}} - \beta \pi_{\mu \nu} \pi^{\mu \nu} \right),
\end{equation}
where $\beta$ is a thermodynamic quantity to be determined \citep{Pratt:2007gj}.
The second term acts as an entropy source and so it must have positive four-divergence 
relative to the dynamical entropy from the Navier-Stokes continuity equation \citep{Romatschke:2009im,Muronga:2001zk}:
\begin{eqnarray}
0 &\le& \frac{1}{T} \pi_{\mu \nu} \nabla^{<\mu} u^{\nu > } - \partial_\alpha \left[ u^\alpha \beta \pi_{\mu \nu} \pi^{\mu \nu} \right], \nonumber \\
0 & \le& \pi_{\mu \nu} \left[  \frac{1}{T} \nabla^{<\mu} u^{\nu > } - 2\beta D \pi^{\mu \nu} - \pi^{\mu \nu} D \beta - \pi^{\mu \nu} \beta \partial_\alpha u^\alpha \right],  \label{eq:sConstraint}
\end{eqnarray}
where $\nabla_{<\mu} u_{\nu > }$ is the traceless part on the velocity gradient and is defined as
\begin{equation}
\nabla_{<\mu} u_{\nu > } = \nabla_\mu u_\nu + \nabla_\nu u_\mu - \frac{2}{3} \Delta_{\mu \nu} \partial_\alpha u^\alpha,
\end{equation}
with 
\begin{equation}
\Delta^{\mu \nu}  = g^{\mu \nu} - u^\mu u^\nu, ~~~~  \nabla^\mu = \Delta^{\mu \nu} \partial_\nu.
\end{equation}
This traceless part of the velocity gradient is just the Navier-Stokes modification to the pressure 
except the factor of the shear viscosity: $\pi^{\mu \nu} = \eta \nabla^{<\mu} u^{\nu >}$.
The constraint in Eq. \ref{eq:sConstraint} must be guaranteed for any expansion with any physical transport coefficients,
which can be achieved if one requires that
\begin{equation}  \label{eq:piSEOM1}
\pi^{\mu \nu} =  \eta T \left[  \frac{1}{T} \nabla^{< \mu} u^{\nu > } - 2 \beta D \pi^{\mu \nu} - \pi^{\mu \nu} D \beta - \pi^{\mu \nu} \beta \partial_\alpha u^\alpha \right].
\end{equation}
This is an equation of motion for the shear tensor, where one should recall that $D$ is the time derivative
in the frame of the matter.  If $\beta$ is taken to be zero, the Navier-Stokes form for the shear tensor is
recovered exactly.  In addition, it can be transformed into a relaxation equation 
with the same structure as Eq. \ref{eq:MaxCatt} via the appropriate
choice for the new transport coefficient $\beta$. 
To show this, one can rewrite Eq. \ref{eq:piSEOM1} as follows
\begin{equation} \label{eq:piSEOM2}
D \pi^{\mu \nu} + \pi^{\mu \nu} \frac{ D \beta}{2\beta} +  \frac{\pi^{\mu \nu}}{2} \partial_\alpha u^\alpha = \frac{ - \left( \pi^{\mu \nu} - \pi^{\mu \nu}_{(NS)} \right) }{ 2 \eta T \beta} ,
\end{equation}
which leads to two equations for the two unknowns:
\begin{equation}  \label{eq:tcConst}
\beta = \frac{2 \eta T}{\tau_\pi},  ~~~~~  \sigma_\eta D \left( \frac{\pi^{\mu \nu} }{\sigma_\eta} \right) = D \pi^{\mu \nu} + \pi^{\mu \nu} \frac{ D \beta}{2\beta} +  \frac{\pi^{\mu \nu}}{2} \partial_\alpha u^\alpha.
\end{equation}
Since the relaxation time is a transport coefficient to be calculated from the underlying 
microscopic theory, the left part of Equation \ref{eq:tcConst} gives a value for $\beta$.
The right part of Equation \ref{eq:tcConst} can be recast as a single, local conservation equation
\begin{equation}
\frac{D ( \sigma_\eta \sqrt{s\beta})}{ \sigma_\eta \sqrt{s\beta}} = 0,
\end{equation}
which is guaranteed if $\sigma_\eta^{-2} = s \beta$.  
Returning to Equation \ref{eq:ViscS} for the viscous entropy current 
and noting that the probability for a given fluctuation of stress-energy
is proportional to $e^{S}$, then the variance of $\pi_{xy}$ in a given volume V is 
\begin{equation} \label{eq:piVariance}
V < \pi_{xy}^2 > = \frac{1}{2\beta} = \frac{s \sigma_\eta^2}{2},
\end{equation}
meaning that $\sigma_\eta$ behaves like the variance of $\pi_{xy}$.

\subsection{Israel-Stewart Hydrodynamics and Kinetic Theory}

The derivation connecting Israel-Stewart hydrodynamics with kinetic theory is rather involved and
the details are beyond the scope of this document, but a quick sketch is helpful for conceptual reasons.
In addition, the form of the phase space distortion shown in this section
will be useful later when connecting the hydrodynamic fireball to the surrounding 
hadronic gas.  

In establishing the connection between 
kinetic theory and ideal hydrodynamics, we discussed the equilibrium phase space density 
and the Boltzmann equation.  In doing so, we discussed the effect of collisions, which are a source (or sink)
to the conservation of phase space density.  We argued that if the collisions happened often enough, then
the system would remain in equilibrium.  This assumption
along with energy-momentum conservation was enough to produce ideal hydrodynamics.

Further, when discussing the Navier-Stokes viscous corrections hydrodynamics, we mentioned
that this can be thought of either as friction in the fluid from a macroscopic perspective, or as the
effects of expansion between collisions.  In particular, if one imagined a box of particles with an arbitrary
distribution of initial momenta but zero total momentum, and then investigated the system at later times, 
the collisions would tend to relax toward the momentum distribution and the system would exponentially
approach the equilibrium distribution 
\citep{Baier:2007ix,Betz:2008me,Muronga:2006zw,Muronga:2006zx,Huovinen:2008te}.  
This can be summarized as 
\begin{equation} \label{eq:collisions}
\it{C} = - p_\mu u^\mu \frac{ f - f_0}{\tau_\pi},
\end{equation}
where $\tau_\pi$ is the relaxation time scale.

Now, hydrodynamics will not generally apply to situations where the system is far from thermal equilibrium
so we concern ourselves only to the case of small deviations of the  phase space density.  Using the fact 
that this deviation should disappear in equilibrium and should only be a function of the available degrees
of freedom, the simplest ansatz was proposed by Grad \citep{ref:grad} who found that the 
distortion to the phase space density should be
\begin{equation} \label{eq:ViscPSDens}
f = f_0 \left[ 1 +  \frac{\pi_{\alpha \beta} p^\alpha p^\beta}{2(\epsilon + P) T^2}   \right].
\end{equation}
This form of the deviation and the equation of motion for the shear tensor come from a detailed 
analysis of higher moments of the Boltzmann equation \citep{Baier:2006um,Betz:2008me,Muronga:2006zw}.

The important revelation from this approach is the possibility of additional terms that 
do not produce any entropy \citep{Romatschke:2009im}.
Such terms would not appear in a derivation based solely on the
form of the entropy current.
These additional terms tend to require even more transport coefficients many of which 
do not have obvious physical descriptions.  The exception is the coupling of the shear tensor
to the vorticity tensor, which in the frame of the matter is
\begin{equation}
\Omega^{ij} = \partial_i v^j - \partial_j v^i,
\end{equation}
would arise naturally from hydrodynamic considerations and does not require additional
transport coefficients.  This term has been included in our calculations but our studies 
confirm others' results that it is not an important driver of dynamics for smooth
initial condition investigations of hydrodynamical flow.

At this point, we have discussed all the important physics related to viscous hydrodynamics
for the purposes of producing a simulation of zero charge heavy ion collisions.  The 
details of the equations of motion to be integrated have been reserved for Chapter 4.

\chapter{Equation of State and Transport Coefficients} 

While hydrodynamics is a theory that applies to a wide variety of systems, both
strongly and weakly coupled, the microscopic details of a fluid need to be
included through the equation of state and the transport coefficients.
For heavy ion collisions, two physical regimes are relevant
as discussed at length in Chapter 1.
At low temperature, the equation of state should be calculated 
as a non-interacting gas of mesons and baryons;
while at high temperature, lattice QCD should be used.
These predictions may not be compatible in the region
near the transition between hydrodynamics and the hadron
resonance gas, and care must be taken to ensure that the 
model is self-consistent at the boundary.  This is especially
important for the equation of state where, for instance, 
discontinuity in the pressure as a function of energy density
would lead to hydrodynamic instability. While this temperature
region is still under active investigation, recent lattice results
suggest that the discrepancy between the two theories is not large
and a composite description is possible.
This is discussed in the first section of this chapter.

In the second section, we discuss the determination of transport
coefficients in linear response theory. At this time, lattice 
calculations are not developed to the point where they can
make predictions about transport coefficients using classic
results like the Kubo relations.  Because of this, our 
results will be from the weakly coupled theory at low 
temperature, the general expectation that shear viscosity should 
be small near the phase transition, and scaling arguments from
scalar field theories of the high temperature phase.

\section{Equation of State}

\subsection{Hadron Resonance Gas}

The equation of state of the hadron resonance gas model is developed as a
gas.  If one assumes that the gas is non-interacting, 
one can simply generate a partition function using all the available particle states,
which are measured experimentally.
This ignores the finite volume or mean field effects that are well known to be important
for lower energy heavy ion physics but are expected to be less important corrections
at energies where zero baryon number hydrodynamics apply.
Data on particle states up to several GeV/$c^2$ in mass are available from the 
Particle Data Group. We use resonances up to masses of $m=2.5$ GeV/$c^2$
as the addition of higher mass resonances no longer affects the equation of state
up to temperatures of $T=175$ MeV.  This choice leads to a nice agreement 
with lattice results in the transition region \citep{Huovinen:2009yb}.

The partition function is calculated from the usual product of available mass states weighted 
by a common temperature. The contribution to the log of the partition function for a 
mass state $m_i$ is given by
\begin{equation}
\frac{1}{V} \ln \textit{Z}_i = \frac{ \mp d_i}{2\pi^2} \int_0^\infty  dk k^2 \ln{(1 \mp z_i e^{-\sqrt{k^2 + m_i^2}/T})}
\end{equation}
where $d_i$ is the spin degeneracy of the state and 
$z_i$ are the fugacities summed over all charges \citep{Borsanyi:2010cj}.
One can then use standard thermodynamic relations to
obtain the equation of state.  For instance the pressure and energy density are 
\begin{equation}
P = \frac{T}{V} \displaystyle\sum\limits_i \ln{ \textit{Z}_i}, ~~ \epsilon = \frac{T^2}{V} \frac{\partial}{\partial T}  \displaystyle\sum\limits_i \ln{ \textit{Z}_i}.
\end{equation}
From these, the entropy density can be calculated using the thermodynamic identity, $sT = \epsilon + P$.


To briefly summarize the results of this procedure, the hadron resonance gas acts like a 
pion gas for temperatures less than 100 MeV.  However, the increasing number of baryon
and meson states play an increasingly important role as the temperature increases.
This produces an order of magnitude increase in the 
trace anomaly (or interaction, $I(T) = \epsilon - 3P$) by 150 MeV.  
In fact, inclusion of the resonances in the mass range 
$1.0~ \text{GeV} < m_i < 2.5~ \text{GeV}$ provides a factor of 3 in the trace anamoly by $T=200$ MeV
\citep{Borsanyi:2010cj, Bazavov:2009zn}.
From a hydrodynamic perspective, this yields a soft region -- with lower speed of sound --
as a heating system would expend part of its energy in changing the degrees of freedom 
instead of increasing the pressure.  This behavior is a milder version of a latent heat which 
would be associated with a first order phase transition, 
where there would be no increase in pressure over a region of increasing energy density.  


\subsection{Lattice QCD}


In Lattice QCD, one also sets out to calculate a partition function as a function of
temperature but within a field theory rather than a non-interacting gas. The details
of this procedure are complex and involve computing path integrals
over the gauge field on a discretized hypercube and extrapolating to the physical
pion mass.  The computational effort required 
for each time step is immense and computations are performed on large clusters designed 
specifically for the task.  Developments in this area now allow for the computation of
twelve time steps for $64^3$ spatial grid points \citep{Borsanyi:2010cj, Bazavov:2009zn}.
Different temperatures are investigated by altering the lattice spacing, and several
spacings must be investigated to quantify convergence and protect 
against discretization effects.


While the lattice has produced some incredibly useful results for finite temperature
QCD matter, it has important limitations.  The ``sign problem" complicates direct investigation
of the equation of state away from zero chemical potential where there is the possibility 
of a critical point. There the phase transition might shift from a smooth cross-over 
seen at zero chemical potential to a first-order phase transition.
Extrapolation toward this critical point from lattice data has been attempted
 via Taylor expansion \citep{Huovinen:2011xc,Schmidt:2006us}.
While one does not expect this expansion to be valid far from zero chemical potential,
especially when approach a critical point, recent results suggest tentative agreement with other theories.

\begin{figure}
\centerline{\includegraphics[width=0.75 \textwidth]{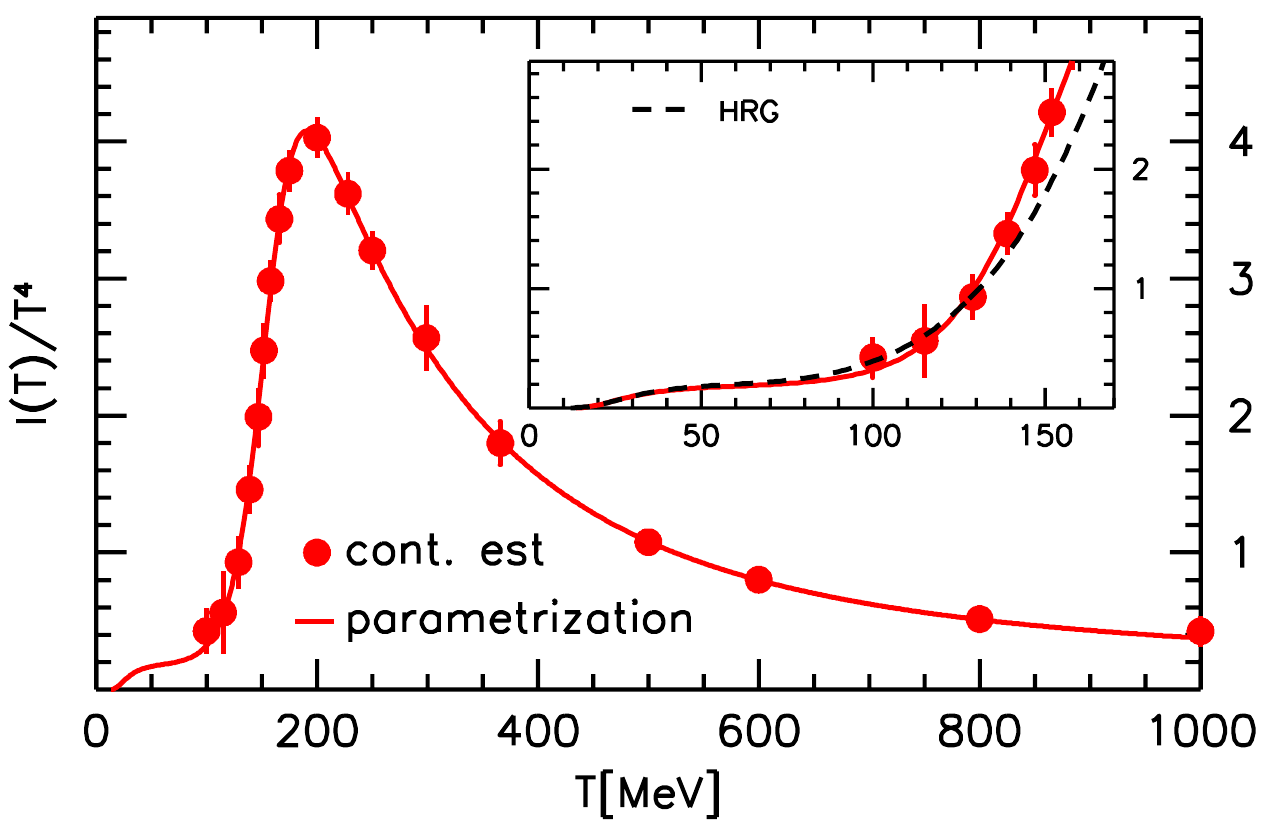}}
\caption{ 
Trace anomaly from the lattice and from the hadron resonance gas model \citep{Borsanyi:2010cj}
showing the location of the critical region and agreement with the Hadron-Resonance gas
equation of state at temperatures well below the critical region.
}
\label{fig:iT-Lattice}
\end{figure}


Furthermore, it is not possible to use a physical quark mass for the up and down (the lightest)
quarks.  The up and down quarks have physical masses of around 3-4 MeV and the strange
quark around 200 MeV, but the typical lattice calculation take the mass of the up and down 
quarks to be a tenth that of the strange quark. This leads to a pion mass that is significantly too
large. This significantly alters conclusions about the pressure in this region
and generally makes interpretation of the lattice results significantly more difficult as one 
must extrapolate to the physical meson and baryon masses in order to  
extract corrected thermodynamic quantities.  Despite of all this, as one finds in Fig. \ref{fig:iT-Lattice},
there is good agreement with the Hadron Resonance Gas model for the temperature region 
in which one hopes both might be valid.

Also included in Fig. \ref{fig:iT-Lattice} is a parameterized fit to the unit-less interaction 
over the full temperature space.  In this case \citep{Borsanyi:2010cj}, the fit function reproduces the 
HRG interaction within 7\% at temperatures below 100 MeV, and calculation of the
pressure via the integral method
\begin{equation} \label{eq:EosIntegralMethod}
\frac{P(T)}{T^4} = \int_{~0}^{T} \frac{dT'}{T'} \frac{I(T')}{T'^4}
\end{equation}
finds deviations of only 2\% and other collaborations find comparable results \citep{Bazavov:2009zn}.

\subsection{ Cross-over Region } 


The general structure of our model is a hydrodynamic region
defined by temperatures above some threshold surround by a gas of 
interacting hadrons.  Practically, dynamic coupling of these two calculations is
difficult as pressure fluctuations in the gas might easily result in extreme hydrodynamic instability. 
Instead, one runs the hydrodynamic simulation for a larger region of configuration space
under the assumption that regions at much lower temperatures do not have undue influence
on the hydrodynamical portion and that near the switching temperature the models
are equivalent.

As a direct consequence, modelers should be fastidious in ensuring that the equation
of state for the gas is used as precisely as possible for those regions.  While
lattice calculations provide a fit that includes data points from the HRG equation of state 
and the lattice, this may not be sufficient to ensure self-consistency.  In fact,
one finds that for $T=165$ MeV, lattice fits can produce 
energy densities that differ from the gas' energy density by 10-20\% without 
fundamentally different behavior in the interaction measure, $I(T)$. 
If left uncorrected, the hydrodynamic code would improperly emulate the gas at the boundary,
and the model of particle production would fail to be self-consistent.


Since one might want to use an arbitrary lattice equation of state that might not reproduce
the gas well enough, we perform a continuous merge between the two 
equations of state over a temperature range starting at the hadronization temperature ($T_H$)
up to the temperature where only lattice results are used ($T_L$).  
In particular, we wish to maintain a continuous speed of sound for hydrodynamic stability, 
which can be written in terms of the unit-less entropy density ($\sigma = s/T^3$) as
\begin{equation}
c_s^2 = \frac{dP}{d\epsilon} = \frac{1}{3 + \frac{T}{\sigma} \frac{d\sigma}{dT}}
\end{equation}
which can be derived using standard Maxwell relations.
This is a clarifying expression as it points out that entropy increasing faster than $T^3$ is 
associated with softness in the equation of state, but also means that $\sigma$ is a 
useful variable for merging equations of state, as continuity and smoothness ensure
the continuity of the speed of sound.

\begin{figure}
\centerline{\includegraphics[width=0.75 \textwidth]{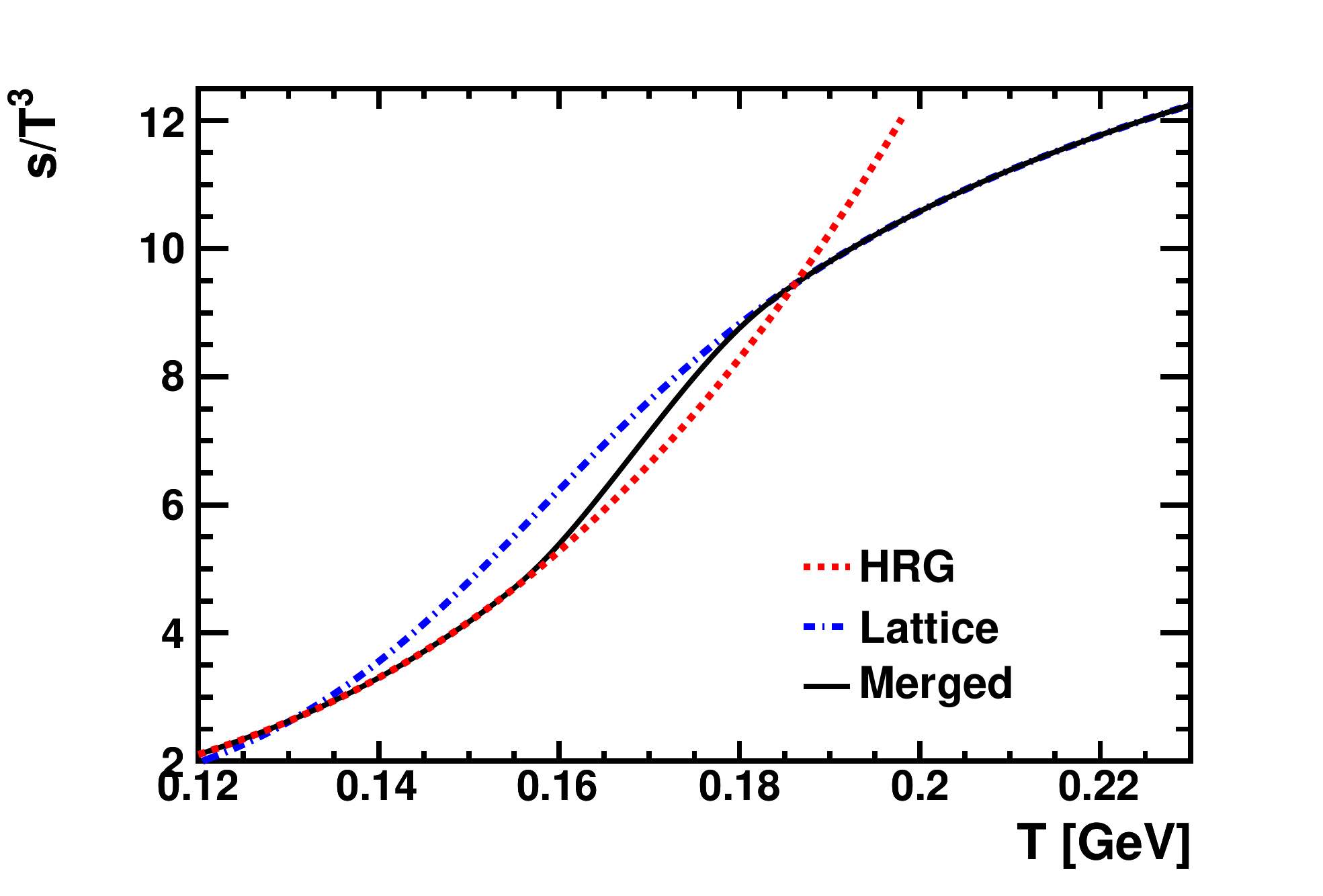}}
\caption{ 
Unit-less entropy density in merging region.  The entropy production follows that 
of the hadron resonance gas up to the hadronization temperature ($T_H = 155$ MeV) and
smoothly produces additional entropy to agree with the lattice entropy 
\citep{Borsanyi:2010cj} at $T_L =185$ MeV.
}
\label{fig:mergeEOS_S}
\end{figure}

We perform the merge in the unit-less entropy density using a linear weight function
that goes smoothly and continuously between the hadron resonance gas entropy ($\sigma_H$)
and the lattice entropy ($\sigma_L$):
\begin{eqnarray}
\sigma(T) &=& \left[1-w(T) \right] \sigma_H(T) +  w(T) \sigma_L(T) \\
w(T) &=& \frac{1}{2} \left[ \tanh{ \left( \tan{  \left[ \frac{\pi}{2} \left(  2\frac{T_L - T}{T_L - T_H} -1  \right) \right]} \right)}  + 1 \right], ~ T_H < T < T_L
\end{eqnarray}
and $w(T)$ is zero for $T<T_H$ and unity for $T_L < T$ and importantly also has zero
slope at both $T_H$ and $T_L$ to maintain continuous speed of sound.
Figure \ref{fig:mergeEOS_S} shows the result of the procedure compared to the original equations of state
and demonstrates that the procedure is continuous and smooth.

\begin{figure}
\centerline{\includegraphics[width=0.75 \textwidth]{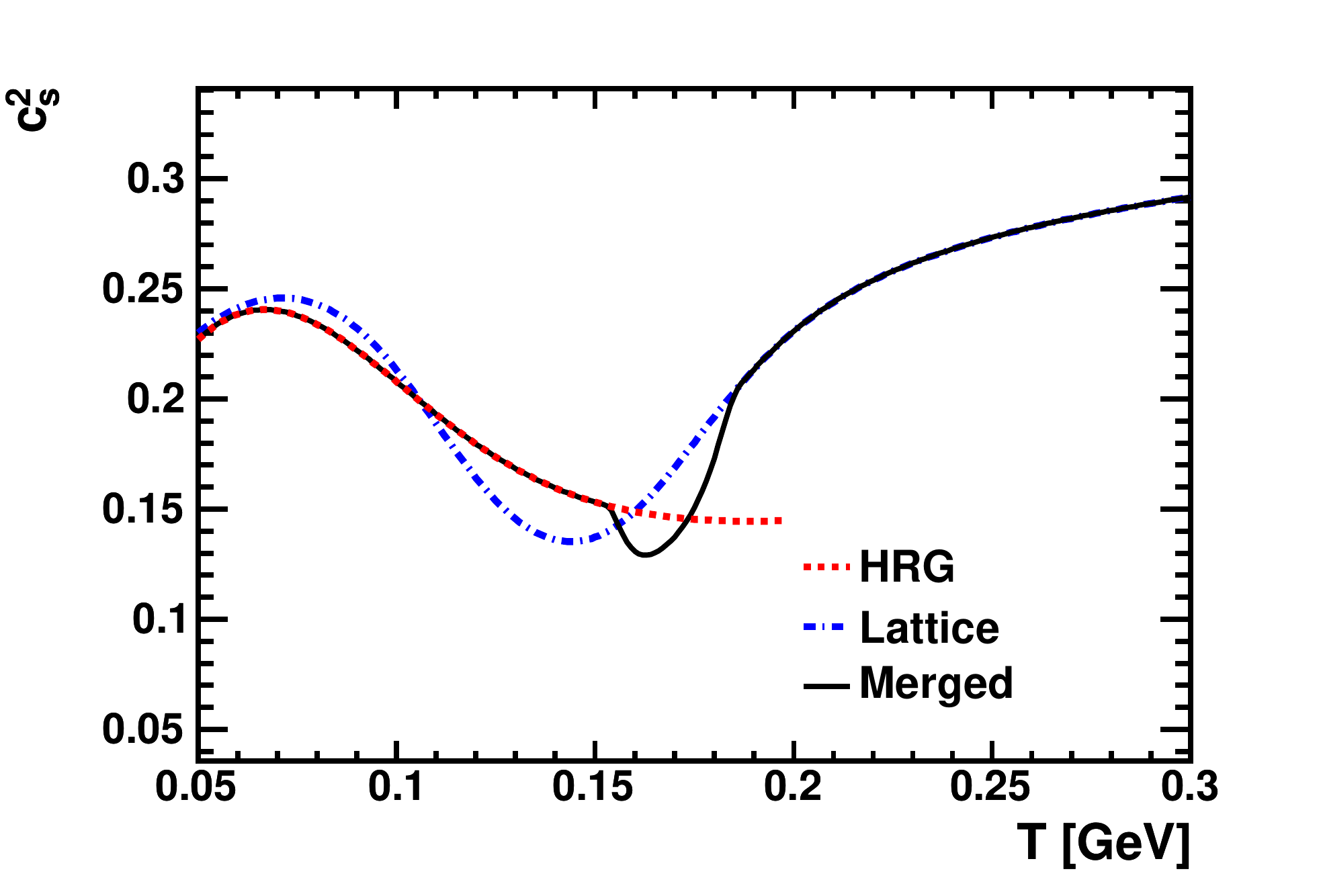}}
\caption{ 
Speed of sound in merging region that results from merging in the unit-less entropy
density via the produce described in the text.  Note that the Lattice data \citep{Borsanyi:2010cj}
does not exactly follow the HRG data below the freezeout temperature due to 
inexact fitting.  Forcing the equation of state to be exactly the HRG below the freezeout
temperature lowers the entropy density at constant temperature and causes a somewhat
more abrupt soft region in the transition region.
}
\label{fig:mergeEOS}
\end{figure}

Like the hadronization temperature,
the lattice-only temperature is a parameter of the model and should not be so large that
gas data is being used at temperatures well above its applicability or so small that the entropy increase
becomes too rapid.  
Figure \ref{fig:mergeEOS} shows the speed of sound squared in the region of the merging
procedure.  Since the entropy of the hadron gas lags behind the lattice, the merging procedure gives a
steeper increase for temperatures just greater than the hadronization temperature.  This manifests as an
even smaller speed of sound in this region.  Lowering the lattice temperature while leaving the hadronization
temperature the same would results in a more extreme dip in the speed of sound squared.  Choosing 
a lattice temperature too close to the hadronization temperature could lead to an extremely soft region
in the equation of state and cause hydrodynamic instabilities. 


For completeness, we note that the entropy density is the temperature derivative of the pressure,
$s = \frac{dP}{dT}$, so knowing the pressure of the hadron gas at $T_H$ and the entropy density
at all temperatures allows one to get the pressure at all temperatures.  Then the energy density is
found from $\epsilon = sT - P$.  
We calculate each quantity at an interval of 1 MeV in temperature and a bookmarked search is 
performed each time a thermodynamic quantity is needed.
Since our model does not solve the Riemann fan and cannot handle shocks, 
we take great care to avoid interpolation errors in the equation of state.  
A cubic spline \citep{Press:1992:NRC:148286} is performed
on each of the thermodynamic variables and the interpolated values from this are used including
the speed of sound squared which must be calculated as part of the splining procedure.  This 
procedure involves a minor amount of preprocessing but the bookmarked search algorithm is unchanged
and a minor part of computational time.

\section{Transport Coefficients} 


As discussed in the previous chapter, viscous hydrodynamics introduces
the effects of non-equilibrium to the collective response of material.  
In the case of shear viscosity, this can be thought of as friction within
the fluid or a finite mean free path, if the underlying theory is a classical
gas.   The form of the correction
is a small change to spatial elements of the stress-energy
tensor due to velocity gradients, for example,
\begin{equation}
\delta T^{xy} = \eta \left( \partial_x v^y + \partial_y v^x \right).
\end{equation}
This can be viewed in terms of Linear Response Theory where
$T^{xy}$ is the operator being varied, $\eta$ is the coefficient
of response, and the velocity gradients are the small field.

For the variation of some operator $A$, the variation of its
expectation value due to the field can be expressed in terms of the variation
of the wavefunction as 
\begin{equation}
\delta \left\langle A \right\rangle = \left\langle \delta \psi | A(0) | \psi \right\rangle + \left\langle \psi | A(0) | \delta \psi \right\rangle = \chi F
\end{equation}
where $\chi$ is a susceptibility and F a small external field.
The change the wave function can be viewed as the result
of some potential, which we assume to be related to the field
by $V(t) = B(t) F$, leading to 
\begin{equation}
| \delta \psi > = \frac{-i}{\hbar} \displaystyle\int\limits_{-\infty}^{0} dt V(t) | \psi_0 > = \frac{-i}{\hbar} \displaystyle \int \limits_{-\infty}^{0} dt B(t) F | \psi_0 > 
\end{equation}
so that the variation of the original operator can be written
\begin{eqnarray}
\delta \left\langle A \right\rangle &=& \frac{-i}{\hbar} \displaystyle\int\limits_{-\infty}^{0} dt \left\langle \left[ A(0), B(t) \right] \right\rangle F = \chi F, \\
\chi &=& \frac{-i}{\hbar} \displaystyle\int\limits_{-\infty}^{0} dt \left\langle [ A(0), B(t) ] \right\rangle. \label{eq:genSuscept}
\end{eqnarray} 

In the case of shear viscosity, we are interested in the response of 
stress-energy to the velocity gradient.  The potential
associated with this is
\begin{equation}
V = \int d^3r T^{0j} r^i \frac{\partial v^j}{\partial r^i},
\end{equation}
where latin indices indicate spatial coordinates.
This form for the potential can be viewed as the change to the energy due to boosting
to the frame of the velocity gradient \citep{Paech:2006st}.
This means that Equation \ref{eq:genSuscept} gives the shear viscosity to be
\begin{equation}
\eta = \frac{-i}{\hbar} \displaystyle\int\limits_{-\infty}^{0} dt \int d^3 r \left\langle \left[ T^{xy}(0), T^{0j}(r,t) \right] \right\rangle r^i.
\end{equation}
Inserting unity in the form $\partial_t t$ and the integrating by parts with the
correlation at minus infinity assumed to be zero, 
\begin{eqnarray}
\eta &=& \frac{-i}{\hbar} \displaystyle\int\limits_{-\infty}^{0}t dt \int d^3 r \left\langle \left[ T^{xy}(0), \partial_tT^{0j}(r,t) \right] \right\rangle r^i, \\
\eta &=& \frac{i}{\hbar} \displaystyle\int\limits_{-\infty}^{0}t dt \int d^3 r \left\langle \left[ T^{xy}(0), T^{xy}(r,t) \right] \right\rangle, \label{eq:kuboComm}
\end{eqnarray}
where we have used the conservation of stress-energy, integrated
by parts in the spatial coordinate disregarding the correlation at 
infinite range, and then noted that symmetry requires that 
some of the stress-energy tensor elements be the same.  
The commutator in Equation \ref{eq:kuboComm} can be eliminated:
\begin{eqnarray}
\displaystyle\int\limits_{-\infty}^{0} t~dt \left\langle \left[ T^{xy}(0), T^{xy}(t) \right] \right\rangle &=& 
\displaystyle\int\limits_{-\infty}^{0} t~dt \left( \left\langle T^{xy}(0) T^{xy}(t) \right\rangle -  \left\langle T^{xy}(0) T^{xy}(-t) \right\rangle \right), \nonumber \\
&=& \displaystyle\int\limits_{-\infty}^{\infty}t~dt \left\langle T^{xy}(0) T^{xy}(t)\right\rangle ,
\end{eqnarray}
where we've ignored the spatial integration for clarity.

Producing the classical Kubo relation from this expression requires
evaluating the thermal average, which is
\begin{eqnarray}
G(t) &=& \left\langle T^{xy}(0) T^{xy}(t) \right\rangle =  \text{Tr} \left[ e^{-\beta H} T^{xy}(0) T^{xy}(t) \right], \nonumber \\
&=& \text{Tr} \left[ e^{-\beta H} ~ T^{xy}(0) ~ e^{iHt/\hbar}~ T^{xy}(0) ~e^{-iHt/\hbar} \right].
\end{eqnarray}
After analytically continuation, $G(z)$ for complex $z$ is symmetric about $z=i\beta \hbar/2$:
\begin{eqnarray}
G(z+ i \beta \hbar/2) &=& \text{Tr} \left[ e^{-\beta H} ~ T^{xy}(0) ~ e^{iHz/\hbar-\beta H/2}~ T^{xy}(0) ~e^{-iHz/\hbar+ \beta H/2} \right],  \label{eq:kubotrace} \\
&=& \text{Tr} \left[ e^{-\beta H} ~ T^{xy}(0) ~ e^{-iHz/\hbar-\beta H/2}~ T^{xy}(0) ~e^{iHz/\hbar+ \beta H/2} \right] = G(-z+i \beta \hbar/2), \nonumber
\end{eqnarray}
where the cyclic symmetry of the trace is used to rotate around $\exp{(-\beta H/2)}$ in Equation \ref{eq:kubotrace}.
Note that this result also means that $G(t+i\beta\hbar) = G(-t)$.  Together these results mean that integration along
the positive real axis can be performed by a complex contour integral.  The integration region is
taken to be rectangular with corners at $z= \{0,i\beta\hbar\}$ and extending to infinity in the positive real
direction.  The region is closed by a curve at infinity which disappears as long as correlations decay 
faster than $t^{-2}$ for large $t$.  The contour integral in full is zero since the contour contains no poles, so
\begin{equation}
0= \oint dz~ (z-i\beta\hbar/2) G(z),
\end{equation}
where the integrand is chosen to be odd about $z=i\beta\hbar/2$.  This leaves only the integral along the 
positive real axis and the return integral along $z=i\beta\hbar$:
\begin{eqnarray}
0 &=& \displaystyle\int\limits_{0}^{\infty} dt~(t- i\beta\hbar/2) G(t) + \displaystyle\int\limits_{\infty}^{0} dt~(t+i\beta\hbar/2) G(t+i\beta\hbar), \nonumber \\
0 &=& \displaystyle\int\limits_{0}^{\infty} dt~(t- i\beta\hbar/2) G(t) - \displaystyle\int\limits_{0}^{\infty} dt~(t+i\beta\hbar/2) G(-t), \nonumber \\
0 &=& \displaystyle\int\limits_{0}^{\infty} dt~t [G(t) - G(-t)] -  \frac{i\beta\hbar}{2} \displaystyle\int\limits_{0}^{\infty} dt~ [G(t) + G(-t)].
\end{eqnarray}
This means that the shear viscosity can be written in terms of an anticommutator instead of a commutator,
\begin{equation}
\eta = \frac{\beta}{2} \displaystyle\int\limits_{0}^{\infty} d^4 r \left\langle \{T^{xy}(0), T^{xy}(r)\} \right\rangle. \label{eq:kuboAntiComm}
\end{equation}

In the classical limit, the stress-energy operators commute and Equation \ref{eq:kuboAntiComm} becomes
\begin{equation}  \label{eq:kubo}
\eta = \beta \displaystyle\int\limits_{0}^{\infty} d^4 r \left\langle T^{xy}(0) T^{xy}(r) \right\rangle,
\end{equation}
which is the familiar Kubo relation.
Larger values of $\eta$ are then due to the persistence of correlations
due to fluctuations and vice versa.  
This is especially clear in the case of a massless gas that loses correlation
exponentially due to a finite time between collisions.  Using 
Equation \ref{eq:ktTxx} for the stress-energy tensor of a massless gas, 
the correlation of stress-energy is 
\begin{equation}
\int d^3x \left\langle T^{xy}(0) T^{xy}(x) \right\rangle = \displaystyle\int\limits \frac{d^3p}{(2\pi)^3}  \left(\frac{p^x p^y}{p_0}\right)^2 e^{-t/t_c-\beta p_0},
\end{equation}
where $t_c$ is the average time between collisions.
This integral and the time integral in the Kubo relation can be performed
analytically and the result is a relationship between the collision time
and the shear viscosity for a massless gas given by $\eta = (4/5) P t_c$
where P is the pressure \citep{citeulike:712984}.  The direct relationship between shear viscosity
and relaxation time confuses the interpretation of the Israel-Stewart equation 
for viscous corrections as a relaxation equation.  Regions where the 
shear viscosity is large are also regions where stress-energy is slow
to respond to changes in velocity gradients.  Other theories predict
different relationships between the shear viscosity and the 
relaxation time, for instance in strongly coupled symmetric Yang-Mills
theory the relaxation time is a factor of two smaller with the same 
temperature dependence.

In principle, Equation \ref{eq:kubo} could be evaluated on the lattice 
to give the shear viscosity of the quark matter for all relevant temperatures
in the hydrodynamic calculation.  Currently, this is
impossible due to numerical noise though such a calculation remains an possibility
given even more computational resources.  
For lower temperatures, it is possible to use kinetic theory
to estimate the shear viscosity.  Viewed in terms of the 
unitless ratio to the entropy density, $\sigma=s/T^3$, this tends to rise very quickly
with decreasing temperatures as is typical for gases below 
their critical temperature -- for instance, jumping from unity
at around T = 100 MeV to five by T = 60 MeV for a massless pion gas \citep{Prakash:1993bt}.
Such temperatures are by design outside of the hot region 
and tend to have a small effect on the output from the hydrodynamic
module.  We investigated the effect of changing the
behavior of shear viscosity at temperatures below the freezeout 
temperature combined with a constant shear viscosity to entropy
density ratio at high temperature,  $\eta/s = 0.2$.
We considered three low temperature scenarios:  constant $\eta/s$,
constant $\eta/\epsilon$ below T=130 MeV, and $\eta/s$ increasing
as a Fermi function up to 0.6 at low temperature.  This is still less 
than the value predicted by the hadron resonance gas model but 
still runs stably for reasonable parameter sets.  None of these
runs showed significant differences in the produced elliptic flow
which justifies the choice to exclude this correction from the set of 
first parameters to investigate.  Other studies in this area have found
that low temperature shear viscosity can have an effect if it is allowed 
to rise for temperatures slightly above the freezeout temperature \citep{Song:2010aq}
most likely due to phase space effects during particle generation as 
others have found that high temperature shear viscosity scaling
matters much more \citep{Niemi:2012ry}.

In the high temperature region, we also expect the shear viscosity
to entropy density ratio to rise.  A scaling argument for this temperature
dependence in perturbative QCD has been calculated to full
leading-order for three flavors of massless quarks
\begin{equation}
\frac{\eta}{s} = \frac{5.12}{ g^4 \ln (2.42/g)}
\end{equation}
which can be combined with a two-loop renormalization group expression
for the running coupling
\begin{equation}
\frac{1}{g^2(T)} = \frac{9}{8\pi^2} \ln \left( \frac{T}{\Lambda_T} \right) + \frac{4}{9 \pi^2} \ln \left( 2 \ln \left( \frac{T}{\Lambda_T} \right)  \right)
\end{equation}
with $\Lambda_T = 30$ MeV \citep{Csernai:2006zz,Huot:2006ys}.
From this we take only the most basic result:  the shear viscosity
to entropy density ratio should increase like the square of the logarithm
of the temperature.  Assuming that this combines continuously 
with the shear viscosity at the critical temperature, which we also
take as a parameter, we add one more parameter for the slope
of increase in shear viscosity to entropy density in the liquid phase
\begin{equation}
\eta/s =  \left. \eta/s \right|_{T_c} + \alpha \ln (T).
\end{equation}
Large values of $\alpha$ have a significant effect on observables and will
be discussed in a later chapter.

A reasonable estimate for the bulk viscosity is more difficult to 
generate.  The methods used to estimate the shear viscosity or
its scaling properties rely on calculations from models either
of massless gases or a collection of mass states
where the bulk viscosity is nearly zero.  However, near the 
critical region, bulk viscosity can be finite due to the changing
of the sigma mass or effectively be non-zero due to loss of 
chemical equilibrium \citep{Paech:2006st}.
 The structure and size of such a peak in the 
bulk viscosity are not well constrained theoretically and 
the width, height, and location of this peak  are introduced 
as parameters to the model.  These parameters are not 
explored within this work but are good candidates for future
exploration.

\chapter{TRISH -- A Three-Dimensional Israel-Stewart Hydrodynamics Algorithm}

In the first two chapters, we developed hydrodynamic equations of motion
and justified their use in the description of the quark matter created
in heavy ion collisions. Since hydrodynamics requires an equation
of state and transport coefficients from the underlying microscopic
theory, these were calculated in the immediately previous chapter.
With these results in hand, we can now discuss the algorithm to evaluate
viscous hydrodynamics for heavy ion collisions can be 
discussed.  This chapter concerns the algorithm itself 
including the coordinate system and integration scheme.
We then discuss the verification of the algorithm by checking conserved quantities
and comparing to known results.  Finally, we discuss the 
process by which the calculation is coupled to the gas 
calculation that will be used at lower temperatures.
This section focuses on the key issues of finding the emission surface and
populating particle states once the surface is known.

\section{Description of Hydrodynamic Algorithm}

\subsection{Coordinates, Variables, and Integration Scheme}

As discussed in Chapter 1, the Bjorken ansatz that the 
initial longitudinal distribution does not depend on rapidity for small rapidity
produces a Hubble-like expansion that hydrodynamics maintains.  Even for a 
finite system where longitudinal density gradients spoil the symmetry, this 
expansion dominates.  For this reason, the algorithm is written in the
coordinate system of this expansion.  To refresh the memory, 
lab time and longitudinal position are replaced by proper time and spatial rapidity,
\begin{equation}
\tau^2 = t^2 - z^2 ,  ~~ \tanh \eta = z/t.
\end{equation}
The integration of the partial differential equations of Israel-Stewart hydrodynamics
will take place on an Eulerian grid, meaning that the grid points are at fixed locations
in the $\eta$-$\tau$ coordinate system.

Since the shear tensor is an independent variable in the Israel-Stewart framework, 
the ten components of the stress energy tensor not given by symmetry are all independent.
In principle, the equations of motion could be written in terms of these.  Instead, we
choose a set of variables more closely related to the original formulation.  
For instance, ideal hydrodynamics is naturally interpreted in terms of the energy density
and three components of collective velocity.
From these, the stress energy tensor can then be computed via
\begin{equation} \label{eq:THydro}
T^{\mu \nu} = (\epsilon + P) u^\mu u^\nu - g^{\mu \nu} P + \Pi^{\mu \nu},
\end{equation}
where $\Pi^{\mu \nu}$ encapsulates both shear and bulk viscous corrections, which 
would be zero for ideal hydrodynamics.
A common alternative in ideal hydrodynamics is to integrate for the temporal part of the stress energy tensor
($T^{\mu t}$), which has the benefit of simple equations of motion and a long history
of numerical methods to solve conservation equations stably and accurately. \citep{shasta,ktHydro}
However, Equation \ref{eq:THydro} for $T^{\mu \nu}$ is not invertible for $u^\mu$, 
so recovering the collective velocity requires a root find.  
This is an acceptable cost if it can be done once and stored but doubles the memory cost.
This makes it attractive for ideal hydrodynamic codes even with large grids but 
the larger set of variables make the memory footprint a more important consideration.

Furthermore, while more complex integration methods for conservation equations
are an integral part of computational fluid dynamics, it is not clear
that they are necessary for a strategy with physical viscosity.
Instead we choose a simple and quick integration scheme that is easily multithreaded
under the assumption that physical viscosity and smooth initial conditions
render such complications unnecessary.  Even for these conditions, shear viscosity
of around half the entropy density often leads to instability as the correction terms
get large.  Also, the peak in bulk viscosity near the phase transition
can produce shock waves which are not treated by our integration method though
for small bulk viscosity the integration is stable.

For the viscous part of the evolution, we choose to track a projected version of the shear
tensor in the frame of the matter.  This complicates the derivation of the equations 
of motion as the evaluation of derivatives in this frame requires Lorentz transformations.
In addition, the usual geometrical methods of computing corrections for the motion
of the coordinate system must be made using small boosts, the details of which are
presented in the next subsection.
The choice of the local tensor is motivated by the more natural comparison with 
ideal hydrodynamics -- shear corrections can be compared to the pressure without kinematic
corrections. This is especially important if one seeks to damp large corrections to 
the shear tensor in the presence of large velocity gradients where corrections should
be restricted to remain smaller than the pressure \citep{Pratt:2007gj}.

The shear tensor has only five independent components due to symmetry, tracelessness,
and orthogonality to the collective velocity.  Therefore, we 
take a projection that is inspired by spherical harmonics,
\begin{eqnarray}
a_1 = \frac{1}{2} \left( \tilde{\pi}^{xx} - \tilde{\pi}^{yy}\right);  && ~~~~  a_2 = \frac{1}{2 \sqrt{3}} \left( \tilde{\pi}^{xx} + \tilde{\pi}^{yy} - 2 \tilde{\pi}^{zz} \right) ; \label{eq:defineAs} \\
a_3 = \tilde{\pi}^{xy};  &&  a_4 = \tilde{\pi}^{xz} ;  ~~~~~ a_5 = \tilde{\pi}^{yz};  \nonumber \\
 b &=&  (1/3) \left[ \tilde{\pi}^{xx} + \tilde{\pi}^{yy} + \tilde{\pi}^{zz} \right] . \nonumber
\end{eqnarray}
In this space, $a_1$ is the difference in effective pressure between the x- and y-directions,
and $a_2$ is the difference between the longitudinal direction and the average of the 
transverse directions.
Therefore, an infinite transverse system undergoing a Bjorken expansion, only $a_2$ is non-zero,
and even when including non-trivial transverse expansion,
$a_2$ will be the strongest shear term since it quantifies the extent to which the 
longitudinal expansion proceeds more rapidly than the transverse expansion.
The remaining three still correspond to the shear along the surfaces of the fluid element.
Since time derivatives in the Israel-Stewart equations always appear as scaled versions
of a fluctuation, we use this scaled anisotropy:
\begin{equation}
\alpha_i = a_i/\sigma_\eta; ~~ \beta = b/\sigma_\zeta. \label{eq:projShear}
\end{equation}
The fluctuations are functions only of the energy density, as discussed 
in Equation \ref{eq:piVariance}, and can therefore easily be calculated
for each fluid cell at the same time as the other transport coefficients.
Since the stress-energy tensor has ten independent elements after symmetry constraints in the viscous case,
these six elements combined with the energy density and the collective velocity vector are 
sufficient to describe the stress-energy tensor completely.
The equations of motion are then written in terms of time derivatives of these ten variables exclusively,
the derivation of which is the thrust of Section 4.1.4.

It is then convenient to rewrite the velocity gradients to which the shear tensor
relaxes in terms of the these projections.  Simply applying evaluating these 
linear combinations of velocity gradients yields
\begin{eqnarray}  \label{eq:projGrads}
\omega_1 = \tilde{\partial}_x n^x - \tilde{\partial}_y n^y, && \omega_2 = \frac{1}{\sqrt{3}} \left[ \tilde{\partial}_x n^x + \tilde{\partial}_y n^x - 2 \tilde{\partial}_z n^z \right], \\
\omega_3 = \tilde{\partial}_x n^y + \tilde{\partial}_y n^x, && \omega_4 = \tilde{\partial}_x n^z + \tilde{\partial}_z n^x, ~~~~ \omega_5 = \tilde{\partial}_y n^z + \tilde{\partial}_z n^y
\end{eqnarray}
so that Equation \ref{eq:MaxCatt} can be rewritten as 
\begin{equation} 
\tau_\pi \tilde{\partial}_t \alpha_i = - \alpha_i - \frac{\eta}{\sigma_\eta} \omega_i.
\end{equation}

\subsection{Evaluation of Local Derivatives}

The design of our program calls for evaluation of local derivatives of the velocity
relative to the expansion of the mesh, $\tilde{\partial}_i u^j$, and of the local shear tensor 
-- $\tilde{\partial}_i \tilde{\pi}^{jk}$ -- where tildes indicate quantities in the local frame.
We leave discussion of the expansion corrections to the next subsection and focus on
corrections for the motion of the fluid frame.  Our goal is to begin with the equations of
motion in the fluid frame, and boost such that we have equations of motion for our chosen
set of variables in terms of derivatives available in the mesh frame.

We will make frequent use of a general form for the boost from a frame that observes 
a velocity $u^\mu$ to one that observes velocity $n^\mu$: 
\begin{equation}
\Lambda^{\mu \nu} (u \rightarrow n) = g^{\mu \nu} + 2 n^\mu u^\nu - \frac{ (u^\mu + n^\mu) ( u^\nu + n^\nu)}{1+u \cdot n}.  \label{eq:genBoost}
\end{equation}
which fulfills the requirement that $\Lambda^{\mu \nu} u_\nu = n^\mu$.

We begin by evaluating the local derivative of the mesh frame velocity.  This requires boosting
both quantities, $\tilde{\partial}_\alpha n^\beta = \Lambda_\alpha^{~\mu} \Lambda ^{\beta \nu} \partial_\mu u^\nu$,
where $n^\mu = $ \{1,0,0,0\} is the local frame velocity and $u^\mu$ is the mesh velocity.
Considering first the boost to the derivatives 
\begin{eqnarray}
\tilde{\partial}_\mu &=&  \Lambda_\mu^{~\nu} \partial_\nu , \nonumber \\
\tilde{\partial}_\mu &=&  \left[  g_\mu^{~\nu} + 2 n_\mu u^\nu - \frac{ (u_\mu + n_\mu) ( u^\nu + n^\nu)}{1+u \cdot n} \right]  \partial_\nu , \nonumber \\
\tilde{\partial}_\mu &=& \partial_\mu + 2 n_\mu u^\nu \partial_\nu - \frac{u_\mu + n_\mu}{1 + \gamma} \left(u^\nu \partial_\nu + n^\nu \partial_\nu \right), \label{eq:boostDerivs}
\end{eqnarray}
where $\gamma = u_0$ is the Lorentz factor.
Equation \ref{eq:boostDerivs} is neatly divided into local time derivatives and local spatial derivatives
\begin{equation}
\tilde{\partial}_t = \gamma \partial_t + u^i \partial_i, ~~
\tilde{\partial}_i = \partial_i - \frac{u_i}{1+\gamma} \left( u^\nu \partial_\nu + \partial_t \right),
\end{equation}
where the latin index is used to indicate a spatial index.  If $u^i u^j$ is small $\forall i,j$, this yields the 
usual comoving derivative expressions 
\begin{equation}
\tilde{\partial}_t = \partial_t + u^i \partial_i, ~~~ \tilde{\partial}_i = \partial_i - u^i \partial_t.
\end{equation}

Likewise, obtaining derivatives of the local velocities just requires a boost
\begin{eqnarray}
\tilde{\partial_\alpha} n^\beta & = & \Lambda^\beta_\nu \tilde{\partial}_\alpha \text{\emph{u}}^\nu ,  \nonumber \\
&=& \left[ g^\beta_\nu + 2 n^\beta u_\nu - \frac{(u^\beta + n^\beta)(u_\nu + n_\nu)}{1+\gamma}  \right] \tilde{\partial}_\alpha u^\nu , \\
&=& \tilde{\partial}_\alpha u^\beta - \frac{u^\beta + n^\beta}{1+\gamma} \tilde{\partial}_\alpha \gamma, \label{eq:localVelDeriv}
\end{eqnarray}
where we have made use of the fact that 
\begin{equation}
2 u_\mu \partial_\nu u^\mu = \partial_\nu \left( u_\mu u^\mu \right) = \partial_\nu (1) = 0 \label{eq:ZeroU2}
\end{equation} 
assuming that the derivative is a total derivative and coordinate system effects (affine connections) are included.
A useful conclusion of Equation \ref{eq:ZeroU2} is that one can convert derivatives of the Lorentz factor into 
derivatives of the velocities via
\begin{equation}
\gamma \partial_\mu \gamma = - u^i \partial_\mu u^i 
\end{equation}
for any complete derivative $\partial_\mu$.
We will tend to leave derivatives of the Lorentz factor in equations and it is understood that they will be converted into 
derivatives of the velocities using this relation.

Combining Equations \ref{eq:boostDerivs} and \ref{eq:localVelDeriv}, we obtain an expression for local spatial derivatives of local velocities
\begin{eqnarray}
\tilde{ \partial}_i n^j & = & \partial_i u^j + u^i   \dot{u}^j + \frac{u^i u^k}{1+\gamma} \partial_k u^j \nonumber \\
&&  - \left( \frac{  u^j u^k}{\gamma ( 1+ \gamma) } \right) \left[ \partial_i u^k + u^i \dot{u}^k + \frac{u^i u^m}{1+\gamma} \partial_m u^k  \right] ,\label{eq:locSpDerivLocVelo}
\end{eqnarray}
where summations over the latin indices are implied and over-dots indicate time derivatives in the mesh frame.

The expansion rate, $\partial_\mu u^\mu$, is a Lorentz scalar and should be conserved under Lorentz transformations.
We calculate the local time derivative of the Lorentz factor, $\tilde{\partial}_0 n^0$, to confirm this.
Using \ref{eq:localVelDeriv}, 
\begin{equation}
\tilde{\partial}_0 n^0 = \tilde{\partial}_0 \gamma - \left[ \frac{1+\gamma}{1+\gamma} \right] \tilde{\partial}_0 \gamma  = 0, \label{eq:n0LF}
\end{equation}
which is expected since, to lowest order in velocity, the Lorentz factor is proportional to velocity squared for small velocities.
In fact, nothing about Equation \ref{eq:n0LF} changes for spatial derivatives, which are also zero.
If one then sums over spatial indices in Equation \ref{eq:locSpDerivLocVelo},
\begin{eqnarray}
\tilde{\partial}_i n^i &=& \partial_i u^i + \left[ \gamma - \frac{u^i u^i}{1+\gamma} \right] \partial_0 \gamma + \left[ -1 - \frac{u^i u^i}{1+\gamma} + \gamma \right] \frac{1}{1+\gamma} u^j \partial_j \gamma \nonumber \\
\tilde{\partial}_i n^i &=& \partial_0 \gamma + \partial_i u^i \label{eq:traceLocVeloDeriv}
\end{eqnarray} 
where the last step makes use of the identity $u^i u^i = \gamma^2 - 1 = (\gamma+1)(\gamma-1)$.
Combining Equations \ref{eq:locSpDerivLocVelo} and \ref{eq:traceLocVeloDeriv} confirms that the expansion rate is invariant.

The only remaining velocity derivative to compute is the time derivative of the spatial components, which are
\begin{equation}
\tilde{\partial}_0 n^i = \dot{u}^i - \frac{u^i}{1+\gamma} \dot{\gamma} + u^j \partial_j u^i - \frac{u^i u^j}{1+\gamma} \partial_j \gamma.  \label{eq:locTempDerivLocVelo}
\end{equation}
For convenience in writing the equations of motion in a subsequent subsection, we will condense the spatial derivatives 
\begin{eqnarray}
\tilde{\partial}_0 n^i &=& \dot{u}^i - \frac{u^i}{1+\gamma} \dot{\gamma} + \left(\tilde{\partial}_0 n^i \right)', \nonumber \\
\tilde{ \partial}_i n^j & = &  u^i \dot{u}^j -  \left( \frac{  u^j u^k}{\gamma ( 1+ \gamma) } \right)  u^i \dot{u}^k  + \left(  \tilde{ \partial}_i n^j \right)'. \label{eq:localSpDeriv}
\end{eqnarray}

Calculating derivatives of the local shear tensor differs somewhat from the previous as the frame of the shear tensor is different
for each fluid element.  Therefore, we introduce a boost from a frame that observes a slightly different collective velocity to the
fluid frame, which we denote $\delta \Lambda^{\mu \nu} (u - \delta u , n) $.  Keeping terms only linear in the difference, we obtain
\begin{eqnarray}
\delta \Lambda^{\mu \nu} (u - \delta u , n) & = & g^{\mu \nu} + 2 n^\mu (u^\nu - \delta u^\nu) \nonumber \\ 
&& - \frac{ (u^\mu - \delta u^\mu + n^\mu) (u^\nu - \delta u^\nu + n^\mu)}{1 + \gamma - \delta \gamma} - \Lambda(u,n) , \nonumber \\
\delta \Lambda^{\mu \nu} (u - \delta u , n)&=& -2 n^\mu \delta u^\nu + \frac{ (u^\mu + n^\mu ) \delta u^\nu}{1 + \gamma} \nonumber \\
&& + \frac{ \delta u^\mu  (u^\nu + n^\nu )}{1 + \gamma}  - \frac{\delta \gamma}{(1+\gamma)^2} (u^\mu + n^\mu) ( u^\nu + n^\nu). \label{eq:SmallBoost}
\end{eqnarray}
This method should be able to reproduce Equation \ref{eq:localVelDeriv} when contracting with the collective velocity
\begin{equation}
\delta \Lambda^\mu_\alpha u^\alpha = \delta u^\mu - \frac{u^\mu + n^\mu}{1+\gamma} \delta \gamma ,
\end{equation}
where we have made frequent use of orthogonality relation $u^\mu \delta u_\mu = 0$.  

We wish to obtain derivatives calculated in the frame of the matter, so this small boost should be boosted back to the frame
of the matter, which gives
\begin{eqnarray}
\delta \Lambda^{\mu \alpha} (u - \delta u , n)  \Lambda_{\alpha}^\nu (n,u)
&=& 2 (n^\mu \delta u^\nu - \delta u^\mu u^\nu) 
+ \frac{2 \delta \gamma}{1+\gamma} \left[u^\mu n^\nu - n^\mu u^\nu \right] \nonumber \\
&& + \frac{1}{1+\gamma} \left[ \delta u^\mu (u^\nu + n^\nu) - (u^\mu +n^\mu) \delta u^\nu \right] .
\end{eqnarray}
To boost a tensor, this must be applied symmetrically to each index.  Again keeping on linear terms,
we apply this to the local shear tensor (for which we will neglect the tildes) yielding
\begin{eqnarray}
\delta \pi^{\mu \nu} &=&  \delta \Lambda^{\mu \alpha}  \Lambda_{\alpha \beta} \pi^{\beta \nu}  + \delta \Lambda^{\alpha \nu} \Lambda_{\alpha \beta} \pi^{\mu \beta} \nonumber \\
\delta \Lambda^{\mu \alpha}  \Lambda_{\alpha \beta} \pi^{\beta \nu} & = &  2  n^\mu \delta u_\alpha \pi^{\alpha \nu} - \frac{2 \delta \gamma}{1+\gamma} n^\mu u_\alpha \pi^{\alpha \nu} \nonumber \\
&&+ \frac{1}{1+\gamma} \left[ \delta u^\mu u_\alpha \pi^{\alpha \nu} - (u^\mu + n^\mu) \delta u_\alpha \pi^{\alpha \nu} \right]
\end{eqnarray}
where we have used the fact that the local shear tensor has no temporal components and so $n_\mu \pi^{\mu \nu} = 0$.

While the local shear tensor has no temporal components, those components have non-zero derivatives.
We find that while $\delta \pi^{00} =0$ as one might expect, the momentum-like term is not:
\begin{eqnarray}
\delta \pi^{0 i} &=& \delta \Lambda^{0 \alpha} \Lambda_{\alpha \beta} \pi^{\beta i} + \delta \Lambda^{i \alpha} \Lambda_{\alpha \beta} \pi^{\beta 0} , \nonumber \\
&=& \delta u_\alpha \pi^{\alpha i} - \frac{\delta \gamma}{1+\gamma} u_\alpha \pi^{\alpha i} = - \pi^i_\alpha \delta n^\alpha. \label{eq:smallBoostPi0i}
\end{eqnarray}
The final step is not useful for the calculation but is a self-consistency check which can be seen by differentiating the orthogonality of
the shear tensor and the local velocity
\begin{equation}
0 = \partial_\mu \left( \pi^{\alpha \nu} n_\alpha \right) = \pi^{\nu}_{~\alpha} \partial_\mu n^\alpha + n_\alpha \partial_\mu \pi^{\alpha \nu},
\end{equation}
which is equivalent to Equation \ref{eq:smallBoostPi0i}.

The same calculation for two spatial indices yields
\begin{equation}
\delta \pi^{ij} = \frac{1}{1+\gamma} \left[ \delta u^i u_\alpha \pi^{\alpha j} + \delta u^j u_\alpha \pi^{\alpha i} - u^i \delta u_\alpha \pi^{\alpha j} - u^j \delta u_\alpha \pi^{\alpha i} \right]. \label{eq:piCorr}
\end{equation}
As a simple sample computation, the Israel-Stewart equation of motion contains time derivatives of the spatial
part of the local shear tensor - $\tilde{\partial}_0 \pi^{ij}$. Even excluding the part of this proportional to spatial 
derivatives, Equation \ref{eq:piCorr} means that the motion of the reference frame gives corrections for time derivatives.
Therefore, one finds the adjustment
\begin{equation}
\partial_0 \pi^{ij} \rightarrow \dot{\pi}^{ij} + \frac{1}{1+\gamma} \left[  u^i \dot{u}^k \pi^{kj} + \pi^{ik} \dot{u}^k u^j - \dot{u}^i u^k \pi^{kj} - \pi^{ik} u^k \dot{u}^j   \right].
\end{equation}
Corrections of this type are needed for all derivatives of the local shear tensor.

\subsection{Expansion Corrections }

The small boost notation developed at the end of the previous section also proves useful for describing 
the effects of the mesh expansion.  Recall that our calculation takes places on a fixed grid in the coordinates
\{$\tau,x,y,\eta$\}, which expand relative to the lab coordinates \{$t,x,y,z$\}.  
One writes equations of motion for the variables as they are observed in a frame
moving with the expanding coordinate system.  The relationship between velocities
observed in the laboratory frame ($v^\mu$) and the mesh frame ($u^\mu$) is
\begin{equation}
v^\mu = \{\gamma \cosh{\eta} + u^z \sinh{\eta}, u^x, u^y, u^z \cosh{\eta} + \gamma \sinh{\eta}  \} ,
\end{equation}
where $\gamma = u^0$ is the Lorentz factor in the mesh frame
Then for a cell at $\eta = \eta + \delta \eta$, $\gamma = \gamma + u^z \delta \eta$ and $u^z = u^z + \gamma \delta \eta$.  
This means that longitudinal derivatives of velocities require correction, for instance $\partial_\eta u^z \rightarrow \partial_\eta u^z + \gamma$.

Using the small boosts, we can write these corrections to mesh frame quantities in a simple form
\begin{equation}
\delta \Lambda^{\mu \nu} = n^\mu \eta^\nu - \eta^\mu n^\nu, ~~n^\mu = \{1,0,0,0\}, ~~ \eta^\mu= \{0,0,0,-\delta \eta\}.
\end{equation}
This form reproduces the above result for the mesh frame velocity
\begin{equation}
\delta \Lambda^\mu_{~\alpha} u^\alpha  = (n^\mu \eta_\alpha - \eta^\mu n_\alpha) u^\alpha = n^\mu (u \cdot \eta) - \gamma \eta^\mu = \delta^{\mu 0} u^z \delta \eta + \delta^{\mu z} \gamma \delta \eta.
\end{equation}
Na\"{i}vely, one might expect that this correction could be applied directly to 
quantities measured in the matter frame to correct for the motion of the mesh.
This turns out not to be the case since the small boost to a nearby mesh frame
does not commute with the boost between the matter frame.

\begin{figure}[fig:boostChart]
\centerline{\includegraphics[width=0.7 \textwidth]{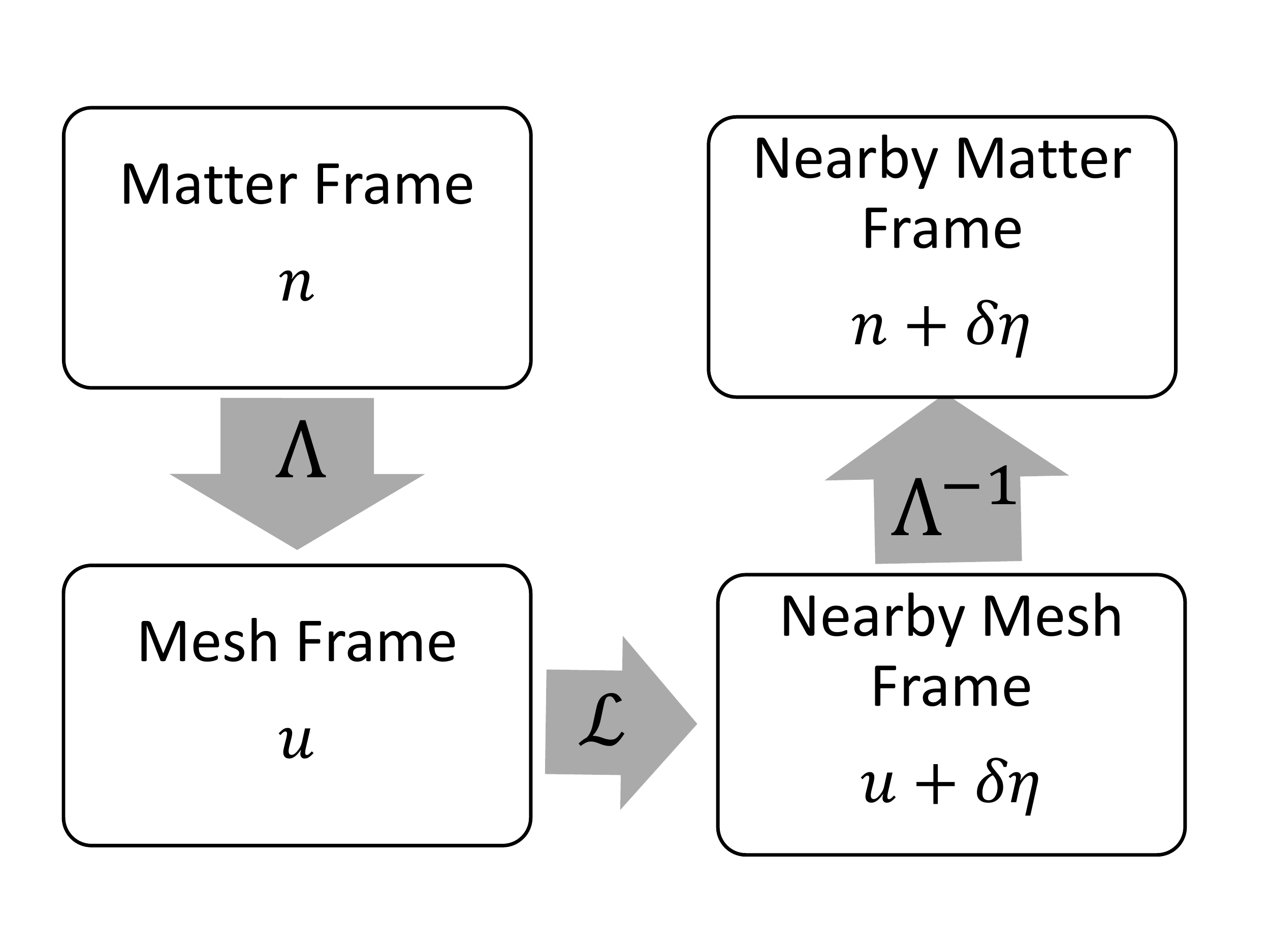}}
\caption{ Schematic outline of the procedure for correcting matter frame quantities for the 
motion of the mesh frame.  The matter frame velocity is first boosted to the mesh frame,
then a small correction for the longitudinal expansion is added, and then the result is boosted
back to the matter frame.  Since $\mathcal{L}$ does not commute with $\Lambda$ all three
must be evaluated to calculate $n + \delta \eta$.
}
\label{fig:boostChart}
\end{figure}

This non-commutation of the frame boosts complicates matters somewhat
and one needs to apply all three boosts sequentially, keeping corrections
to first order in $\delta \eta$.  This procedure is described in Figure \ref{fig:boostChart}
for mesh expansion corrections to the fluid velocity.  If $\mathcal{L}$ commuted 
with $\Lambda$ to first order in $\delta \eta$, one could apply $\mathcal{L}$ directly 
to $n$ to obtain the expansion corrections.

For derivatives of quantities in the fluid frame, we need to boost the correction,
$\mathcal{L} = \delta \Lambda$, into the fluid frame.
Using again Equation \ref{eq:genBoost},
\begin{eqnarray}
\Lambda^{\mu \alpha}(u,n) \Lambda^{\nu \beta} (u,n) \delta \Lambda_{\alpha \beta} &=& \Lambda^{\mu \alpha} \Lambda^{\nu \beta} (n_\alpha \eta_\beta - \eta_\alpha n_\beta) , \nonumber \\
(\Lambda \Lambda \delta \Lambda)^{\mu \nu} &=& (2 \gamma n^\mu - u^\mu ) ( \eta^\nu + 2 (u \cdot \eta) n^\nu - \frac{u \cdot \eta}{1+\gamma} [u^\nu + n^\nu] ) \nonumber \\
&& - (\eta^\mu + 2 (u \cdot \eta) n^\mu - \frac{u \cdot \eta}{1+\gamma} [ u^\mu + n^\mu] ) (2 \gamma n^\nu - u^\nu) .
\end{eqnarray}
Note that $\Lambda^{\nu \beta}(u,n)$ boosts from the mesh frame to the matter frame 
and that the boosts are not written in the order of application.

Applying this to the matter frame velocity, we obtain several results
\begin{eqnarray}
(\Lambda \Lambda \delta \Lambda)^0_{~\alpha} n^\alpha &=& \gamma (u \cdot \eta) - \gamma (u \cdot \eta) = 0, \\
(\Lambda \Lambda \delta \Lambda)^x_{~\alpha} n^\alpha &=& - u^x (u \cdot \eta) + \frac{\gamma}{1+\gamma} u^x (u \cdot \eta) = \frac{-u^x}{1+\gamma} u^z \delta \eta, \label{eq:longCorrDzNx} \\
(\Lambda \Lambda \delta \Lambda)^z_{~\alpha} n^\alpha &=& - u^z (u \cdot \eta) - \gamma \left[ \eta^z - \frac{u^z}{1+\gamma} (u \cdot \eta) \right] = \left[ \gamma - \frac{(u^z)^2}{1+\gamma} \right] \delta \eta,
\end{eqnarray}
and a few results for the shear tensor
\begin{eqnarray}
(\Lambda \Lambda \delta \Lambda)^x_{~\alpha} \tilde{\pi}^{\alpha \nu} &=& - u^x \delta \eta \tilde{\pi}^{z \nu}, \\
(\Lambda \Lambda \delta \Lambda)^z_{~\alpha} \tilde{\pi}^{\alpha \nu} &=& - (u^z \tilde{\pi}^{z \nu} -  u_\alpha \tilde{\pi}^{\alpha \nu}) \delta \eta = (u^x \tilde{\pi}^{x \nu} + u^y \tilde{\pi}^{y \nu} ) \delta \eta.
\end{eqnarray}
Applying the small boost to each index is not necessary as this would produce a second-order correction.
As was the case for the collective velocity, these correction factors alter the longitudinal derivatives
of the local shear tensor.  For example, Equation \ref{eq:longCorrDzNx} means the longitudinal
derivative of the matter frame transverse velocity, $\partial_\eta n^x$, should be calculated by
the finite difference in the longitudinal direction in addition to the mesh factor: $-u^x u^z/(1+\gamma)$.

\subsection{Equations of Motion} 

Since we wish to use the local shear tensor, our derivation of the equations of motion
begin in the fluid frame.  In this frame, the conservation equations take the form
\begin{eqnarray}
\tilde{\partial_t} \epsilon + (\epsilon+P) ~\tilde{\partial_i} n^i + \tilde{\pi}^{ij} ~ \tilde{\partial}_i n^j  &=& 0 ,\\
(\epsilon + P) ~ \tilde{\partial_t} n^i + \tilde{\pi}^{ij} ~ \tilde{\partial_t} n^j + \tilde{\partial}_i P + \tilde{\partial_j} \tilde{\pi}^{ij} &=& 0 ,
\end{eqnarray}
where we have excluded terms that are explicitly zero in this frame.
For reasons explained in the following subsection,
we expand all time derivatives but leave some spatial derivatives in the form of Equation \ref{eq:localSpDeriv}.
We exclude corrections for bulk viscosity from the conservation equations for simplicity as the 
corrections alter the pressure but do not add any interesting structure.  

Expanding energy conservation in this way produces
\begin{eqnarray}
\gamma \dot{\epsilon} + \frac{(\epsilon+P) u^i}{\gamma} \dot{u}^i + (u^i \tilde{\pi}^{ij})  \dot{u}^j - \frac{ (u^i u^j \tilde{\pi}^{ij}) u^k}{\gamma(1+\gamma)} \dot{u}^k &&\\
+ u^i \partial_i \epsilon + (\epsilon + P)  \partial_i u^i + \tilde{\pi}^{ij} ( \tilde{\partial}_i n^j )' &=& 0  \nonumber
\end{eqnarray}
where the second term of the first line comes from $\dot{\gamma}$ and final two terms of the first line come from $\tilde{\partial}_i n^j$.

The same procedure for momentum conservation gives
\begin{eqnarray}
 u^i c_s^2 \dot{\epsilon} + (\epsilon+P) \gamma \dot{u}^i - \frac{ (\epsilon+P) u^i u^j}{1+\gamma} \dot{u}^j + \gamma \tilde{\pi}^{ij} \dot{u}^j && \\
+ \frac{ u^k }{1+\gamma} \left[ u^i \tilde{\pi}^{kj} - 2 \tilde{\pi}^{ik} u^j  \right]  \dot{u}^j+ (\gamma -1) \tilde{\pi}^{ij} \dot{u}^j   + u^j \dot{ \tilde{\pi}}^{ij} && \nonumber \\
+ (\epsilon + P) (\tilde{\partial}_t n^i)' + \tilde{\pi}^{ij} (\tilde{\partial}_t n^j)' + \partial_i P + \frac{u^i u^k}{1+\gamma} \partial_k P  + (\tilde{\partial}_j \tilde{\pi}^{ij})' &=& 0 \nonumber.
\end{eqnarray}
where $(\tilde{\partial}_j \tilde{\pi}^{ij})' $ includes the effects of boosting the derivative and the effects of the collective velocity on the shear tensor in
nearby fluid cells.  This form of the momentum conservation equations include time derivatives of the local shear tensor, 
which are not among the integrated variables.  Conversion of these to time derivatives of $\alpha_i$ makes the equations even more difficult
to parse, and the conversion takes the simple form
\begin{equation}
\dot{\pi}^{ij} = \partial_\tau \left[ \sigma_\eta P_{ijk} \alpha_k \right] = P_{ijk} \alpha_k \frac{\partial \sigma_\eta}{\partial \epsilon} \dot{\epsilon} + \sigma_\eta P_{ijk} \dot{\alpha_k}
\end{equation}
where $P_{ijk}$ projects out the correct $\alpha_k$'s according to the inverse mapping corresponding to Equation \ref{eq:defineAs}.

The remaining equations of motion are the relaxation equations for the shear tensor.  We rewrite
Equation \ref{eq:MaxCatt} in terms of the scaled, projected moments of the shear tensor,
\begin{equation} 
\tau_\pi \tilde{\partial}_t \alpha_i = - \alpha_i - \frac{\eta}{\sigma_\eta} \omega_i
\end{equation}
where $\omega_i$ is the projection of the Navier-Stokes velocity gradients described in Equation \ref{eq:projGrads}.
As an example, for the first projected shear stress element this is
\begin{equation}
\gamma   \dot{\alpha}_1 +u^i \partial_i \alpha_1  + \frac{\alpha_1}{ \tau_\pi} = \frac{ -\eta}{\sigma_\eta \tau_\pi} (\tilde{\partial}_x n^x - \tilde{\partial}_y n^y).
\end{equation}
Expanding this to the form of Eq. \ref{eq:MatrixEq} yields
\begin{eqnarray}
\gamma \dot{\alpha}_1 &-& \frac{2}{(1+\gamma)\sigma_\eta} \left[ \left( u^x \tilde{\pi}^{xj} - u^y \tilde{\pi}^{yj} \right) \dot{u}^j   - u^j \tilde{\pi}^{jx} \dot{u}^x + u^j \tilde{\pi}^{jy} \dot{u}^y   \right]   \\
&+& \frac{\eta}{\sigma_\eta \tau_\pi} \left( (u^x \dot{u}^x - u^y \dot{u}^y) - (u_x^2 - u_y^2) \frac{u^k \dot{u}^k }{\gamma(1+\gamma)}  \right) , \nonumber \\
&=& -u^i \partial_i \alpha_1 - \frac{\eta}{\sigma_\eta \tau_\pi} [ (\tilde{\partial}_x n^x)' - (\tilde{\partial}_y n^y)' ] \nonumber ,
\end{eqnarray}
where $\partial_i \alpha_1$ still contains corrections due to velocity gradients.

\subsection{Time Integration}

As stated before, the algorithm is written on a fixed grid in the expanding coordinate system.
The integration variables, denoted $V_i$ are the natural logarithm of the energy density -- $\varepsilon = \ln \epsilon$,
the collective velocity relative to the expansion -- $\vec{u}$, and the scaled, projected local shear 
tensor -- $\alpha_i$ as given by Equation \ref{eq:projShear}.

The choice of the logarithm of the energy density is motivated by the shape of the fireball.
Models that we consider for the initial energy density tend to have tails that are either
Gaussian or exponential.  Taking the logarithm means that second-order finite difference derivatives
will be exact for such profiles.  

Furthermore, the structure of the source term from the Bjorken expansion suggests that one 
should not use linear proper time at early times. 
Consider a (0+1)-dimensional system under going a Bjorken expansion.  For such a system, the evolution of
the energy density is given by
\begin{equation}
\frac{d \epsilon}{d \tau} = \frac{-(\epsilon+P)}{\tau}.
\end{equation}
If one used a first order upwind solver (often called Euler integration) for this application, the source term would
always be too large since the proper time would be smaller when the term was evaluated.  This error
goes like $e^{-\tau}$ and was found to produce integrated errors on the order of 1\% in simple situations whereas
the final code produces total integrated errors on the order of 0.1\% for more complex situations.

These simple integration errors suggest that one should integrate in the logarithm of proper time, $x=\ln{\tau}$.
This changes the energy density evolution to
\begin{equation}
d\epsilon = -(\epsilon+P) \frac{d \tau}{\tau} = - (\epsilon + P) dx.
\end{equation}
While this reduces integration errors at small times, for a fixed $\Delta x$ the size of the proper time steps
grows without bound.  This introduces numerical instability by violation of the Courant condition.  A pleasant
compromise defines $x = \ln \sinh \tau$, which causes the time integration to proceed slowly at first getting 
the Bjorken corrections right but limits the steps to a fixed maximum size.  Results for the equations of
motion will ignore this distinction.

Spatial derivatives are evaluated using centered second-order finite differences of the integration variables
\begin{equation}
\partial_j V_i(x^\mu) = \frac{ V_i(x^\mu+\triangle x^j) - V_i(x^\mu-\triangle x^j)}{2 | \triangle x^j | } , \label{eq:CentFD}
\end{equation}
which assumes that the $V_i$ is locally parabolic.  The hot region in a heavy ion collision obviously
has no hard boundary to couple to, instead we must deal with coupling to the vacuum at the 
edges of the computational region.  There, we use an asymmetric second-order finite difference
\begin{equation}
\partial_j V_i(x^\mu) = \frac{ 3 V_i(x^\mu) - 4 V_i(x^\mu-\triangle x^j) + V_i(x^\mu-2\triangle x^j)}{2 | \triangle x^j | } , \label{eq:nonCentFD}
\end{equation}
for the positive $x_j$ boundary and the equivalent for the negative $x_j$ boundary.
Equations \ref{eq:CentFD} and \ref{eq:nonCentFD} produce the same result if the function is exactly parabolic in 
the region $x^\mu - 2 \triangle x_j < x^\mu < x^\mu + 2\triangle x_j$.
Using the same order derivatives at the boundary proved important for modeling systems whose tails were not 
purely exponential.

Time integration is done simultaneously for all integration variables
using the second-order Runge-Kutta method, sometimes called the predictor-corrector method.
This method uses spatial derivatives at a fixed proper time to compute proper time derivatives at that time.
These are used to move the system forward by a half-step in time by finite difference
and the results are stored in a separate 
mesh.  Spatial derivatives computed are then computed on the advanced mesh, from which time derivatives
are computed.  These time derivatives are used to advance the original mesh a full step.  For a one-variable
integration, this amounts to
\begin{eqnarray}
V_i (t+\triangle t/2) &=& V_i(t) + \frac{\triangle t}{2} \left. \frac{\partial V_i}{\partial t} \right|_{t} , \nonumber \\
V_i (t+\triangle t) &=& V_i(t) + \triangle t \left. \frac{\partial V_i}{\partial t} \right|_{t+\triangle t/2}.
\end{eqnarray}
This procedure is repeated until all of the fluid cells are below the freezeout temperature.

The equations of motion are coupled in the sense that the continuity equation contains time derivatives
of the energy density and the velocities though not the shear tensor. Decoupling the equations
is tedious algebraically and it is not clear that such a procedure would be computationally beneficial.
Instead, we find the time derivatives on a mesh by considering the coefficients to all the time derivatives
as a matrix equation
\begin{equation}
C_{ij}(V_m) \dot{V}_i = b_j(\partial_m V_n, V_i)  \label{eq:MatrixEq}
\end{equation}
where $C_{ij}$ is the coefficient matrix that depends on the coordinates and integration variables but $b_j$ 
also depends on their derivatives.  $C_{ij}$ can be inverted for each fluid cell to find the 
time derivatives.  In this form, the coefficient matrix is invertible unless the viscous corrections are larger
than the energy density where Israel-Stewart theory would not be valid in any event.  Such situations 
do not seem to arise unless the system begins in such a condition, though this happens frequently if the system begins
with viscous corrections from the Navier-Stokes theory which will be discussed later.

The code is also designed to run in modes that reduce the integration region using system symmetries
that are frequently present for smooth initial conditions.  In many cases the gradients of the energy density
can be aligned with the coordinate axes and have reflective symmetry with respect to the coordinate planes.
This allows one to integrate a single octant of configuration space.
Since the velocity vector arises from the gradient of the energy density, it is anti-symmetric in the 
sense that $u^x(\delta x) = - u^x(-\delta x)$ but $u^x(\delta y) = u^x(-\delta y)$. 
The shear tensor gets it symmetry structure from the velocity gradients that produce it. This means that
on-diagonal components are even in all coordinates, and off-diagonal components are odd in their 
coordinates and even otherwise, which we summarize as
\begin{equation}
\pi^{xx} (\delta \vec{r}) = \pi^{xx} (- \delta \vec{r}); ~~  \pi^{xy}(\delta x) = - \pi^{xy}(-\delta x); ~~ \pi^{xy}(\delta z) = \pi^{xy}(-\delta z) \label{eq:piSymmetries}.
\end{equation}
This could be violated if the initial conditions for the shear tensor are not proportional 
to velocity or density gradients, though the source of such conditions is not immediately clear.

\section{Verification of Hydrodynamic Algorithm}

This section focuses on efforts to verify that the code solves the correct set of 
equations and solves them to sufficient accuracy for application to heavy ion 
collisions.  The general approach is to confirm that the ideal hydrodynamics algorithm 
conserves entropy, that the viscous hydrodynamics algorithm conserves $T^{00}$,
that results from lower dimensional codes are qualitatively reproduced.

\subsection{Conserved Quantities}

As discussed in Chapter 2, ideal hydrodynamics locally conserves entropy.  For 
an infinite system, the volume integral of the entropy density over the whole space 
would be a constant.  Since most hydrodynamic algorithms are unstable in regions
where the velocity gradients are large and the energy density is small, the integration
region is cutoff at temperatures around 50 MeV.  Over the length of the calculation, 
the entropy flux from the integration region is not negligible at the level of conservation
that we hope to achieve.   As such, it must be calculated in order to ensure proper conservation.

In the $\eta$-$\tau$ coordinate system, there is an additional term due to the work
done by the expanding fluid cell:
\begin{eqnarray}
0 &=& \int d^4x \tau d_\mu s^\mu = \int d^4x \tau \partial_\mu s^\mu + \int d^4x s\gamma
 = \int d\tau dx dy d\eta \left[ \partial_\tau (\tau \gamma s) + \tau \vec{\nabla} \cdot (\vec{u} s) \right] , \nonumber \\
C &=& \int dx dy d\eta (\tau \gamma s) + \int \tau d\tau \int s \vec{u} \cdot d\vec{A} , 
\end{eqnarray}
where $C$ is a constant, $d\vec{A}$ is a surface element, and the velocity is relative to the expanding mesh.
Note that the expansion of the fluid cell has been incorporated into the first integral which is over the volume at fixed proper time.
At each time step, the integral over all cells and the flux integral over the surface are calculated.
Time integrals are evaluated via second order Runge-Kutta, using the surface integrals evaluated
on the intermediate grid.
The result is compared to the same integral done in the first step.
For typical ideal calculations, the entropy conservation integral was found to be conserved at the level
of 0.1\%.  The importance of changes to the integration scheme as described above improved
this conservation by an order of magnitude for calculations beginning at proper times of $\tau = 0.1$ fm/c,
though this is earlier than typical calculations.

Furthermore, viscous hydrodynamic calculations were checked to ensure that 
entropy increases monotonically throughout.
More satisfying would be to check a conserved quantity and
a useful choice for the viscous case is $T^{00}$, since $T^{0i}$ conservation
can be verified by symmetry.
We provide the same calculation as for entropy conservation
\begin{eqnarray}
0 &=& \displaystyle\int d^4x d_\mu T^{\mu 0} = \displaystyle\int d\tau d\eta dx dy \left[ \partial_\tau T^{00} + \partial_i T^{0i} + \frac{T^{00} + T^{zz}}{\tau} \right] , \nonumber \\
C &=&  \displaystyle\int  T^{00} \tau d\eta dx dy   + \displaystyle\int d\tau \left[ \displaystyle\oint T^{0i} ds^i  + \displaystyle\int T^{zz} d\eta dx dy \right] , \label{Eq:consT}
\end{eqnarray}
where $T^{0i} ds^i$ is the outgoing momentum flux.
Here the extra terms due to the expansion cannot be completely avoided and an integral over all
time and space must be tracked.  This integral tends to be about 25\% of the total by the end of 
calculation. 
Still we find that this integral is conserved at the 0.1\% level for typical
initial conditions and cell densities.

\subsection{Comparison to Known Results} 

The equations of motion for second-order viscous hydrodynamics are quite complex and
verifying that the code solves the correct set of equations is not trivial.  
Since the conservation equations involve
first-order time derivatives and first-order spatial derivatives, one natural approach
to find an analytic solution for verification is to consider an exponential dependence
of the density for a static system
\begin{equation}
\epsilon = \epsilon_0 \cdot e^{-x/R} , ~~~~~ \vec{u} = 0 , \label{expHydro_IC}
\end{equation}
where $\epsilon_0$ and $R$ are constants.
For verification purposes we arbitrarily take $R=3$ fm; and the initial energy density,
$\epsilon_0$, was taken at many values to ensure numerical accuracy for a practical
range of values, but presented results will be normalized by this factor.
We will investigate 
Such a system has infinite energy and is therefore not physical except locally,
but we emphasize that this is a test case.
The collective velocity begins at the same value everywhere
and the absence of velocity gradients proves a stable point, which will be clear when we arrive at
the equations of motion for the velocity and energy density.  
A constant speed of sound is a necessary condition for this behavior.

For an ideal system, there are only two equations to be found - for the velocity and energy
density - which come from the continuity equations and one Euler equation.
We begin by investigating the case without a longitudinal expansion and return to that
such conditions later.
From energy conservation, we obtain
\begin{eqnarray}
\nonumber \partial_\mu T^{\mu 0} &=& 0 , \\
(1+c_s^2) \gamma u \frac{\epsilon}{R}  &=& 2\epsilon(1+c_s^2) u \dot{u} + [ \gamma^2 + c_s^2(\gamma^2 -1) ] \dot{\epsilon}, \label{eq:expHydro_cont}
\end{eqnarray}
and from momentum conservation,
\begin{eqnarray}
\nonumber \partial_\mu T^{\mu x} &=& 0 , \\
\left[ u^2 (1+c_s^2) - c_s^2 \right]  \frac{\epsilon}{R} &=& \frac{(1+c_s^2) \epsilon}{\gamma} [ u^2  + \gamma^2 ] \dot{u} + (1+c_s^2) \gamma u \dot{\epsilon}. \label{eq:expHydro_mom}
\end{eqnarray}

\begin{figure}[fig:expU]
\centerline{\includegraphics[width=0.7 \textwidth]{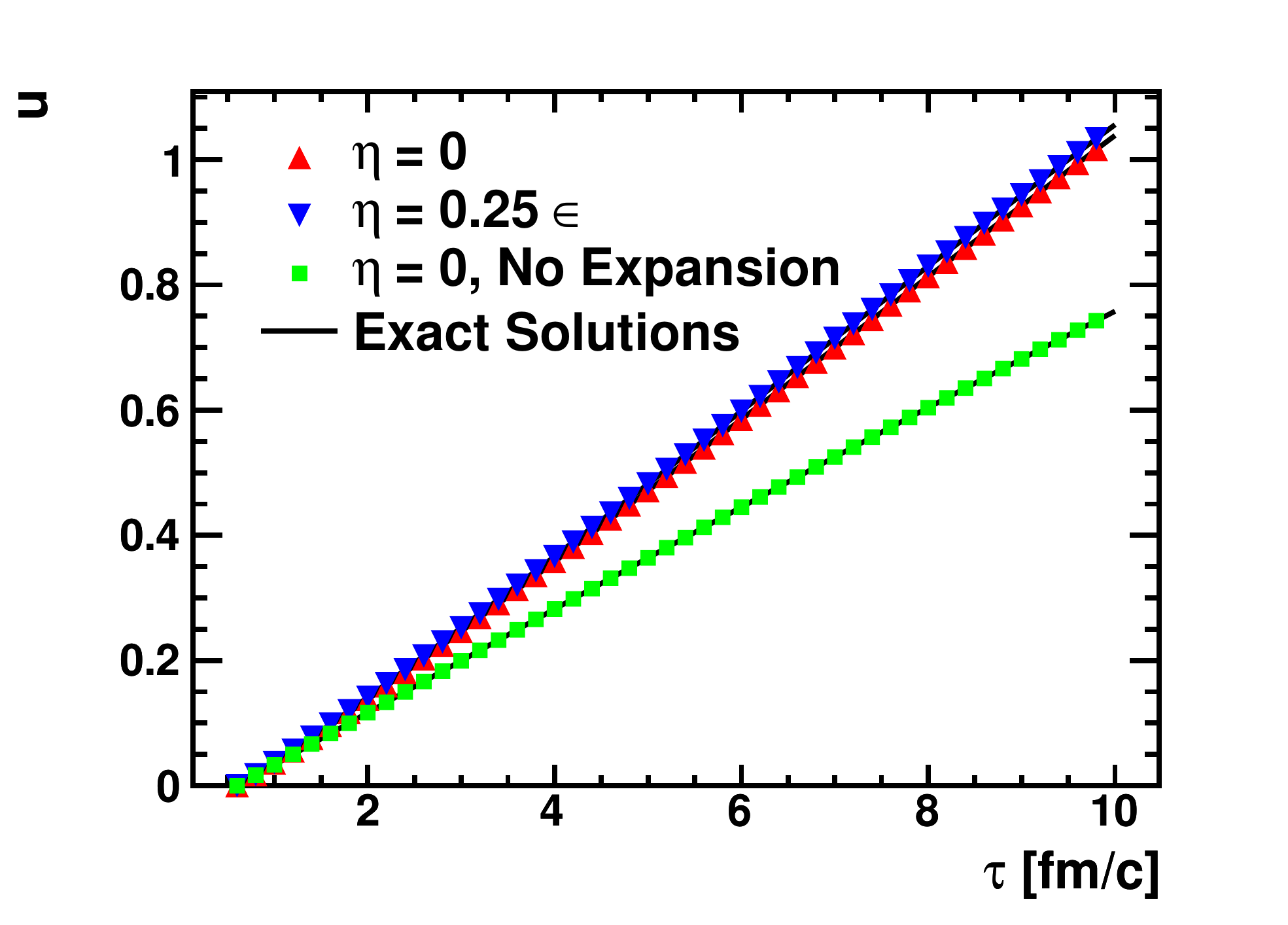}}
\caption{ (color online)
The collective velocity at all points for a exponential energy density profile with no Bjorken expansion (blue squares), 
a transverse profile with Bjorken expansion (red stars),
and a longitudinal profile with Bjorken expansion (green triangles).
In each case the velocity calculated by the full code agrees with the correct value.
}
\label{fig:expU}
\end{figure}

The system of Equations \ref{eq:expHydro_cont} and \ref{eq:expHydro_mom} can be solved for 
$\dot{\epsilon}/\epsilon$ and $\dot{u}$, where the former
is only possible due to the choice of the density gradient as proportional to the density.
This separability proves that the absence of velocity gradients is a stable condition for 
this initial condition.
The equation for the velocity is
\begin{equation}
\dot{u} = \frac{ \gamma}{ \gamma^2 - c_s^2 u^2} \cdot  \frac{c_s^2}{R(1+c_s^2)} , \label{eq:expHydro_UD}
\end{equation}
where we have used the constant speed of sound.
For early times when the collective velocity is small, the solution to Eq. \ref{eq:expHydro_UD} is linear,
\begin{equation}
u(t) \propto \frac{c_s^2}{R(1+c_s^2)} t \label{eq:expHydro_FOU}.
\end{equation}
While Eq. \ref{eq:expHydro_UD} is exactly integrable, the solution is not invertible
\begin{eqnarray}
\frac{c_s^2}{R(1+c_s^2)} \int dt  &=& \int du ~~ \frac{1+u^2(1-c_s^2)}{\sqrt{1+u^2}} \label{expHydro_uInt} , \\
 \frac{c_s^2}{R(1+c_s^2)}t &=& \frac{1}{2} \left[ (1+c_s^2) \sinh^{-1}(u) + (1-c_s^2)u\sqrt{u^2+1} \right] \label{expHydro_AnalyticU}
\end{eqnarray}
which limits its utility.
Figure \ref{fig:expU} shows that the agreement of the code with this solution.

The addition of the Bjorken expansion does not lead to an analytic solution, but
the computation provides a stricter test on the code and can be done with a much
less complex numerical routine.
The conservation equations are modified as follows: for energy conservation 
\begin{eqnarray}
\nonumber \partial_\mu T^{\mu 0} + (T^{00} + T^{zz} )/\tau &=& 0 , \\
(1+c_s^2) \gamma u \frac{\epsilon}{R} + \gamma^2(\epsilon+P) / \tau &=& 
2\epsilon(1+c_s^2) u \dot{u} + [ \gamma^2 + c_s^2(\gamma^2 -1) ] \dot{\epsilon} , \label{expHydro_ECB}
\end{eqnarray}
and for momentum conservation
\begin{eqnarray}
\nonumber \partial_\mu T^{\mu x} + T^{0x}/\tau &=& 0 , \\
\left[ u^2 (1+c_s^2) - c_s^2 \right]  \frac{\epsilon}{R} + \frac{u \gamma (1+c_s^2)\epsilon}{ \tau}
&=& \frac{(1+c_s^2) \epsilon}{\gamma} [ u^2  + \gamma^2 ] \dot{u} + (1+c_s^2)\gamma u \dot{\epsilon} \label{expHydro_MCB} .
\end{eqnarray}
Solving these equations for $\dot{u}$ yields
\begin{equation}
\dot{u} = \frac{ \gamma c_s^2}{ \gamma^2 - c_s^2 u^2} \left[ \frac{\gamma u}{\tau} + \frac{1}{R(1+c_s^2)} \right] \label{expHydro_UDB} ,
\end{equation}
where the first term in the brackets comes from the Bjorken expansion and differentiates it from Eq. \ref{eq:expHydro_UD}.
The inseparability of the equation makes the prospects of an analytic solution unlikely, but the 
expression can be used for comparison with a simple and easily verified numerical code.
Comparison to the larger code again validates the larger (3+1)-dimensional code which
is again shown in Figure \ref{fig:expU}.
Note that the presence of the Bjorken expansion actually significantly
increases the transverse expansion rate.
Shear viscosity will further increase the transverse expansion rate as viscosity tends to 
resist the asymmetric expansion inherent to the Bjorken initial condition.

Also shown in Figure  \ref{fig:expU} is the same result if the exponential distribution
is in the longitudinal direction where $\epsilon \propto e^{\eta/\sigma}$, which is also reproduced.  While in the transverse case
the expansion due to the infinite exponential profile continues forever, in the longitudinal case,
the expansion of the mesh means that the effective gradients get weaker over time.  The inclusion
of the mesh expansion can be accounted for by making the substitution $R \rightarrow \tau \sigma$ in 
the final results of Eq. \ref{expHydro_UDB}, which comes from correcting the spatial derivative $\partial_z \rightarrow \tau^{-1} \partial_\eta$.

\begin{figure}[fig:expE]
\centerline{\includegraphics[width=0.7 \textwidth]{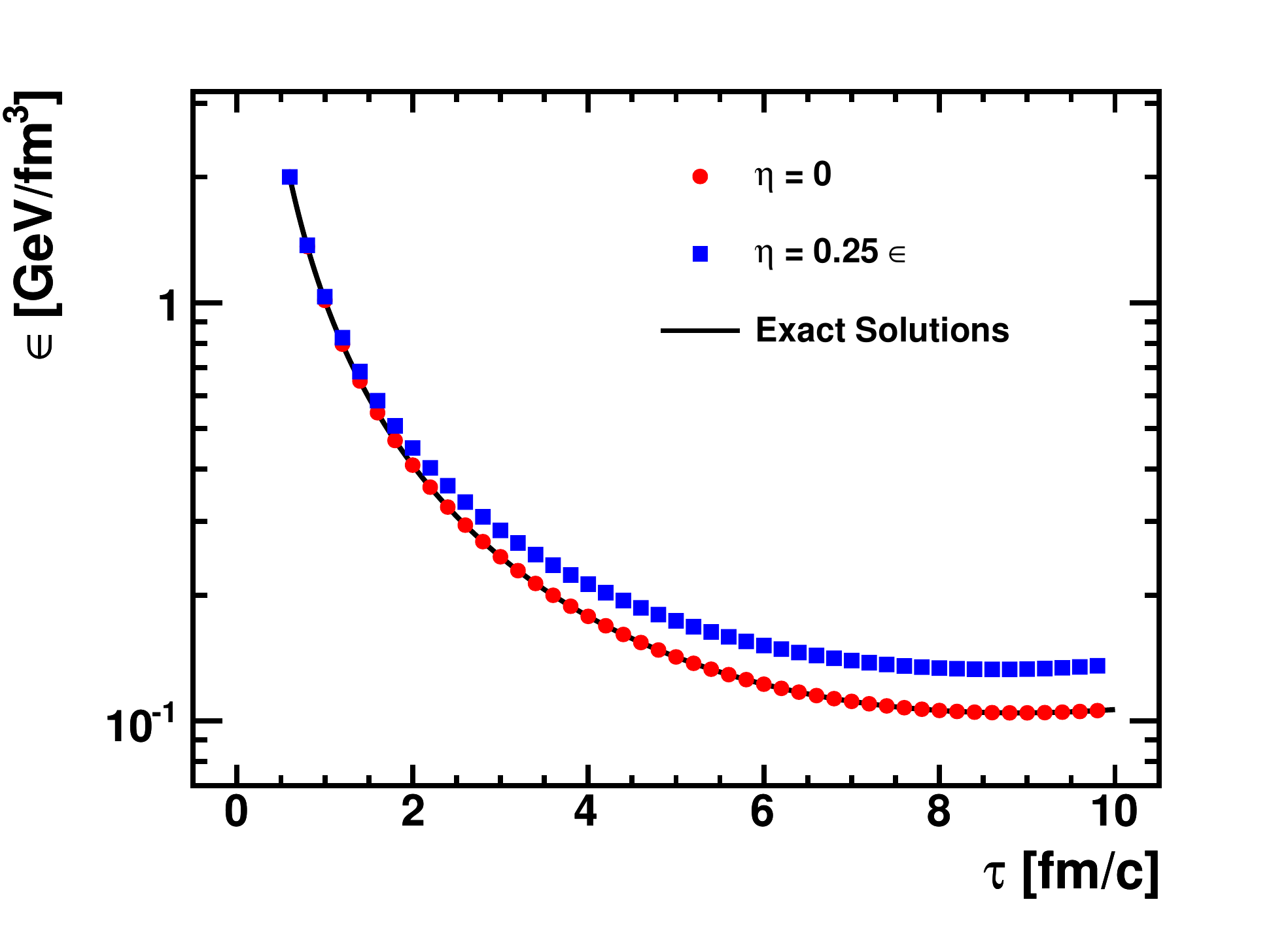}}
\caption{ (color online)
The ratio of the energy density to its initial value for a exponential energy density profile 
with a Bjorken expansion in the ideal case (red squares) and with shear viscosity equal
to one quarter of the energy density.  In either case the calculation produces the correct energy density.
}
\label{fig:expE}
\end{figure}

The equations for the collective velocity are easy to work with as they involve only the velocity itself.
For the energy density, the logarithmic time derivative can be written as a function of the collective velocity.
 For the case without the Bjorken expansion, one simply
uses the result in Eq. \ref{eq:expHydro_UD} along with Eq. \ref{eq:expHydro_cont} or \ref{eq:expHydro_mom}.
This results in 
\begin{equation}
\frac{\dot{\epsilon}}{\epsilon}= \frac{ \gamma u(1-c_s^2) }{R (u^2(1-c_s^2) +1 ) }.
\end{equation}
While this expression is not so pleasing, it can be integrated to obtain $\epsilon(u)$ 
using a change of variables:
\begin{eqnarray}
\int \frac{ d\epsilon}{\epsilon} &= & \int du \frac{dt}{du} \frac{d\epsilon}{dt} ,  \nonumber \\
\ln \left[ \frac{\epsilon}{\epsilon_0} \right] & = & \int_0^u du ~~ \frac{ R (1+c_s^2)(u^2(1-c_s^2) +1)}{\gamma c_s^2} \cdot \frac{ u \gamma (1-c_s^2) }{R (u^2(1-c_s^2) +1 ) } ,\nonumber \\
   & = & \frac{1-c_s^4 }{c_s^2} \int_0^u u ~~ du , \nonumber \\
\ln \left[ \frac{\epsilon}{\epsilon_0} \right] & = & \frac{1-c_s^4 }{2 c_s^2} \cdot u^2 , \nonumber \\
\epsilon & = & \epsilon_0 \cdot \exp \left[ \frac{1-c_s^4 }{2 c_s^2} u^2 \right] \label{expHydro_EU}.
\end{eqnarray}
To the extent that $u$ is linear for small $t$, $\epsilon$ grows like $e^{t^2}$.  Again we emphasize
that this is not intended to be a physical solution and mainly the super-exponential growth is 
due to the original exponential density distribution everywhere.
Also note that while this expression appears not to depend on the length scale from the
density distribution, but this information enters through the velocity.
Eq. \ref{expHydro_EU} was also used to verify the full code. 

The Bjorken expansion can be added again. The result can be arrived at by using the result
for the time derivative of the velocity with either conservation equation, or by returning to the
conservation equations and resolving.  The result of either is
\begin{equation}
\frac{\dot{\epsilon}}{\epsilon} = \frac{\gamma}{u^2(1-c_s^2) +1} \cdot \left[ \frac{(1-c_s^2)u}{R} - \frac{(1+c_s^2)\gamma}{\tau} \right] \label{expHydro_EB}.
\end{equation}
An analytic solution to this equation was not attempted, but the equation was numerically
integrated using a small code and again confirmed the full code shown in Figure \ref{fig:expE}
Note that at large times the expansion due to the infinite exponential initial condition eventually
overcomes the Bjorken expansion.  However, the lifetime of the matter created in $\sqrt{s} = 200$ GeV Au+Au
collisions tends to be less than 10 fm/c and up until around this time the evolution of the energy 
density is dominated by the Bjorken expansion even where the density gradients are significant.
Therefore, we expect the lifetime of the system to be roughly independent of the transverse details.

The above scenarios demonstrate that the conservation equations are being correctly solved in 
the absence of shear viscosity.  Adding shear viscosity would tend to spoil the constancy of the 
collective velocity simply by introducing a new scale to the problem.
However if the shear viscosity also scaled with the energy density and the relaxation time 
was a constant the scaling solution would not be broken.  This violates the Second Law
of Thermodynamics and is unphysical based solely on units, but it is useful for 
testing purposes.

\begin{figure}[fig:expHydroA]
\centerline{\includegraphics[width=0.7 \textwidth]{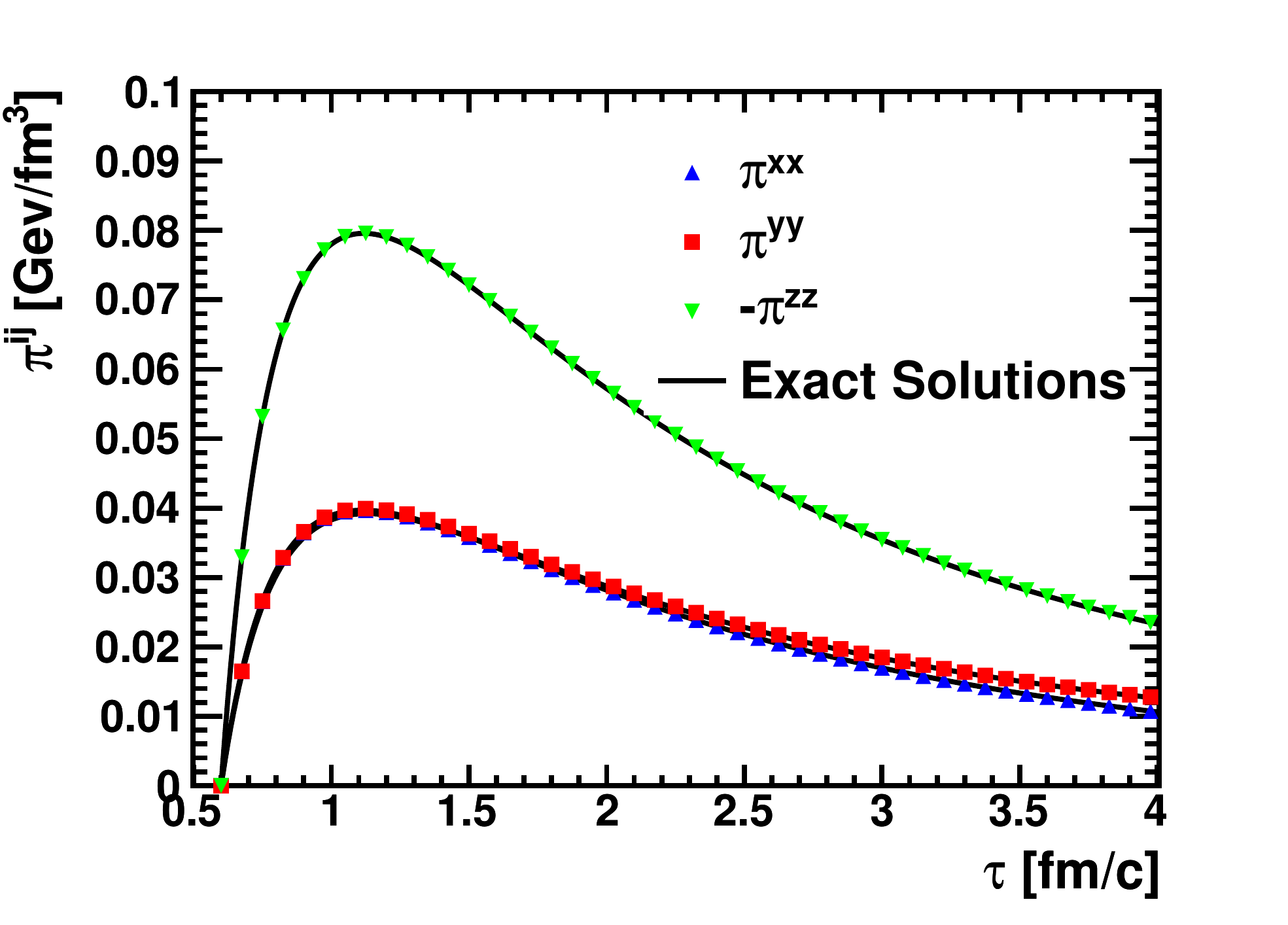}}
\caption{ (color online)
Viscous corrections to the pressure along each axis for the transverse exponential test 
of the hydrodynamics code for a viscous system undergoing a Bjorken expansion.
This demonstrates that TRISH reproduces analytic results for viscous hydrodynamics
as well as ideal hydrodynamics
}
\label{fig:expHydroA}
\end{figure}

Deriving the equations of motion for the shear tensor in addition to the conservation equations
is a bit tedious and not particularly enlightening.  For completeness, the equations of motion to be
integrated are
\begin{eqnarray} \label{eq:expTestFull}
\gamma \dot{\epsilon} + \frac{u}{\gamma} \left( \epsilon + P + \frac{\pi^{xx}}{\gamma^2} \right) \dot{u} &=&
\frac{u \epsilon}{R} - \frac{ \gamma}{\tau} \left( \epsilon + P + \pi^{zz} \right), \\
\gamma u c_s^2 \dot{\epsilon} + \gamma \left( \epsilon + P + \frac{1-u^2}{\gamma^4} \pi^{xx} \right) \dot{u} + \frac{u}{\gamma} \dot{\pi}^{xx} &=& 
\frac{c_s^2 \gamma^2 \epsilon + \pi^{xx}}{R} + \frac{u}{\gamma \tau} \left( \gamma^2 \pi^{zz} - \pi^{xx} \right), \nonumber \\
\frac{2u}{3} \left( \frac{2\gamma \eta }{\tau_\pi} - \frac{\pi^{xx}}{\gamma} \right) \dot{u} + \gamma \dot{\pi}^{xx} &=& 
\frac{-\pi^{xx}}{\tau_\pi} + \frac{u \pi^{xx}}{R} + \frac{2 \gamma }{3 \tau} \left(  \frac{\gamma^2 \eta}{\tau_\pi} - 2 \pi^{xx} \right), \nonumber \\
 \frac{2u}{3\gamma} \left( 2 \pi^{zz} - \frac{\eta}{ \tau_\pi} \right) \dot{u}  + \gamma \dot{\pi}^{zz} &=& 
\frac{-\pi^{zz}}{\tau_\pi} + \frac{u \pi^{zz}}{R} - \frac{4 \gamma}{3 \tau} \left( \frac{\eta}{\tau_\pi} + \pi^{zz} \right). \nonumber 
\end{eqnarray}
This system of equations has the structure of the full set of equations in that it contains the dependence
of time derivatives of the shear components on the conservation equations and vice versa.  This means that
 this test case would identify errors in the matrix inversion procedure.  In addition, frequently terms are
not expressed in this set of equations in the same way that they are expressed in the full code and a large
fraction of terms are represented and checked.

The results of the full code and a small test program are shown in Figure \ref{fig:expU} for collective velocity,
in Figure \ref{fig:expE} for the energy density, and in Figure \ref{fig:expHydroA} for components of the shear tensor,
each in the viscous case.
Only two components of the shear tensor need to be calculated due to orthogonality with the collective velocity
and tracelessness of the shear tensor.  The test code is written to integrate only $\pi^{xx}$ and $\pi^{zz}$, whereas 
the full code integrates $\alpha_i$ meaning that some numerical effects are different between the two codes.
These normalized deviations were found to be on the order of $10^{-4}$ or less.

\begin{figure}[fig:FOSL]
\centerline{\includegraphics[width=0.7 \textwidth]{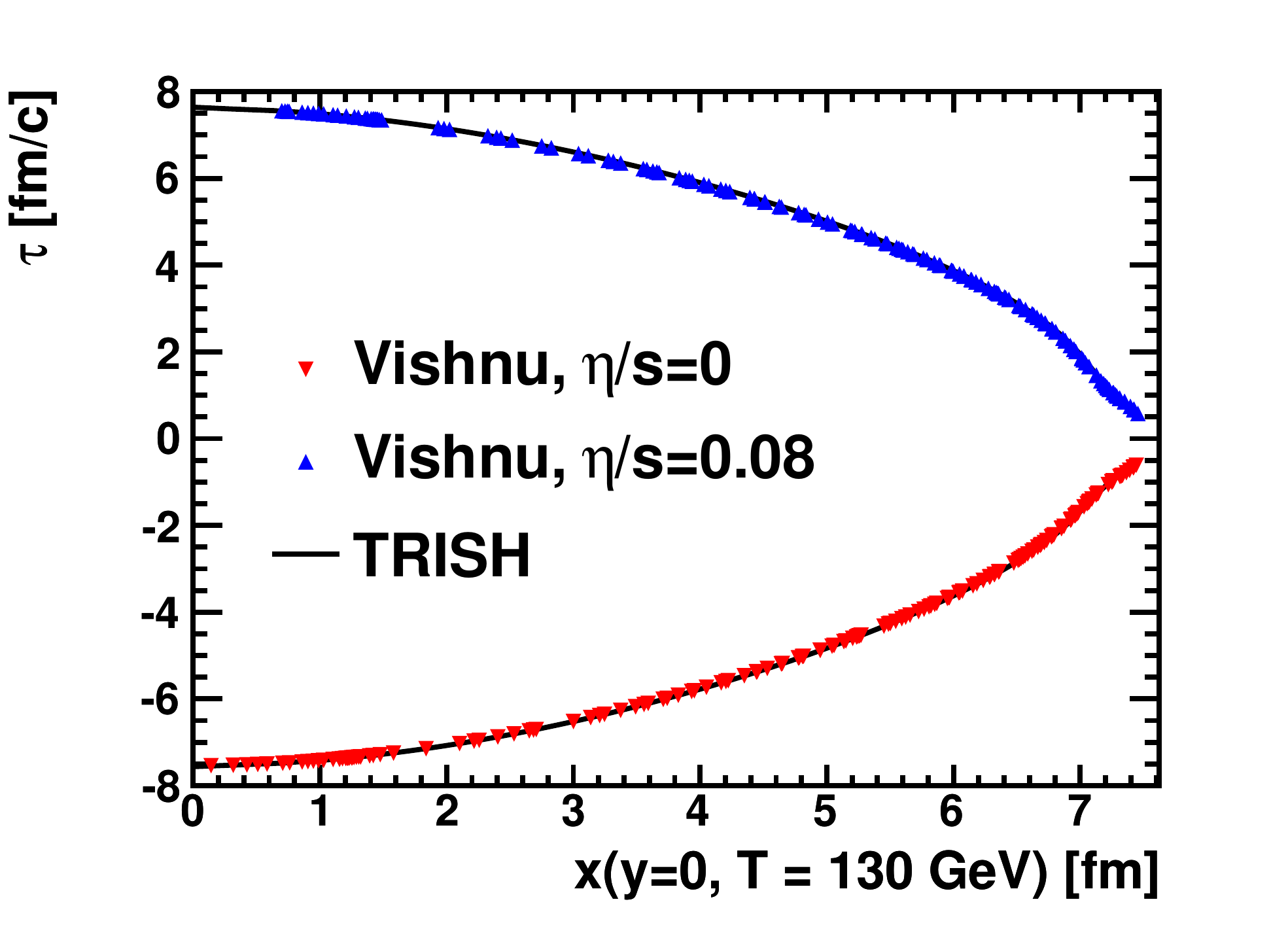}}
\caption{ (color online)
The location of the $T=130$ MeV isotherm along the line $y=0$ for comparison
with another numerical code, where the ideal results have been reflected $\tau \rightarrow - \tau$ 
for the case, $\eta/s = 0$.   Comparison shows that the hydrodynamic codes 
produce the same freezeout surface for the case of zero or small shear viscosity.
}
\label{fig:FOSL}
\end{figure}

Comparisons for more realistic conditions were made in the interest of validation.
Particularly of interest were calculations of the motion of a constant temperature surface, as these are the
principle output of the hydrodynamic calculations.  The chosen initial conditions were energy density scaled 
to the number of wounded nucleons (see Chapter 5), zero initial collective flow, and shear corrections
as given by Navier-Stokes for viscous simulations. The equation of state is from a massless gas, 
and the shear viscosity was scaled to the entropy density.

Numerical results for the initial conditions were provided by the TECHQM collaboration for
code verification. These codes are restricted to boost invariant problems and so the following
comparisons will be done for boost invariant initial conditions.
For the ideal case, the initial conditions were matched exactly.  In the viscous
case, the choice of Navier-Stokes shear corrections has the problem that the corrections become much
larger than the pressure at low temperature, because the shear viscosity is proportional to $T^3$ and
the pressure is proportional to $T^4$, so the ratio $\eta/P \propto T^{-1}$ diverges at low temperature. 
While physically this is not concerning as the simulation is only valid 
if the motion of the surface is not strongly influenced by the details of
the evolution at temperatures much lower than the freezeout temperature, large viscous corrections
make hydrodynamic calculations unstable. 

The TECHQM collaboration chose to unphysically compensate by making the shear viscosity decrease 
at large distances from the center of the hot region.  TRISH was
instead run with the initial shear corrections proportional to the pressure at all temperatures with 
the corrections in the center of the calculation roughly the same as the Navier-Stokes value.
It has been found that this does not significantly affect observables in general, and in this case
we find that it has little effect on the hydrodynamics.

The motion of the point of constant temperature ($T=130$ MeV) 
along the line $\eta = y = 0$ as a function of the proper time is shown in Figure \ref{fig:FOSL}.  
The differences in integration strategy mean that the calculations make predictions at different times, 
making a differential comparison reliant on interpolation strategy. Still, we found
qualitative agreement between the codes in both the ideal and viscous cases.  This includes
the general feature that viscosity increases the transverse pressure at early times, which causes
the temperature isosurface to remain at larger distances from the center at moderate times, 
before collapsing more suddenly at roughly the same final time.  

\begin{figure}[fig:FOSSST]
\centerline{\includegraphics[width=0.7 \textwidth]{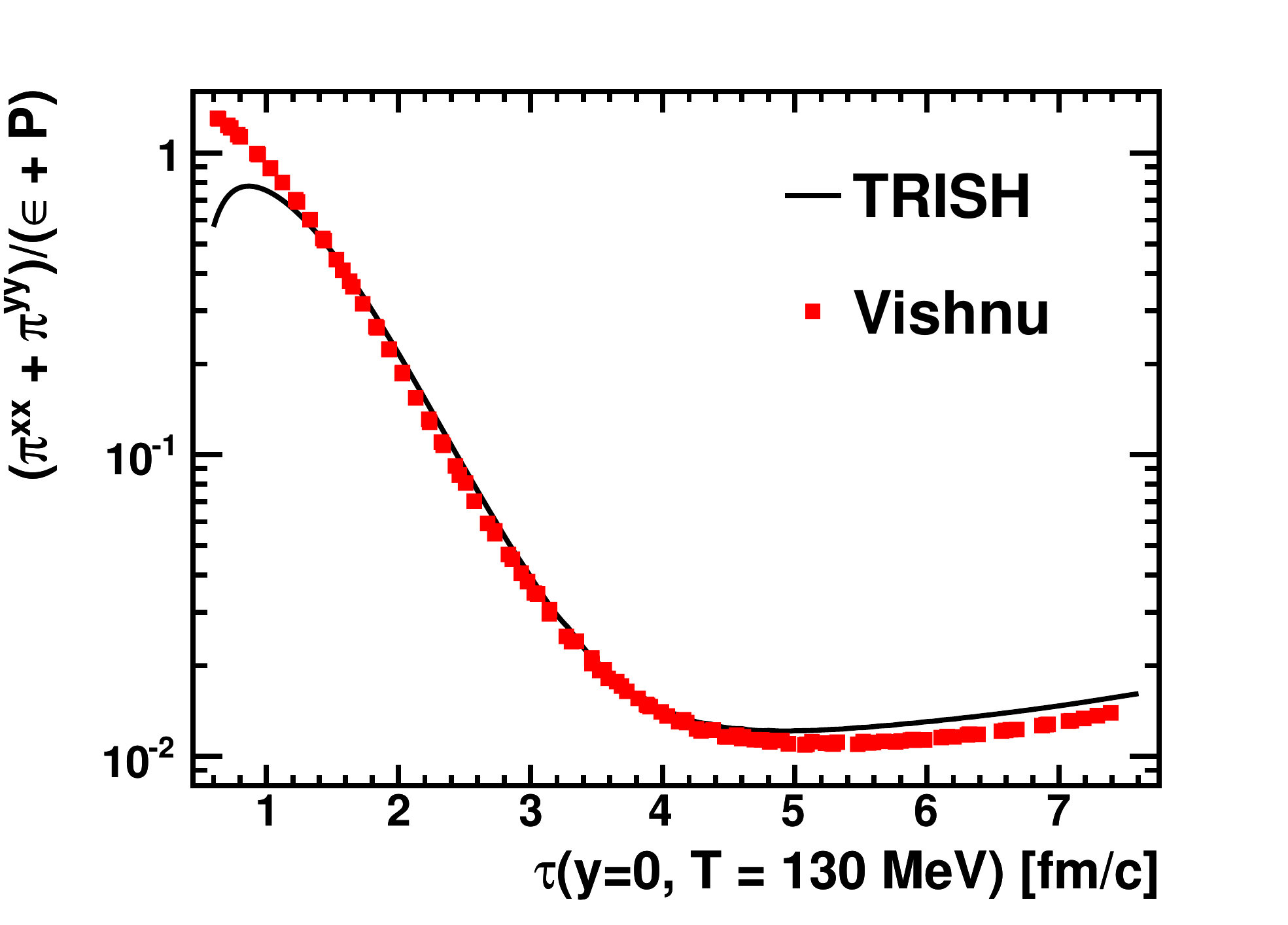}}
\caption{ (color online)
The sum of the transverse shear corrections to the pressure as a function of proper time
for the isotherm $T=130$ MeV and along the line $y=\eta=0$.  While the initial value differs
due to scaling with the pressure, the corrections have the same general behavior. 
}
\label{fig:FOSSST}
\end{figure}

In addition to the evolution of the freezeout surface, the shear correction to the phase space
density has a significant effect on emitted particle momentum distributions. 
This is most notable in elliptic flow which is reduced by the correction to the 
phase space density as described in Equation \ref{eq:ViscPSDens}.
To confirm that our predictions for the shear correction at the freezeout surface 
Figure \ref{fig:FOSSST} shows its evolution as a function of proper time along the line $y=\eta=0$.  
The initial value of the correction is smaller for our calculation due to
the choice of initial state but the two solutions converge rapidly to the same behavior.  
While deviations at larger times are on the order to 10\%, it is not clear if this is due 
to differences in the initial state or differences in integration scheme.

\section{Liquid-Gas Coupling}

\subsection{Particle Creation}

The low temperature regions of the calculation cannot be calculated 
as a liquid even within the Israel-Stewart framework.  In these regions,
the mean free paths of the particles become larger than the size of the
system and local kinetic equilibrium can no longer be maintained.  
Furthermore, differences between particle species become important
as the pions begin to flow faster and cool slower than heavier particles
such as protons.  Other groups have attempted to address this
by tracking fluctuations of baryon number and other conserved charges
and their associated fugacities below some chemical freezeout temperature,
while assuming that ideal hydrodynamics still governs the kinetic behavior
until some lower temperature.  This approach is motivated by the 
data which display softer spectra than the particle ratios would
suggest, but strain credibility by assuming local equilibration with long
mean free paths and low Reynolds numbers.

A more satisfactory approach is to couple the hydrodynamic calculation
to a gas calculation that treats particles as the important degrees of 
freedom.  In principle, a simulation of both the fluid and the gas
phases could proceed simultaneously with the fluid receiving information
about the pressure in the gas and the gas receiving information about 
the emission of particles from the fluid.  In practice, this is a very demanding
calculation and the fluid likely does a good job of approximating the
motion of the gas for temperatures near to the phase transition assuming
the equation of state is consistent.  The reverse is also of concern; 
if this temperature is chosen to be too high, the hadrons cannot be treated
as independent, incoherent objects and corrections akin to the classical finite
volume correction become important.  Fortunately for temperatures
around $140 \le T \le 165 $MeV the two descriptions mesh nicely.  \citep{Pratt:2010jt}

In general, the goal is to generate a distribution of particles such that
the stress-energy tensor is the same in the freezing fluid cell as it 
is for the particles that are now in that same volume.  In the gas phase,
the stress-energy tensor is given by
\begin{equation}
T^{ij} = \displaystyle\sum\limits_s \displaystyle\int \frac{d^3 p}{(2\pi)^3} \frac{p^i p^j}{E(\vec{p})} f_s (\vec{p}), 
\end{equation}
where $s$ indexes the particle species.
While this is at least a well-defined task if the phase space density is
given, in viscous hydrodynamics the phase space density is altered
by the presence of shear viscosity.
As discussed in Chapter 2, viscous corrections are related to the distortion
of the phase space density.  In Equation \ref{eq:ViscPSDens}, we took 
the relationship of the shear tensor to the phase space distortion
as the Grad ansatz, which has the general form
\begin{equation}
f ( p^\mu, r^\nu) = f_{\text{eq}} ( p^\mu, r^\nu) \left[ 1 + C(p) p_\alpha p_\beta \pi^{\alpha \beta}_{(s)} \right],
\end{equation}
where the momentum dependence of the correction coefficient, C(p),
may be different for different species.
This form has the difficulty that the correction term can become arbitrarily large
and certainly will do so at large momentum leading to negative phase
space density \citep{Huovinen:2012is,Holopainen:2012id,Holopainen:2010gz}. 

An alternative approach \citep{Pratt:2010jt} would be to evaluate the equilibrium
phase space density at an altered momentum by shifting linearly
in the form
\begin{equation}
p_i = p_i' + A \tilde{\pi}^{ij} p_j' ,
\end{equation}
where A is a coefficient to be determined, $\tilde{\pi}^{ij}$
is the shear correction in the fluid frame.
This allows one to generate particles according to a 
static thermal distribution and then adjust the momentum
to reproduce the local shear tensor before applying the
boost from the fluid frame to the lab frame.

This linear form for the 
correction can be understood in terms of the Navier-Stokes 
theory where $\tilde{\pi}^{ij}$ is the velocity gradient.  
Given that viscous hydrodynamics allows for a non-zero mean
free path ($\Delta r$) and collision time ($\Delta t$), particles
observed locally have originated nearby. Then the momentum
observed locally is given by
\begin{equation}
p_i = p_i' - E \Delta v_i = p_i' - E \partial_j v_i \frac{p_j \Delta t}{E} = p_i' - \Delta t \partial_j v_i p_j',
\end{equation}
where the velocity gradients are to be evaluated in the fluid frame.
More general considerations based on requiring fixed energy
and particle densities show that this rough argument holds well 
and the general form should be
\begin{equation}  \label{eq:momShifting}
p_i = p_i' + \lambda_{ij} p_j' = p_i' + \frac{\Delta t}{2 \eta} \tilde{\pi}_{ij} p_j'.
\end{equation}
In principle, this procedure can be done separately for each particle
species, the result of which would be that the pions receive larger phase
space corrections than the protons though this is not done at present.
Furthermore, $\Delta t$ in Equation \ref{eq:momShifting} could be different 
for each particle species or could depend on momentum.

These results allow one to calculate the phase space density.  Combining
this with information about the volume elements comprising the freezeout
hypersurface, $d\Sigma^\mu$, one generates particles according to the Cooper-Frye
prescription \citep{CooperFrye}
\begin{equation}
dN = \frac{f(p \cdot u) d^3 p}{(2\pi)^3 E(p)} (p_\nu d\Sigma^\nu) \Theta(p_\nu d\Sigma^\nu).
\end{equation}
The direction of $d\Sigma^\mu$ is orthogonal to the breakup surface. It points
in the time direction for sudden breakup and in the spatial direction for 
emission from a static shock front.
Discussion of calculating the hypersurface elements follows in Section 5.2, but
it suffices to understand that the vector points outward, from
higher energy density to lower, and has length proportional to the volume of 
the fluid element undergoing the transition to gas.  
The step function, $\Theta(p \cdot d\Sigma)$, ensures that we generate only particles
that are emitted from the surface rather than particles that are moving further 
into the fluid.
The corollary to this point is that particles within the cascade that move inside 
of the freezeout surface should be removed from that calculation.  For 4d 
calculations determining surface collisions becomes rather expensive,
and for 2d calculations it was found that roughly 1\% of the particles actually
re-enter the surface due to a combination of significant radial and 
the rapid shrinking of the hot region.  Absorbed particles could be a significantly large
effect with fluctuating initial conditions and a more thoughtful consideration
in the gas phase would be needed.

To conclude, we summarize the methodology.  One first calculates the
total number of particles to be emitted by the surface using only the
surface volume, temperature and chemical potentials.  For each particle,
one randomly selects a momentum from a static thermal distribution, 
which is then scaled to reproduce the stress-energy tensor via 
\ref{eq:momShifting}.  The momentum is then boosted from the fluid
frame to the lab frame.  In this frame, one can determine whether 
the momentum is into or out of the surface by keep or reject Monte
Carlo with keep probability  $p_\mu d\Sigma^\mu / (E |d\Sigma|)$.
This set of particles are then treated with the gas model.

\subsection{Surface Finding}

While the hydrodynamic calculation extends to rather low temperatures,
the model uses the low temperature evolution only as a proxy for the 
evolution of the gas phase.  The output of the hydrodynamic module 
is in fact the evolution of the boundary between the liquid and gas 
phases, which is taken to be a temperature isosurface,
in terms of the hypersurface vector $d\Sigma^\mu$ which characterizes
the location and orientation.  Generally the strategy \citep{Huovinen:2012is} is to break the
surface into tetrahedra embedded in the four-dimensional space. For each tetrahedron, the surface 
vector is proportional to the generalized cross-product of the three four-vectors
from one vertex to the other three:
\begin{equation} \label{eq:surfaceVector}
d\Sigma^\mu = (1/6) g^{\mu \nu} \epsilon_{\nu \alpha \beta \gamma} x_1^\alpha x_2^\beta x_3^\gamma
\end{equation}
where $\epsilon_{\nu \alpha \beta \gamma}$ is the totally anti-symmetric 
tensor.  This only defines $d\Sigma^\mu$ up to a sign and one must 
ensure that the vector points outward.  The simplest cases of Equation
\ref{eq:surfaceVector} are when the tetrahedron's vectors are along 
the coordinate axes.  For instance, if the tetrahedron lies along the 
spatial axes, then the surface vector points in the temporal direction.
This means that the surface is collapsing rapidly in time as would 
happen near the end of the collision when many fluid elements
freezeout simultaneously.  In contrast, if one of the elements is exactly
along the time axis and the other two along spatial axes, the surface
is static in time and emission happens outward along the final
spatial direction.

The hydrodynamic equations of motion are evaluated on a fixed grid
in configuration space in the moving coordinate system described in 
previous chapters.  This means that the mesh is made up of hypercubes
that intersect the temperature isosurface.  Determining that the isosurface
passes through a given hypercube is not difficult -- if there is at least
one corner above and at least one below the freezeout temperature, then
the surface passes through this cube. Rather all of the subtlety is in
determining the orientation and volume of surface within the hypercube.
This problem has been studied in great detail in three dimensions for 
the purpose of surface finding for graphical applications  \citep{marchingCubes}.  
On of the most famous and successful algorithms is called Marching Cubes,
which uses the cubic symmetry of the mesh and breaks down all possible
combinations of vertices inside and outside the surface exhaustively.  

\begin{figure}
\centerline{\includegraphics[width=0.6 \textwidth]{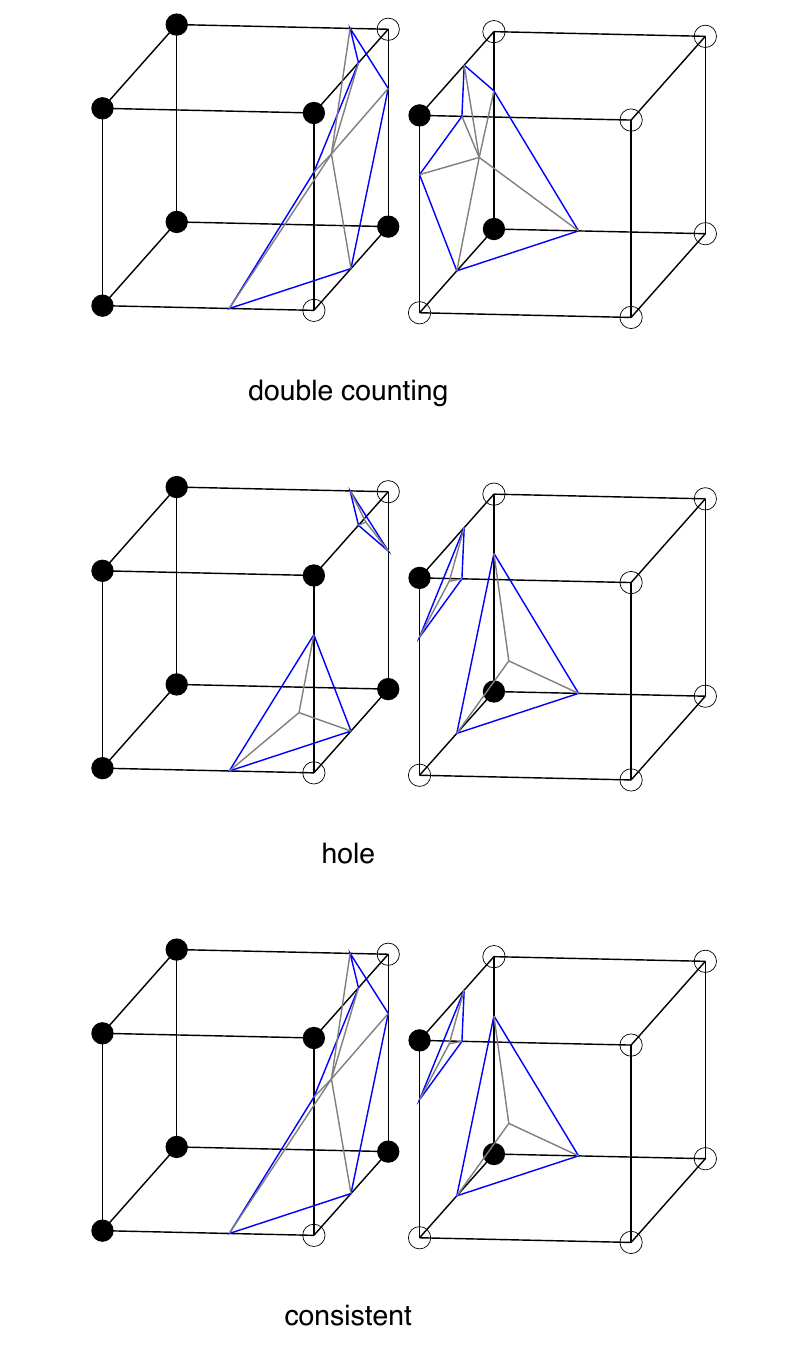}}
\caption{ 
Shows the freezeout surface passing through two adjacent cubes in a three
dimensional space.  The top figure demonstrates double counting resulting
from assuming that the surface is a single sheet in each cube.  The 
middle panel demonstrates the possibility that the surface could 
develop a hole, while the bottom panel shows a consistent solution
as determined by the Marching Cubes algorithm.\citep{Huovinen:2012is} 
}
\label{fig:surfCons}
\end{figure}

In three-dimensions, Marching Cubes produces a triangulation of the surface within 
each cube.  One of the key difficulties of resolving each cube individually is the relationship
between surfaces in adjacent cubes. For instance, in the two-dimensional
case (a square) when opposite corners are on one-side of the surface, 
there are two configurations of the lines that would make up the surface --
one with the center above the surface and one below.  
A self-consistent approach to determining which of these configurations is 
to be used is required.  In this case, whether the center of the face 
is inside or outside the surface is determined by the average
of all four values, though more complicated approaches exist \citep{Huovinen:2012is} .
Failure to account for this possibility leads to significant 
overlaps or holes in the surface even for relatively simple configurations, 
for instance Figure \ref{fig:surfCons} shows possible errors and a 
self-consistent solution for adjacent cubes.  The difficulty here is that
the surface should have two pieces in the right cube and one piece 
in the left cube.  Use of the average along that cube face allows the 
algorithm to determine that the center is on the white side of the surface,
and make the correct choice of tilings in each cube.

Such problems arise any time there is more than one piece to the surface in any
particular hypercube.  In three dimensions the general approach has been
to exhaustively determine by hand all possible cases, up to available symmetries,
and then use a lookup table to calculate surface elements in that configuration.  Since
this relies on human calculation and has factorial scaling with dimension, this is
not done in four dimensions.  Instead, one can algorithmically determine how
to combine elements of the surface in lower dimensions to form surface elements
in higher dimensions.  One begins by interpolating points on the surface on every
grid edge between grid points inside and outside the surface.  Edges are constructed
by connecting surface points on each face.  This is done independently for each
of the eight cubes comprising the hypercube, six of which can be constructed by connecting
each cube face in space with its corresponding face at the other time.  If the cube 
contains less than six edges, there can be only one surface element within that cube 
and the surface with triangles can proceed.  The easiest way to do this is to calculate
the centroid of the surface element and generate triangles using the centroid and
each edge.  This is especially useful since the location of the centroid can be used 
as the point at which to interpolate hydrodynamic quantities for the Cooper-Frye
procedure defined in the previous section.  

If the surface in each cube consists of six or more edges, there might or might 
not be two distinct faces in this cube.  The edges must be then be sorted in
sequence to determine whether surface element are distinct and if so which
edges belong to which surface element.  If there are distinct elements, each
element is tiled individually, the centroids are calculated separately, and 
particles would be generated from each.  Once this two dimensional surface
has been calculated, the construction of the three-dimensional surface can
begin.  This proceeds analogously.  If there are few enough faces within
each cube, the polyhedron can immediately calculated from the centroid
of all of the surfaces and the sum of all the tetrahedra between the centroid
and the triangular faces.  If there is a possibility that there might 
be distinct surface elements, the lower dimensional surface elements need
to be ordered and separated into distinct polyhedra.  

Surface finding algorithms of this complexity were designed and developed
for hydrodynamic codes that include fluctuations, and as such are likely 
much more robust than is strictly necessary.  In fact, earlier versions of the
boost-invariant model used a surface finding algorithm that assumed the
surface was a well-defined function of the azimuthal angle and radius at each time 
step.  If this is the case, one can interpolate the surface radius at evenly spaced 
samplings in azimuthal angle at each time step.  These emission elements 
are easy to work with as there are no tiling issues or ambiguities related
to surface divisions between cells.  However, certain initial conditions
can cause the hot region to divide into two symmetric pieces in the 
transverse plane even for smooth initial conditions and the surface 
would violate our assumptions.  

This necessitated the integration of
these more robust surface finding algorithms.  Particle spectra calculated
from surface elements output from both algorithms were compared 
and agreed to within a few percent in the average transverse momentum,
though some of this could be attributed to rather large time-like elements
from the final, rapid collapse of the surface and associated interpolation
schemes.  The four-dimensional version of that algorithm was used 
for the full hydrodynamics code.  For test cases that we examined,
there were actually no grid cells that contained more than one surface
element.  This is partly due to our choice to generate the surface 
on the same density grid as the hydrodynamic evolution and partly
to the smoothness of our initial conditions.  In a sense this means that
our results are not dependent on the finest details of surface generation
and these methods are more than sufficient for our purposes.

\chapter{ Initial Conditions}

At this point, we have developed an algorithm for
viscous hydrodynamics for making predictions 
in heavy ion collisions.  We have discussed the 
equation of state and transport coefficients 
for the quark matter created and the hadron
gas that surrounds it, and we have shown
how this can be coupled to a gas calculation.
Hydrodynamics predicts the evolution of the density 
and collective velocity deterministically for a given system,
but it requires an initial condition, which is the final
ingredient in creating our model. The most common method
of initializing a hydrodynamics code is to provide the state of a system at a given time, 
though the equations do not demand that particular hypersurface.  

In the case of relativistic heavy ion collisions, the procedure for producing 
these initial conditions is highly uncertain and a principle source of uncertainty 
in interpreting observables.  For the initial energy density, there is 
ambiguity in the contribution of each nucleon-nucleon collision
to the hot phase of the collision.  Several models and the variation
in their predictions is discussed in the first section of this chapter.
In addition, it is likely that some
time elapses between the crossing of the original nuclei and the
onset of hydrodynamic behavior which should be accounted for.
Regardless of the microscopic description of the matter during 
this pre-equilibrium phase, we find that flow should be present
at the onset of hydrodynamics. This result is explained in the second section.
Finally, there is significant uncertainty in the initialization of the 
shear stress tensor ($\pi^{\mu \nu}$). While many groups have chosen
to use the Navier-Stokes values, it is not at all clear that this is
well motivated and can lead to aberrant behavior.
The issue of initializing the shear tensor
is discussed in the final section of this chapter.

\section{Independent Nucleon Models}

We begin by assuming that the initial stage of a heavy ion collision 
behaves like a collection of nucleon-nucleon collisions \citep{Miller:2007ri}.  Therefore,
we begin considering the density distribution of nucleons within each
nucleus, which is generally taken to be a Woods-Saxon 
\begin{equation} \label{eq:WoodsSaxon}
\rho(r) = \rho_0 \left[1 + \exp^{(r-R)/\xi} \right]^{-1},
\end{equation}
where the nuclei are assumed to be spherically symmetric and described by a radius (R), 
diffusiveness ($\xi$), and normalization ($\rho_0$). 
The energy density generated in the transverse plane of the hot region
should be proportional in some way to the areal density of nucleons which
is often called the thickness function and is given by
\begin{equation} \label{eq:thickness}
T_A(x,y) = \displaystyle\int_{-\infty}^{+\infty} dz \rho_A (x,y,z)
\end{equation}
where the subscript A indicates that this is the thickness of nucleus A.
If one thought that each pair of colliding nucleons contributed equally to the energy
density, then the energy density would be proportional locally to the 
product of the two thickness functions.
Another possibility is that each nucleon that participates in any collision 
contributes the same energy density to the hot region.  

To underline the difference between the participant and collisional scaling,
consider a region much smaller than the cross-section for interaction
in the transverse plane where there are four nucleons
present in one nucleus and three nucleons in the other nucleus.  If 
one counts by collisions there are twelve binary collisions that would take
place in this region, since each nucleon will collide with all possible
counterparts in the other nucleus.  In the alternate method of counting,
there are only seven participants in the collision.  For this reason,
participant scaling produces smaller energy densities at small
impact parameters where there are regions where many nucleons 
are present in each nucleus.

Locally, one can then calculate the energy density in both the 
collisional and participant scaling models.  For collisional scaling,
the energy density should be proportional to the cross-section and 
the product of the two thickness function:  $\epsilon(x,y)  = K \sigma_{\text{inel}} T_A T_B$,
where $\sigma_{\text{inel}}$ is the free nucleon-nucleon inelastic cross-section and
$K \propto (dE/dy)_{pp}$ is proportional to the energy contribution of each collision.
On the other hand, while participant scaling is easy to understand in a picture where there 
are a finite number of nucleons distributed throughout a nucleus,
 it can also be used for smooth density distributions.  
To reproduce participant scaling, the density should be scaled to
\begin{eqnarray} \label{eq:partScaling}
\epsilon(x,y) &\propto& T_A \left[ 1 - \left(1 - \frac{\sigma T_B}{B} \right)^B \right] + T_B \left[ 1 - \left(1 - \frac{\sigma T_A}{A} \right)^A \right], \\
\epsilon(x,y) &\approx& \frac{K}{2} \left[ T_A \left( 1 - e^{-\sigma T_B} \right) + T_B \left( 1 - e^{-\sigma T_A} \right) \right]
\end{eqnarray}
where A and B are the number of nucleons in each nucleus and are assumed to be large.
The factor of two emphasizes that if both $T_A$ and $T_B$ are small, the normalization
should be the same as for binary collision scaling.

The second part of  Equation \ref{eq:partScaling} suggests
that the function behaves like symmetric penetration of the thickness functions -- that is,
$T_A$ is attenuated as it interacts passing through nucleus B and vice versa.
This is not exactly the case. The addition of these two terms means that if $T_A >> T_B$, 
one of these terms still behaves like $\sigma_{\text{inel}}T_A T_B$ while the other 
is suppressed.  This means that doubling the larger thickness roughly doubles
the output density.  While this is what one expects from the wounded nucleon
picture, it seems likely that doubling the larger thickness would have a reduced 
effect.  

\begin{figure}
\centerline{\includegraphics[width=0.75 \textwidth]{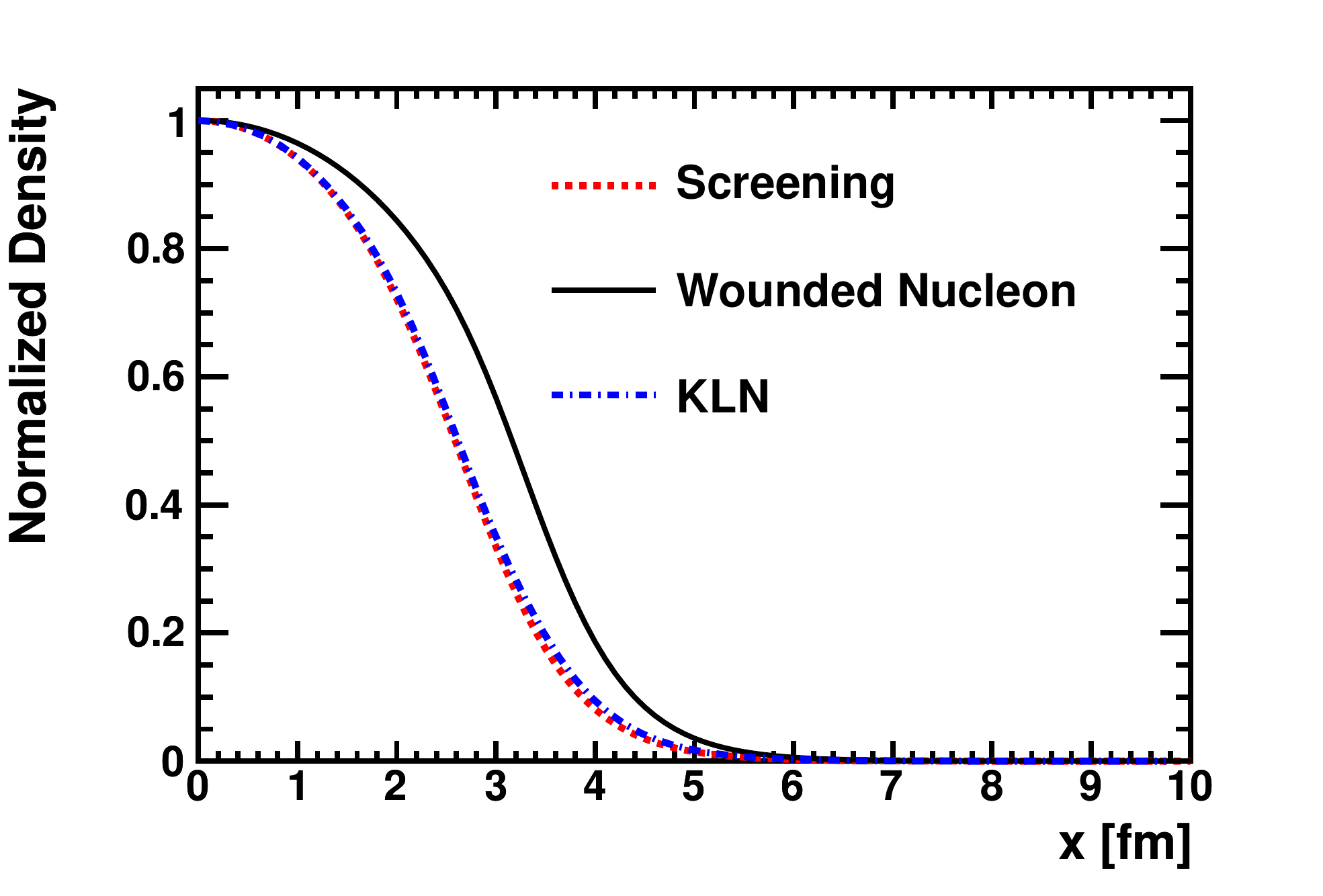}}
\caption{ 
Comparison of the normalized energy density along the short axis 
between wounded nucleon (black solid), KLN \citep{Drescher:2006pi} (blue dot-dash), and screening
(red dotted) models.  The wounded nucleon predicts the largest extent corresponding
to the smallest source eccentricity, while the screening model produces roughly
the same prediction as KLN.
}
\label{fig:ICDensX}
\end{figure}

Therefore, in order to more naturally take into account the expected screening,
we propose the following form for the energy density: 
\begin{equation} \label{eq:simpleSatIC}
\epsilon(x,y) = K \frac{\sigma_{\text{inel}}}{\sigma} \frac{2T_A T_B}{T_A+T_B} \left[ 1 - e^{-\sigma(T_A+T_B)/2} \right]
\end{equation}
where $\sigma$ is the cross-section associated with screening and should not be larger
than the full inelastic cross-section.
This form no longer has the property that it behaves like collisional scaling when
one density is much larger than the other but instead reduces the energy density
based on whichever nucleon density is smaller. This model achieves the same scaling 
in the limit that both nucleon densities are diffuse. In the higher density region
the scaling described in Equation \ref{eq:simpleSatIC}
behaves like participant scaling instead of collisional scaling and,
in fact, in regions where the thicknesses are similar
the model is the same as the wounded nucleon model.  In regions where
the thicknesses are very different, such as where the tails of one nucleus
overlap with the center of the other nucleus in the transverse plane, the
scaling described in Equation \ref{eq:simpleSatIC} 
falls off much faster than wounded nucleon scaling (Equation \ref{eq:partScaling}) 
as seen in Figure \ref{fig:ICDensX}.
This increases the eccentricity of the initial source which one na\"{i}vely
expects should increase the produces elliptic flow with other parameters constant.

Up to this point, we have assumed that the nuclear thickness functions are smooth
and interact probabilistically.  One might also force the nucleons to interact locally
and with precise locations through Monte Carlo methods. 
This is done by distributing nucleons in each nucleus by drawing from the
distribution $4 \pi r^2  \rho(r)$ for radial positions of the nucleons and flat distributions
for the azimuthal angle and the cosine of the polar angle.  Draws are performed once 
for each nucleon in each nucleus involved in the collision.  
The nucleons are then assumed to be black discs in the transverse plane of radius given by 
the total proton-proton inelastic cross-section for the collision energy
of interest.  Collisions occur any time black discs overlap in the transverse plane with 
the longitudinal distribution ignored.  One generally counts both the total number of
pairwise collisions, the number of binary collisions ($N_{BC}$),
and the number of nucleons participating in any collision, the number of wounded nucleons ($N_{WN}$).
To generate an energy density profile, one can place Gaussians at the location of the 
participants or collisions with arbitrary normalization and width, though the width should
certainly be no larger than the radius of the proton.  This procedure produces 
lumpy and fluctuating profiles, but profiles without significantly different trends
from the smooth variety of the wounded nucleon described in Equation \ref{eq:partScaling}
 \citep{Miller:2007ri}.  Furthermore, the design of the hydrodynamic model makes the 
study of fluctuations impossible.

Other improvements to the Glauber model have been considered at impressive 
length by the PHOBOS collaboration including variation of nuclear density
profile parameters, the inelastic nucleon-nucleon cross-section, hardcore nucleon
repulsion within the original nucleus, and several experimental considerations
that would affect interpretation of the multiplicity results \citep{Miller:2007ri}.  For reasonable variations
of the parameters, they find only $\pm 5\%$ influence on results for collisions mid-central
and central collisions for Au+Au collisions at $\sqrt{s} = 200$ GeV where
the largest deviations are for the experimental considerations.  Also of note
is that corrections for very peripheral collisions can be considerably larger, up to $\sim 20\%$.
Furthermore, they investigate differences between using smooth density profiles 
and using the average of many Monte-Carlo events to determine multiplicity scaling.  
They find that for mid-central collisions, smooth profiles predict $\sim 10\%$ fewer
particles produced by the 30-40\% centrality bin for Au+Au collisions at $\sqrt{s} = 130$ GeV.

One difficulty in comparing theoretical predictions to experimental data is that
the impact parameter is not known for any particular event, though the smallest
impact parameters should correspond to the highest multiplicity events.  Geometrically 
it is clear that the change in probability of a particular impact parameter scales with the impact parameter,
$d\sigma  = 2 \pi b ~ db$, like circular shells for integration in polar coordinates.  
Integrating this equation and assuming that no collisions take place beyond a
maximum impact parameter, the total cross-section is related to the maximum
impact parameter by $b_{\text{max}} = \sqrt{\sigma_{T}/\pi}$.
Experimentally events are separated into centrality classes which are the percentile
of the total number of particles produced in a particular collision but reversed such 
that the 0-5\% centrality class corresponds to the 5\% of all events containing the
most particles.  As long as particle production falls monotonically with impact
parameter, the relationship between the centrality class and the impact parameter
should be 
\begin{equation}
 \frac{\% {\text{cent}}}{100} = \frac{\pi b^2}{ \sigma_T} = \left( \frac{b}{b_{\text{max}}} \right)^2.
 \end{equation}
This allows one to calculate the average number of participant nucleon pairs for
each centrality class, which one can compare to the total number of charged particles
produced. Experimentally, the number of charged particles observed at 
midrapidity scales with the number of participants up to center of mass
collision energies around $\sqrt{s} = 100$ GeV \citep{Abelev:2008ab}.  
At $\sqrt{s} = 200$ GeV the deviation
from this scaling is $\sim5\%$ faster than linear, there are more observed charged particles
than predicted by the number of wounded nucleons, from 0-40\% centrality.
Some groups have suggested that this implies an increasing importance of hard processes
that might scale with the number of binary collisions and find that a mixture of 75\% participant
and 25\% collisional scaling reproduces data at this energy \citep{Kolb:2001qz}.

A related theory that has had success predicting longitudinal rapidity distributions
in d+Au systems is often called the Color-Glass-Condensate (CGC) 
\citep{Kharzeev:2002ei,McLerran:1993ni,Lappi:2006xc,Drescher:2006pi}.   The assumption
is that the gluon density distribution falls like the transverse momentum squared above a
saturation scale ($Q_s^2$) but remain constant below it, an ansatz which proved very
useful in describing the data collected in the HERA experiment \citep{Iancu:2003xm}.  The saturation scale depends 
on the density of partons in the original nuclei, 
generally taken to be the density of participants as defined in the Glauber model.
There are many versions of CGC \citep{Drescher:2006pi,Lappi:2006xc,Drescher:2006ca} that contain these features but there are not so many
direct constraints on such inputs as the apparent gluon density distribution at some
momentum scale.  This leads to quite some number of unique models.  
We present the equations for generating CGC initial conditions in one particular model \citep{Drescher:2006pi}.
The scattered gluon density is given in terms of the original gluon distributions ($\phi_{A_1}$ for nucleus 1) as
\begin{equation}
\frac{d^3N}{d^2p_T dy} \propto \frac{1}{p^2_T} \displaystyle\int^p_T dk_T^2  \alpha_s (k_T) \phi_{A_2} (x_1, k^2_T) \phi_{A_2} (x_2, (\vec{k}_T - \vec{p}_T)^2),
\end{equation}
where  $x_{1,2} = (p_T/\sqrt{s}) e^{\mp y}$ and y in the longitudinal rapidity.
The unintegrated gluon distributions and saturation scale are given by
\begin{eqnarray}
\phi_{A} ( x, k_T^2) &=& \frac{1}{\alpha_s(Q_s^2)} \frac{ Q_s^2}{\max{(Q_s^2, k_T^2)}} \left( \frac{n_{\text{part}}^A}{T_A} \right) (1-x)^4, \\
Q_s^2(x) &=& \frac{2 T_A^2 \text{GeV}^2}{n_{\text{part}}^A} \left( \frac{ \text{fm}^2}{1.53}  \right) \left( \frac{0.01}{x} \right)^{0.288},
\end{eqnarray}
where each is implicitly a function of the transverse position, rapidity, and nucleus.  
Also, $n_{\text{part}}^A$ is the participant
density for nucleus A as given by the appropriate term of Equation \ref{eq:partScaling}.
In this sense, the CGC is a reweighting of Glauber densities based on 
classical chromodynamics arguments about the saturation of
gluon momentum distributions.  Finally, we note that CGC models include
rapidity dependence, which is one of its great successes notably for d+Au collisions 
\citep{Kharzeev:2002ei}
that is not predicted by Glauber models where that would be added ad hoc.

\begin{figure}
\centerline{\includegraphics[width=0.75 \textwidth]{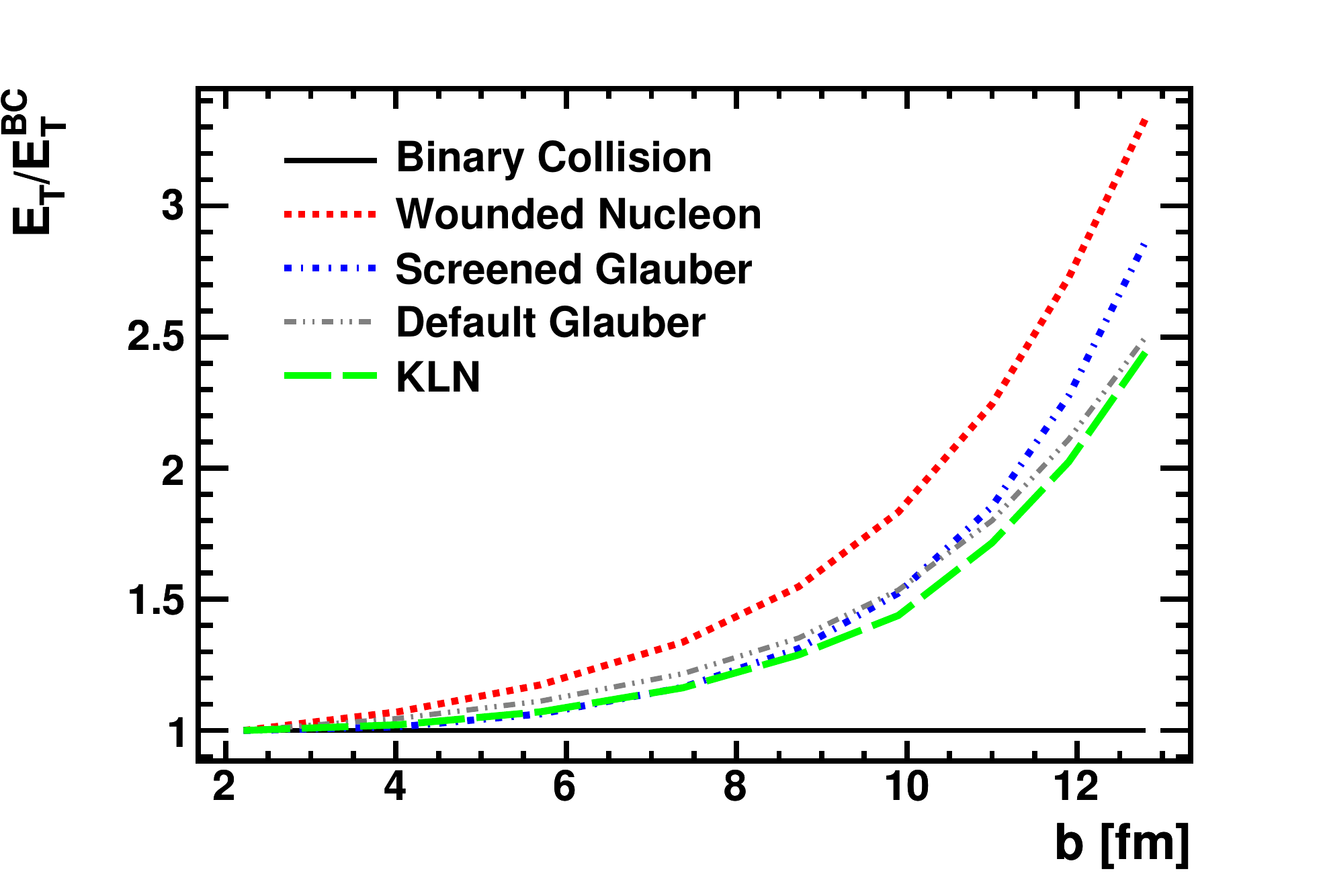}}
\caption{ 
Comparison of the normalized transverse energy in initial conditions from
KLN (green dashed), wounded nucleon (red dotted), 
binary collision (black solid), screening (blue dot-dash), and 
a mixture wounded nucleon and binary collision (gray double dot-dashed) 
chosen to reproduce experimental multiplicity scaling.  Since the total overlap
area scales with total particle production and normalization is set
by the most central bin, all curves are defined to be unity at b=2.23 fm before
dividing by the binary collision result at each impact parameter.
The wounded nucleon model therefore predicts the slowest scaling
of multiplicity with impact parameter while binary collision scaling would
produce the most rapid scaling.  The screening model and KLN predict
similar scaling with impact parameter as the best fit mixture of wounded
nucleon and binary collision scaling, therefore we expect these models
to follow experimental data in this respect.
}
\label{fig:ICNComp}
\end{figure}

\begin{figure}
\centerline{\includegraphics[width=0.6 \textwidth]{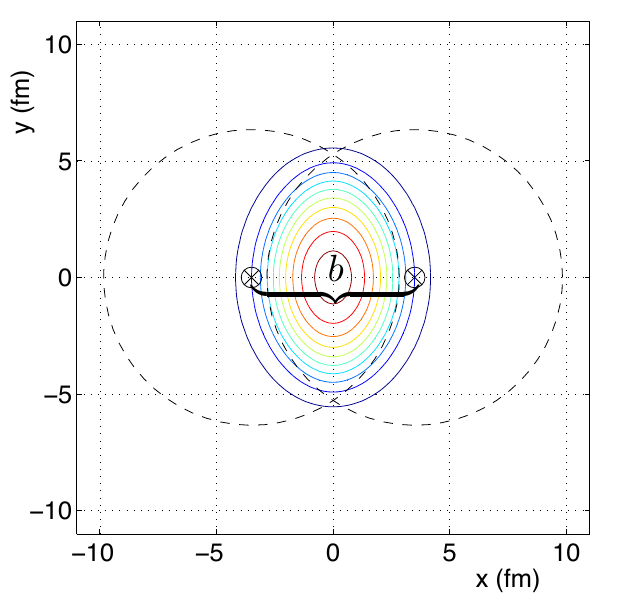}}
\caption{ 
Schematic of source of initial eccentricity as it originates directly from the finite
impact parameter \citep{kolbThesis}
}
\label{fig:ICEccentricity}
\end{figure}

\begin{figure}
\centerline{\includegraphics[width=0.75 \textwidth]{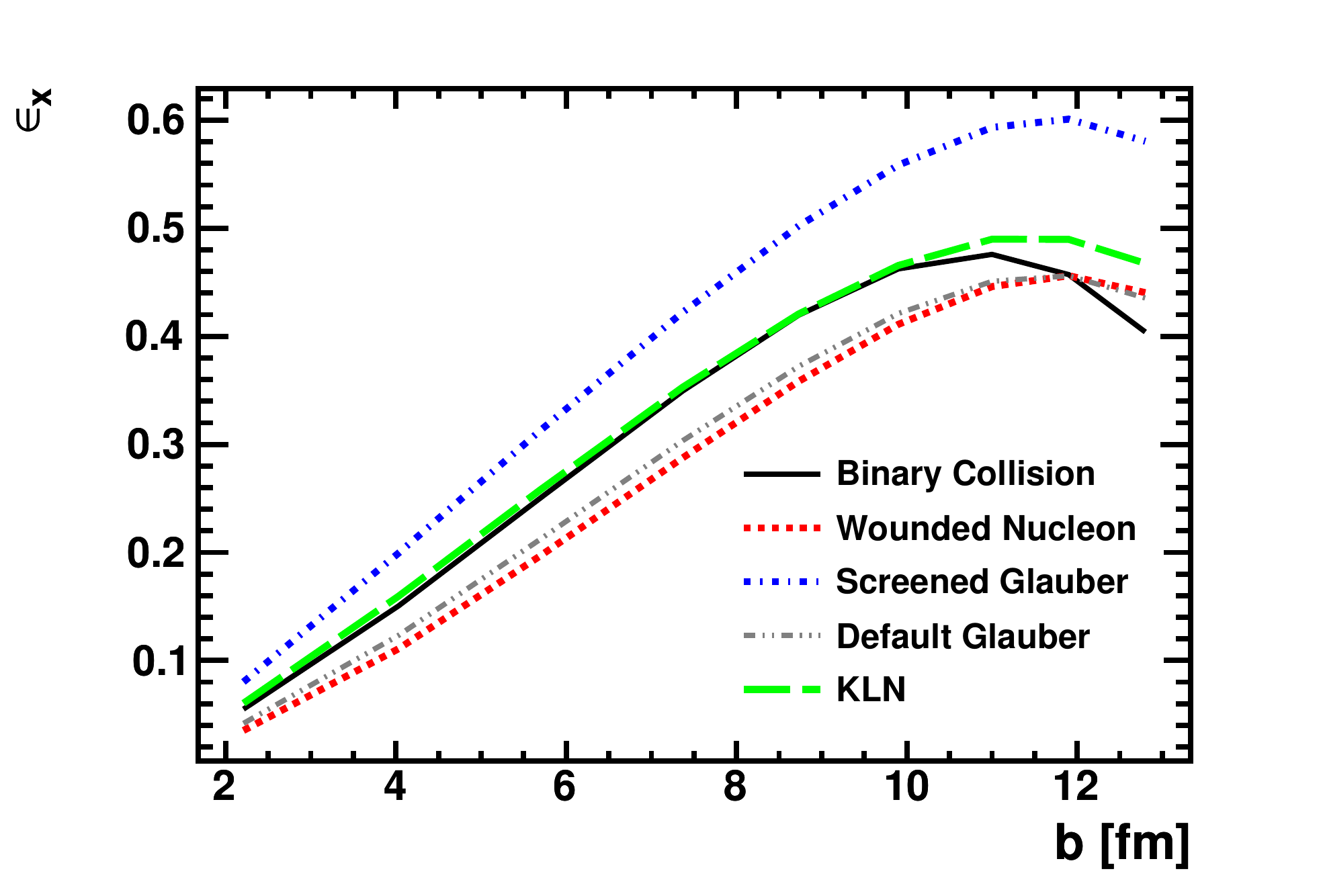}}
\caption{ 
Comparison of initial source eccentricity between KLN (green dashed), 
wounded nucleon (red dotted), binary collision (black solid), screening (blue dot-dash), 
and a mixture wounded nucleon and binary collision (gray double dot-dashed) 
chosen to reproduce experimental multiplicity scaling.  Wounded nucleon
scaling predicts the smallest initial source eccentricity and is not greatly increased
by the inclusion of some binary collision scaling to reproduce multiplicity scaling.
The screening model predicts the largest eccentricity of the models considered here,
including KLN, though other CGC calculations of initial conditions predict eccentricities
this large.  This suggests theoretical uncertainty of roughly 50\% in the source 
eccentricity that will propagate to uncertainty in the shear viscosity.
}
\label{fig:ICEpComp}
\end{figure}

As stated before, there are many varieties of CGC and few of the parameters are well known
so we consider only the energy density profile predictions.  These profiles tend to 
predict higher initial source eccentricities and somewhat faster total density scaling
with impact parameter than participant scaling in the Glauber model. 
To compare the scaling across models we compare the integrated 
energy density at midrapidity as a function of impact parameter.
Each is normalized to give
the same total number for the average central collision impact parameter
with b=2.21 fm since this normalization is a free parameter in all of the models.
For ease of interpretation, in Figure \ref{fig:ICNComp}, 
we divide by the binary collision result.  All the curves lie above one, 
meaning that collisional scaling predicts the most rapid
scaling with impact parameter, while participant scaling varies the most slowly.
The double-dot dashed (green) line shows the linear combination of collisional and participant 
scaling found to reproduce the data.  Note that both the CGC variety and the simple
saturation model given in Equation \ref{eq:simpleSatIC} predict similar scaling
to this linear combination, and we therefore expect that these models will all
predict similar multiplicity scaling with impact parameter.

Another important experimental observable affected by the initial condition
is the elliptic flow.  In the hydrodynamic picture, elliptic flow is the result 
of initial source eccentricity. Hydrodynamics generates flow in proportion 
to pressure gradients leading to faster collective expansion in the short direction 
of the source. Viewed in the transverse plane, the initial eccentricity 
comes from the overlap geometry of the overlapping nuclei.
As shown in Figure \ref{fig:ICEccentricity}, for impact parameters of roughly
the nuclear radius, the overlap region is shorter along the impact parameter than
in the orthogonal direction. 

In addition to predicting different multiplicity scaling, models of initial conditions
predict different initial eccentricities.  Figure \ref{fig:ICEpComp} shows the 
comparison of the eccentricity predicted by the various models
for initial densities, as defined by
\begin{equation} \label{eq:eccentricityDef}
\epsilon_x = \frac{ < y^2 > - < x^2> }{ < y^2 > + < x^2> }
\end{equation}
where the averages are over the transverse plane weighted by the energy density.
Participant scaling turns out to produce the least source eccentricity for this set of 
models with the mixture of collisional and participant scaling that reproduces multiplicity
scaling being almost as small.  The screening model produces the most eccentricity
of these models, but this is more a function of the choice of CGC model; others predict
eccentricities as large or larger compared to the saturation model \citep{Drescher:2006pi}.

\section{Initial Flow}

\subsection{Early Flow Model}

The energy density alone is not enough to initialize even ideal hydrodynamics
as the collective velocity is not determined.  The early assumption of modelers
was that the system should exhibit no collective behavior prior to thermalization
and that the fluid should begin at rest in the transverse direction
\citep{Kolb:2000sd,Song:2007ux,Hirano:2002hv,Schenke:2010nt} .  This is in
direct contrast to the longitudinal initialization, where boost invariance
means that the system is borne with strong velocity gradients despite the
lack of acceleration in any comoving frame. If the longitudinal treatment 
were translated to the transverse case, 
that would assume that particles free-streamed prior to thermalization, 
though this is not strictly a requirement for the longitudinal
assumptions to hold.

Unfortunately, we find that the conclusions one would draw about the 
structure of the matter are sensitive to the assumptions one makes about 
the dynamics of the system prior to equilibration \citep{Vredevoogd:2008id}.  
In fact, early hydrodynamic models using a very stiff equation of state
predicted that the system would have to thermalize after only 0.1 fm/c
to achieve the correct amount of flow in the total collision time suggested 
by the longitudinal extent of the system \citep{Heinz:2002un}.
This result is not difficult to understand: the development
of elliptic flow is initially parabolic for a system starting from rest so,
as in a sprint, a difference in starting time is often more critical than top speed.

\begin{table}
\centering
\begin{tabular}{|c|c|c|c|c|}\hline
{\bf Model} & $T^{xx}=T^{yy}$ & $\tau^2 T^{\eta \eta}$ & $T^{\tau \tau}$ \\ \hline
Longitudinal electric field & $T^{\tau \tau}$ & $-T^{\tau \tau}$ & $\sim$constant\\ \hline
\parbox[c]{3.3in}{Free streaming massless particles,\newline
two-dimensional relativistic gas\newline
or fields from incoherent longitudinal currents}
 & $T^{\tau \tau }/2$ & $0$ & $\sim 1/t$\\ \hline
Ideal hydrodynamics of massless gas & $T^{\tau \tau}/3$ & $T^{\tau \tau}/3$ & $\sim 1/t^{4/3}$\\ \hline
\end{tabular}
\caption{\label{table:Tij}
Elements of the stress energy tensor for simple models at early times \citep{Vredevoogd:2008id}.  
A wide variety of reasonable values for the stiffness of the transverse equation of state are represented by the models.
}
\end{table}

This puts a premium on understanding the model dependence of the initial conditions
to the hydrodynamic phase.
We consider models in which the stress-energy is conserved, $ \partial_\mu T^{\mu \nu} =0$,
and $T^{0i}(\tau=0) = 0$ for spatial indices due to symmetry.
The models differ in terms of the their assumption about the relationship between
$T^{00}$ and $T^{xx}$ and $T^{yy}$.
As described in Table \ref{table:Tij}, this framework can describe a model based around 
longitudinal fields like the CGC, which would predict very strong transverse 
pressure and negative longitudinal pressure to ideal hydrodynamics, 
with complete transverse-longitudinal symmetry in terms of a single parameter $\kappa$
given by
\begin{equation} \label{eq:KappaDef}
\kappa = \frac{T^{xx} + T^{yy}}{2 T^{\tau \tau}} = \frac{T^{rr} + r^2 T^{\phi \phi}}{2 T^{\tau \tau}} ,
\end{equation}
where we include the equivalent polar form for use later.
The value of $\kappa$ depends on the microscopic description of the matter,
for instance $\kappa = 1/3$ in ideal hydrodynamics while $\kappa = 1$ for
longitudinal classical fields.
We assume that the stress-energy tensor is traceless which allows us to 
determine all of the elements of the shear tensor.
We further assume that $\kappa$ varies with proper time but not with spatial coordinates,
which implies that the important degrees of freedom do not depend on location though
they may depend on time.

The evolution of the system is given by the conservation of the stress-energy tensor
and the system of equations are closed by the value of $\kappa$.
Since we are primarily interested in physics at midrapidity and in the early generation
of flow, we will consider a one-dimensional system that varies only 
in the transverse direction undergoing a boost invariant expansion.  For 
such a system, only the elements of the stress-energy tensor and $T^{\tau x}$ will be 
non-zero where $x$ is the direction of the density gradient.

For small velocities and neglecting the longitudinal expansion, the Euler equation
is exact and 
\begin{equation} 
\label{eq:zerolongexp}
\frac{\partial v^x}{\partial t}=\frac{-\partial_x T^{xx}}{T^{tt}+T^{xx}} = \frac{-\kappa}{1+\kappa} \frac{\partial_x T^{tt}}{T^{tt}}.
\end{equation}
Usually this implies that $T^{xx}$ behaves as the effective transverse pressure,
and larger values of the effective pressure lead to more rapid expansion.
The natural conclusion would be that models with larger values of $\kappa$ would
lead to larger acceleration.  The longitudinal expansion turns out to include a very important 
correction to this conclusion.

To see this, consider the equation of momentum conservation in polar-Bjorken coordinates
\begin{equation}
\label{eq:consmom0}
\partial_\tau T^{\tau r}=-\partial_r T^{rr} - \partial_\phi T^{\phi r} - \partial_\eta T^{r \eta} - \frac{T^{\tau r}}{\tau} - \frac{1}{r} \left( T^{rr} - r^2 T^{\phi \phi} \right).
\end{equation}
The final three terms are derived from affine connections but have simple explanations in
Cartesian coordinates.
The longitudinal expansion means that $v^z = z/t$ and therefore $T^{zx} \approx (z/t) T^{tx}$
for small $z$.
Furthermore, if the system has rotational invariance, then 
\begin{equation}
T^{xy} (y=0) \approx (y/x)  (T_{xx} - T_{yy}) =(y/x) (T^{rr} - r^2 T^{\phi \phi} ) .
\end{equation}
Initially and to first order in $u^r$ for the hydrodynamic case, 
$T^{rr} = r^2 T^{\phi \phi}$.
For the hydrodynamic case,
both are roughly the pressure though there is a second order correction.
More generally, the term is explicitly zero initially and
we will assume that this term remains small.

Similar considerations for energy conservation yield the equations of motion
\begin{eqnarray}
\label{eq:conse}
\partial_\tau T^{\tau \tau} &=& \frac{-1}{\tau}(T^{\tau \tau}+ \tau^2 T^{\eta \eta})-\left(\partial_r+\frac{1}{r}\right)T^{r \tau}, \\
\label{eq:consmom}
\partial_\tau T^{\tau r} & \approx & -\partial_r T^{rr}-\frac{T^{r \tau}}{\tau}. 
\end{eqnarray}
where the approximation in Equation \ref{eq:consmom} is ignoring the term
from rotational symmetry.
The quantity that is of interest is the ratio $T^{\tau r}/T^{\tau \tau}$.  In the 
hydrodynamic picture, this quantity scales linearly with collective velocity
for small velocities and generally should be thought of like a normalized 
momentum density.  We can easily calculate the proper time derivative to be
\begin{eqnarray}
\partial_\tau \left( \frac{T^{\tau r}}{T^{\tau \tau}} \right) &=& \frac{\partial_\tau T^{\tau r}}{T^{\tau \tau}} - \frac{ T^{\tau r} \partial_\tau T^{\tau \tau}}{ (T^{\tau \tau})^2}, \\
\partial_\tau \left( \frac{T^{\tau r}}{T^{\tau \tau}} \right) &\approx & \frac{-\partial_r T^{rr}}{T^{\tau \tau}} + \frac{\tau^2 T^{\eta \eta} T^{\tau r}}{ \tau (T^{\tau \tau})^2},
\end{eqnarray}
where we ignore second order terms,
$(T^{\tau r} / T^{\tau \tau})^2$, including the term proportional to $\partial_r T^{r \tau}$.

For small times the change in this ratio is linear, so we can solve for the acceleration $\alpha(r)$,
\begin{eqnarray}
\frac{T^{\tau r}}{T^{\tau \tau}} &=& \alpha(r) \tau, \\
\alpha(r) &=& \frac{-\partial_r T^{rr}}{T^{\tau \tau} - \tau^2 T^{\eta \eta}},
\end{eqnarray}
where $\tau^2 T^{\eta \eta} = T^{zz}$ is not small even for small proper times.
We now employ all of our assumptions which are that 
\begin{itemize}
\item the radial and azimuthal pressure remain the same --  $T^{rr} = r^2 T^{\phi \phi} = \kappa(\tau) T^{\tau \tau}$,
\item the anisotropy of the stress-energy tensor is the same everywhere -- $\kappa(r,\tau) = \kappa(r)$, and 
\item a traceless stress-energy tensor -- $g_{\mu \nu} T^{\mu \nu} = 0$.
\end{itemize}
This leads to 
\begin{eqnarray}
\alpha(r) &=& \frac{ - \kappa \partial_r T^{\tau \tau} }{2(T^{rr} + r^2 T^{\phi \phi} )},\\
\alpha(r) &=& \frac{ - \partial_r T^{\tau \tau}}{2 T^{\tau \tau}}. 
\end{eqnarray}
This means that the development of momentum density at small times,
\begin{equation} \label{eq:earlyFlowResult}
\frac{T^{\tau r}}{T^{\tau \tau}}  = \frac{ - \partial_r T^{\tau \tau}}{2 T^{\tau \tau}} \tau,
\end{equation}
does not depend on $\kappa$.  Therefore classical, longitudinal
fields produce the same amount of flow as free streaming particles by this measure.
This is exactly the opposite conclusion suggested by the Euler equation, Eq. \ref{eq:zerolongexp},
where we anticipated that the development of flow would be directly proportional
to $\kappa$.  Note that the boost invariant expansion is critical to this result.  
Furthermore, the result to first order is exactly twice the acceleration that one would
predict from Eq. \ref{eq:zerolongexp} which one might have anticipated from the 
tests performed in Chapter 4.

The choice to measure the pre-equilibrium flow as ratios of the temporal 
components of the stress-energy tensor was made with an eye toward thermalization.
When the system undergoes a change in the underlying degrees of freedom, for example in
the sudden decay of the longitudinal fields, conservation of the stress-energy tensor,
$\partial_\mu T^{\mu \nu} =0$, should remain valid.  
As before, we assume that this transition is sudden in proper time,
meaning that the transition hypersurface can be defined by a normal vector, 
$n^\mu = (1, 0, 0, 0)$.
Integrating the conservation equation across this hypersurface gives
\begin{eqnarray}
0 &=&\int_{\tau-\delta \tau}^{\tau+\delta \tau} \left(\partial_\tau T^{\tau \alpha}+\partial_i T^{i \alpha}\right), \\
\nonumber
0 &=&T^{\alpha \tau}(\vec{r},\tau+\delta \tau)-T^{\alpha \tau}(\vec{r},\tau-\delta \tau),
\end{eqnarray}
where the term proportional to a spatial derivative is zero because $\delta \tau$ is 
small and that term is not divergent.
This result means that, as long as the transition between descriptions is time-like,
the temporal components of the stress-energy tensor cannot change suddenly.  
In the case that the system is suddenly thermalizing into an ideal hydrodynamic
or Navier-Stokes viscous hydrodynamic system, these four components are sufficient 
to determine all the remaining components of the stress-energy tensor.  
The resulting flow profile could be rather different if one assumed thermalization
to Navier-Stokes viscous pressures instead of the ideal pressures,
though all of the early models would still produce similar flow profiles.  Also, 
Israel-Stewart hydrodynamics does not prescribe the value of the six
independent components of the shear tensor, $\pi^{\mu \nu}$, and therefore
one would need a prescription to determine these values with possibilities
including maintaining all the components of the stress-energy tensor.

If the transition is not time-like in the laboratory
frame but is time-like in the comoving frame or even space-like, different 
elements of the stress-energy tensor might be conserved instead.  This more 
general case can be summarized in terms of $n^{\mu}$, the surface normal fourvector,
by
\begin{equation}
\delta ( n_\mu T^{\mu \nu} ) = 0,
\end{equation}
and is often called the Rankine-Hugoniot relation when discussed in 
the context of hydrodynamic shock fronts.  In the case of static surface facing the x-direction,
the components $T^{x \mu}$ are conserved from which ideal hydrodynamics
can be initiated as in the time-like case.

We are particularly interested in the influence of this flow precursor on the anisotropic
momentum distribution observed in the final state, which is the second-Fourier
component of the azimuthal momentum distribution
\begin{equation} \label{eq:v2Def}
v_2 \equiv \left\langle \cos{ 2(\phi - \phi_{\text{RP}}) } \right\rangle,
\end{equation}
where $\phi_{\text{RP}}$ is the azimuthal angle of the reaction plane, which should
roughly lie along the impact parameter though fluctuations may mean that 
it deviates from this.
This requires running a full hydrodynamic calculation to generate an emission
isosurface, generating millions of particles as emitted from the surface, and following those 
particles through their many interactions before one can calculate the final effect.
It is much more convenient is to use a hydrodynamic substitute that can be 
calculated directly from the stress-energy tensor.  This measure has been
proven useful as a proxy for the final observable 
\citep{Teaney:2001av,Kolb:2000sd} and is often called the 
momentum-space anisotropy or eccentricity by analogy to Eq. \ref{eq:eccentricityDef}. 
It is defined as
\begin{equation} \label{eq:momAniDef}
\epsilon_p \equiv \frac{ \int d x_T (T^{xx} - T^{yy} )}{ \int d x_T (T^{xx} + T^{yy} )}
\end{equation}
which for free particles would be the same as $v_2$ up to a factor of roughly two \citep{Luzum:2008cw}.
Note that since the discontinuity to our system is time-like, $T^{xx}$ and $T^{yy}$
are not continuous and this measure of the expected $v_2$
can change dramatically during thermalization.

\subsection{Three Scenarios}

The previous subsection points out that one should expect the same development
of flow from models with rather different underlying physics.  We now consider
those models in more detail to confirm that our findings hold for the range
of hydrodynamic starting times considered possible.  To do this, we calculate
the predictions of each model and monitor the development of our flow measure
and our anisotropic flow measure.  

The models we consider are from the range $1/3 \le \kappa \le 1$.  The smallest 
value of $\kappa$ is for ideal hydrodynamics of a conformal fluid.  Since the energy
density is singular at $\tau = 0$, it is convenient to transform to equations of motion
for flow and scaled energy density,
\begin{eqnarray}
w&\equiv&\frac{T^{\tau r}}{T^{\tau \tau}},\\
\nonumber
U&\equiv&T^{\tau \tau}\tau^{4/3},
\end{eqnarray}
where the choice of $\tau^{4/3}$ motivated by the expected rate of energy density decay
for longitudinal ideal hydrodynamics.
The equations of motion for $w$ and $U$ are derived from Equations \ref{eq:conse} and \ref{eq:consmom}
and are solved numerically for a fixed radial mesh.

The largest value of $\kappa$ is meant to mimic the effects of the CGC by considering 
a classical, coherent longitudinal electric field.  This is not exactly what one expects
from CGC models, as the fields should be coherent only on length scales corresponding
to the saturation scale.  On the other hand, this model provides an extreme example
of how strong the effective transverse pressure could be with $T^{\tau \tau} \approx T^{rr}$ 
and the maximum value of $\kappa = 1$.
For a single pair of oppositely charged particles created at $x=y=0$, the Lienart-Wiechart 
form for the vector potential can be used to generate the electric and magnetic fields,
\begin{eqnarray}
\label{eq:maxwellpoints}
A_z(r,\tau)&=&2q\int_0^\infty d\tau' \delta(x^2+y^2-(\tau-\tau')^2),\\
\nonumber
E_z(r,\tau)&=&4q\delta(r^2-\tau^2),\\
\nonumber
B_\phi(r,\tau)&=&-E_z(r,\tau).
\end{eqnarray}
Then for the charge density $\rho(x,y)$, one can integrate over the charge density
\begin{eqnarray}
E_z(x,y,\tau)&=&2\int d\phi~\rho(x-\tau \cos\phi,y- \tau \sin\phi),\\
\nonumber
B_x(x,y,\tau)&=&2\int d\phi~\rho(x-\tau \cos\phi,y-\tau \sin\phi)\sin\phi,\\
\nonumber
B_y(x,y,\tau)&=&-2\int d\phi~\rho(x-\tau \cos\phi,y-\tau \sin\phi)\cos\phi.
\end{eqnarray}
One can then calculate elements of the electromagnetic stress-energy
tensor from
\begin{eqnarray}
T^{\tau \tau} &=& \frac{1}{8\pi} \left[ E^2 + B^2 \right] , \\
T^{\tau i} &=& \frac{1}{4\pi} \left( \vec{E} \times \vec{B} \right)^i , \\
T^{ij} &=& \frac{1}{4\pi} \left( E^i E^j + B^i B^j - \frac{\delta^{ij}}{2} \left[ E^2 + B^2 \right] \right).
\end{eqnarray}

Since the region between the nuclei is charge-free, the evolution of the
fields can be calculated by the conservation of the stress-energy tensor.
The initial state of the charge density was chosen such that the resulting
energy density profile would be the same as for other models. For the 
Gaussian profiles that we will consider in the next section, this means
increasing the charge density radius by a factor of $\sqrt{2}$.

If instead the fields are generated by incoherent sources, that is arbitrarily
oriented charged pairs receding from one another at the speed of light,
Equation \ref{eq:maxwellpoints} still describes the fields due to each
source. The arbitrary orientation means that the fields from each source
will be arbitrarily oriented as well, but the elements of the stress-energy
tensor are all the same because the energy density always moves
outward from each pair.  In this case the longitudinal electric field and
the azimuthal magnetic field have the same strength which gives 
$T^{zz} = 0$ and $\kappa = 1/2$.  This model thus behaves exactly
the same as massless particles free-streaming in the transverse
plane.  This is not surprising as both models are of point sources
emitting outgoing transverse waves, and the distinction between
an electromagnetic pulse and partons is unimportant in the limit 
that the partons are massless.

These two electromagnetic models are meant in some sense as thought
experiments but should bound a more realistic theory.  The limit 
that the sources are point particles will break down at these length
scales, which are roughly the length and time scales where the uncertainty
principle is relevant. In low momentum QCD this coherence length
is generally called the saturation scale and is approximated by 
$\lambda \sim 1/E$ where E is the typical energy of a parton.  For time and length scales
that are smaller than $\lambda$, the stress-energy tensor should
be like the coherent limit of longitudinal classical fields, while at longer scales 
the incoherent limit should apply.  This does not affect the result in Equation \ref{eq:earlyFlowResult}
as $\kappa$ is allowed to vary with time as long as it does not depend
on location which is just what one would expect from these considerations.

\subsection{Model Results}

We initialize all three models -- ideal hydrodynamics, incoherent field, 
and coherent fields -- with Guassian profiles for the effective energy density
\begin{equation}
T^{\tau \tau}(x,y)\propto \exp\left\{-\frac{x^2}{2R_x^2}-\frac{y^2}{2R_y^2}\right\}.
\end{equation}
Our focus will be the effect on the radial flow that would be 
present upon initializing an ideal hydrodynamics model with the equation
of state of a massless gas -- $P = \epsilon/3$.  Recall that
the time-like transition means that time-like elements of the stress-energy
tensor are conserved -- $T^{\tau \mu}$.  This means that one can calculate
the collective flow observed after the transition, denoted $u_r'$,
 just from the ratio $T^{\tau r}/T^{\tau \tau}$:
 \begin{equation}
 \label{eq:T0alphamatch}
T^{\tau \tau}=\frac{\epsilon}{3} \left(4 \gamma^{\prime 2}- 1 \right), ~~
T^{\tau r}=\frac{4\epsilon}{3}\gamma 'u'_r.
 \end{equation}
For the model that assumes ideal hydrodynamic evolution for the whole
time, there is no change in the velocity and $u_r = u_r'$. 
In other models this will not be the case.
We investigate transitions at 
a few times that encompass the assumed range of starting times, 
$\tau = \{ 0.3, 0.6, 1.0\} $ fm/c, and we take the radii to be equal
$R_x = R_y = 3.0$ fm.  

\begin{figure}
\centerline{\includegraphics[width=0.55 \textwidth]{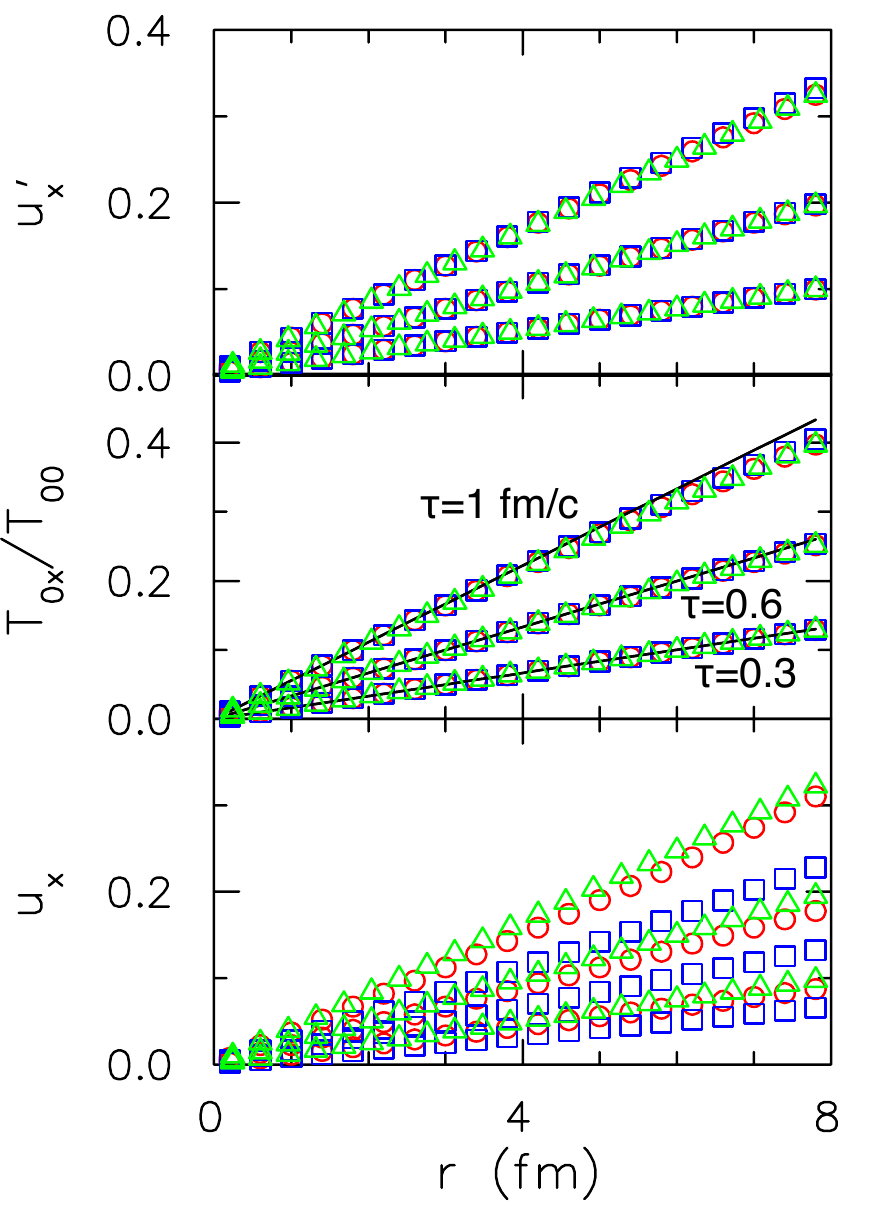}}
\caption{ 
Bottom panel:  The velocity of the frame in which $T^{0x}=0$
for coherent electromagnetic fields (blue squares), for incoherent 
fields or a non-interacting gas (red circles), and for an ideal liquid 
(green triangles) shown at three time steps, $\tau = \{0.3, 0.6, 1.0\}$ fm/c.
The models predict significantly different amounts of 
radial flow, though in the opposite hierarchy suggested by the transverse
pressure due to the effects of the Bjorken expansion. This quantity may
change during thermalization. \newline
Middle Panel: The ratio of $T^{\tau r}/T^{\tau \tau}$ at the same times
as the bottom panel, a measure of the 
flow developed in the early stage that will be maintained upon transition
to ideal hydrodynamics.  All the models predict the same evolution of this 
quantity up to 1 fm/c, which is roughly the domain of the model,
and all are fairly well described by the linear approximation. \newline
Top Panel:  The collective velocity that would be apparent following transition to
ideal hydrodynamics at the same three times as a function of radial position.  
Confirms that all of the models 
make the same prediction for the initialization of hydrodynamics.
}
\label{fig:InitFlowUs}
\end{figure}

For Gaussian profiles and to first order in time, Equation \ref{eq:earlyFlowResult}
gives the evolution of the flow
\begin{equation}
\label{eq:lineartime}
\frac{T^{\tau r}}{T^{\tau \tau}}\approx \frac{r \tau}{2R^2},
\end{equation}
which using Equation \ref{eq:T0alphamatch} can be used to calculate
the velocity before and after transition
\begin{equation}
\label{eq:analyticflow}
u'_r \approx \frac{3r\tau}{8R^2},
\end{equation}
which are accurate to first order in time only but are useful to
benchmark our expectations.

Figure \ref{fig:InitFlowUs} shows the results of all three models in
terms of the velocity observed with the model itself (bottom panel),
the conserved flow conserved by the transition to hydrodynamics (middle panel),
and the flow observed within the hydrodynamic model following the 
transition (top panel).  The results for the velocity observed within 
each model show considerably different results depending on
the effective transverse pressure of that model.  Somewhat counterintuitively,
the ideal hydrodynamic model predicts the largest amount of apparent
pre-equilibrium collective velocity despite having the smallest pressure, 
$\kappa = 1/3$, while the coherent field model, $\kappa = 1$,
predicts the largest value. However, even for times up
to $\tau \approx 1.0$ fm/c, the ratio $T^{\tau r}/T^{\tau \tau}$ is quite
well approximated by the linear estimate in Equation \ref{eq:analyticflow},
shown as a solid line, and all the models track with one another 
even more closely.  Finally, we include a demonstration that this 
produces the same flow within the hydrodynamic model 
though this is guaranteed by the temporal elements
of the stress-energy tensor being the same.

\begin{figure}
\centerline{\includegraphics[width=0.6 \textwidth]{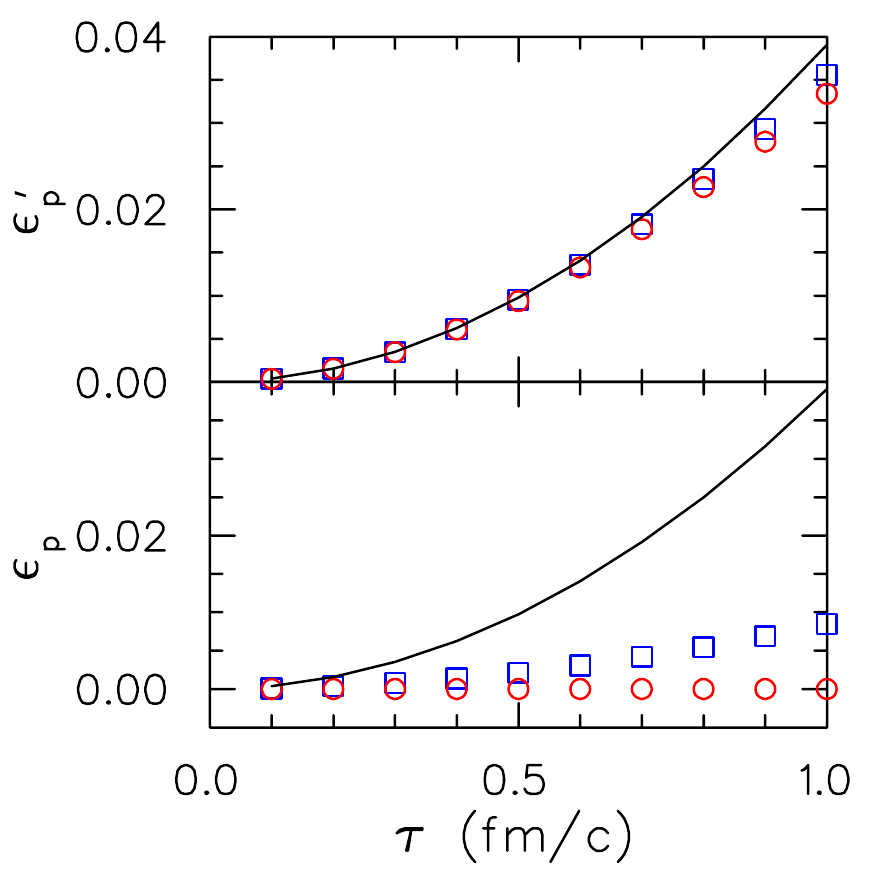}}
\caption{ 
(Color in electronic version)
Bottom Panel shows the momentum space anisotropy apparent within
each model.  Free-streaming particles (red circles) produce no apparent
anisotropic flow while coherent fields (blue squares) produce much less than the analytic
result of Eq. \ref{eq:earlyEpAnalytic} which is accurate to first order in time
for ideal hydrodynamics (black line).  \newline
Top Panel shows the momentum space anisotropy observed in an ideal
hydrodynamic model if the early flow model thermalized suddenly at that
time.  All of the models produce the same elliptic flow in the hydrodynamic
source at very early times though they differ slightly by $\tau = 1.0$ fm/c.
This helps to explain why the matter created at RHIC appears to thermalize
instantly.
}
\label{fig:InitFlowEp}
\end{figure}

While the development of radial flow is somewhat important in determining 
the average transverse momentum and can have some effect on the 
collision time, the elliptic flow is of greater concern.  This is for two reasons,
early times affect elliptic flow more dramatically and the discriminating power
of elliptic flow is expected to be larger for interesting theoretical parameters
like the shear viscosity.  Since anisotropic flow is the result of eccentricity of
the initial density, we alter the initial profile by change the radius in the 
y-direction to be $R_y =2 $ fm instead of 3 fm.  We then run the same 
calculation as above for the coherent and incoherent fields in the transverse
plane.  Development of anisotropic flow is calculated in terms of the 
difference of stress-energy tensor elements as defined by $\epsilon_p$ 
in Equation \ref{eq:momAniDef}.  Figure \ref{fig:InitFlowEp} shows the 
apparent anisotropic flow within the model itself (bottom panel) and 
contrasts that with the anisotropic flow that would be apparent in the 
hydrodynamic model if it were initialized at that time.  Models based 
on free-streaming particles or incoherent fields (red circles) do not 
generate any anisotropic flow at all, while coherent fields generate
considerably less than an estimate of the flow generated by ideal
hydrodynamics.  And yet, whenever the system undergoes 
thermalization, the conservation of the time-like components 
of the stress-energy tensor mean that flow immediately appears 
regardless of the previous description.

For short times, the flow generated in either direction
was given by Equation \ref{eq:analyticflow} to be
\begin{equation}
u_x\approx \frac{3}{8}\frac{x \tau}{R_x^2},~~~u_y\approx \frac{3}{8}\frac{y \tau}{R_y^2}.
\end{equation}
One can then calculate the momentum anisotropy directly using a few Gaussian integrals to be
\begin{equation} \label{eq:earlyEpAnalytic}
\epsilon_p\approx\frac{9\tau^2}{32}\left(\frac{1}{R_x^2}-\frac{1}{R_y^2}\right)
\end{equation}
which is plotted as a solid line in Figure \ref{fig:InitFlowEp}.  For short times,
this approximation describes all of the models well, though it begins
to overestimate the effect by 1.0 fm/c due to the asymmetric expansion
of the source reducing the eccentricity of the transverse energy density.

The effects of this on final elliptic flow will be discussed at greater length below, but
we emphasize that the argument's main thrust is that a reasonably large class of models
all produce the same predicted input for ideal hydrodynamics in terms of collective 
flow.  Non-zero flow at the onset of hydrodynamics explains why early hydrodynamics 
models overestimated the longitudinal size of the source, 
when they used a reasonable time, $\sim 1.0$ fm/c, for the onset of hydrodynamic behavior.
Including this pre-equilibrium flow causes the system to dissipate more quickly and 
decreases the longitudinal size.
Also of note is that our conclusion is independent of whether or not 
the model for early times appears to contribute to anisotropic 
flow, instead it only depends on the transition to ideal hydrodynamics.
We expect that this conclusion would remain true for a transition
to Navier-Stokes hydrodynamics since the same ratio, $T^{0x}/T^{00}$ is sufficient 
to determine the collective velocity in that case as well.

\section{Initial Shear Corrections}

The remaining six variables to be determined for initializing a viscous hydrodynamics
code are components of the shear tensor.  In Israel-Stewart hydrodynamics, 
each of these moments is an independent variable
that cannot be determined from other hydrodynamic information such as the energy density and velocity gradients.
If there were an underlying gas description prior to the time of thermalization, 
one could calculate the full stress-energy tensor from moments of the momentum distribution. 
Lacking that, if the system were very near to equilibrium, 
one might assume that the Navier-Stokes equation would give a good approximation
to the shear elements.
This turns out to be difficult for reasons discussed above -- namely, the 
corrections do not get small as rapidly as the pressure. The shear viscosity
is proportional to the entropy density and thus diverge slowly, like $\epsilon^{-1/4}$.
This means that for any initial condition with a longitudinal velocity gradient
there will be some small temperature below which corrections to the pressure
are larger than the pressure.  

This can happen even for moderately small
viscosities ($\eta/s \approx 0.16$) at relevant temperatures ($T \approx 100$ MeV),
where the value of $|\pi^{zz}|$ approaches not just the pressure but the
energy density.  If $\epsilon + P + \pi^{zz} = 0$, the conservation of longitudinal
momentum equation is singular and the resulting correction to the longitudinal
velocity diverges.
For Navier-Stokes conditions this is true when
\begin{equation}
0=(\epsilon + P) \left[ 1 - \frac{4\eta} {3 T_b s \tau } \right], ~~ T_b = \frac{4 \eta}{3 \tau s},
\end{equation}
where $T_b$ is the temperature at which the hydrodynamic equations of motion 
can no longer be solved.
Since $\eta/s$ is often taken to be a constant, any choice of $T$ and $\tau$
will eventually have this divergence issue.
The problem is even worse if the shear viscosity to entropy density ratio
rises rapidly in the gas phase as one would expect when moving away 
from a phase transition.

Extreme viscous corrections are not always a dynamic problem in the Israel-Stewart framework:
if the system does not begin with excessively large corrections
they will not develop.  The relaxation time is proportional to shear viscosity,
meaning that in regions where large corrections would be expected in
the Navier-Stokes theory, the Israel-Stewart theory allows the system
to approach such large corrections only slowly.  Of course, this also means that if large 
corrections exist in the system at the onset of hydrodynamics they will remain
large and if the density falls faster in the tails due to the longitudinal expansion
than it rises due to transverse flow the problem will worsen.
Regardless of these details, while the picture that Israel-Stewart theory 
is a relaxation toward the Navier-Stokes theory is somewhat na\"{i}ve, 
it does tend to prevent aberrant behavior outside of the freezeout surface
for not unphysical initial conditions.

Previous studies considering the potentially large corrections
in the tails have taken a variety of tacks.  A
popular choice is to initialize the shear tensor to zero.  This reduces the
overall effect of shear viscosity on the fluid and thus in comparing 
produced anisotropic flow to data one finds an upper bound on the shear viscosity  
at the expense of the attempting to determine the actual value.  
Another choice is to treat the problem as one of regularization 
and therefore treat the region outside of the freezeout surface differently, 
for instance the comparison code used in Chapter 4
applies a position dependent suppression of the 
shear viscosity.  Such an approach defies the natural structure of 
the code and might need to be adjusted for each parameter set
which would become especially problematic when exploring systems
at different impact parameters.

Another approach that would give proper reduction of the corrections 
in the tails would be to allow the shear viscosity to scale with the energy density
below some temperature.  This choice reduces corrections in the tails and
prevents dynamical issues that occasionally arise for cases with large 
shear viscosity.  For the boost invariant case, allowing the shear viscosity
to scale with the energy density instead of the entropy density
did not affect model predictions as long as this rescaling temperature 
was more than 20 MeV less than the freezeout temperature.
This was expected since the output from the hydrodynamic module
is the freezeout surface itself and not the entire evolution, and one 
expects the motion of matter from inside to outside to be more important
than the reverse.

In addition to allowing the shear viscosity to decrease at low temperatures,
we also scale the initial shear tensor to the pressure everywhere.  This 
measure helps to keep corrections sane everywhere but also allows 
for the possibility that the system is too far from equilibrium at the 
beginning of the calculation for Navier-Stokes to be a good description.
For instance, if the system were described by the coherent fields 
described above and the longitudinal direction were roughly pressure-free
at the beginning of the evolution, this could be modeled.  The coefficient
scaling the initial corrections to the pressure is introduced as an input
parameter to the model and its influence will be investigated in the next chapter.

\chapter{Results and Conclusions}

\section{Longitudinal Flow Results}

The unique feature of the code discussed here is the inclusion
of both viscosity and non-trivial longitudinal expansion.  Generally,
viscous hydrodynamic codes were written assuming a 
boost invariant expansion making the non-trivial evolution take place in only
two dimensions though the system does expand longitudinally.  
We begin by investigating the approximation
that the features of the expansion away from mid-rapidity are 
not important.  
To do this we choose a typical set of initial conditions 
and compare the evolution of hydrodynamic quantities 
at midrapidity with the full expansion to the same conditions
under the assumption of boost invariance. 

We begin the hydrodynamic evolution at $\tau_0 = 0.8$ fm/c.
The initial transverse energy density distribution is taken from the Glauber model
with a mixture of 85\% wounded nucleon scaling 
and 15\% binary collision scaling as described in Chapter 5 with
the initial energy density for the central collision, b = 2.21 fm, as
$\epsilon(0) \approx19.5 $ GeV/fm$^3$, to reproduce pion multiplicities 
and scaling respectively for $\sqrt{s_{\text{NN}}} = 200$ GeV Au+Au collisions.  
In the longitudinal dimension, we 
choose a Gaussian with scale $\sigma_\eta = 1.4$ units in spatial rapidity;
that is, $\epsilon(\vec{r}) = \epsilon(x,y) \cdot \exp (-\eta_s^2/2\sigma_\eta ) $.
While this is in contrast to other models that 
choose a long flat region of several units of rapidity with
steep half-Gaussian tails \citep{Schenke:2010nt,Hirano:2002hv},
a Gaussian profile is motivated by the experimental data which shows
little deviation from this profile over a wide range of beam 
energies \citep{Bearden:2004yx}.  An improvement to the model
would account for the angular momentum of the hot region
momentum due to the non-zero impact parameter and for
the breaking of longitudinal symmetry away from the symmetry
axis between the nuclei.
These improvements would be required to model directed flow away 
from midrapidity for any finite impact parameter,
but for simplicity both were ignored in these tests.
The initial flow was taken from the universal flow model discussed in
the previous chapter, but only as half of the result predicted by
Equation \ref{eq:earlyFlowResult}.  In the longitudinal direciton,
we assume that the matter is not moving collectively relative 
to the boost invariant expansion.  And finally, we take the initial longitudinal
pressure correction to be half the pressure, $\tilde{\pi}^{zz} = P/2$,
and the transverse components to be equal to each other and to half the 
longitudinal value so that the shear tensor remains traceless.
All off-diagonal elements are taken to be zero in the frame of the matter.
The shear viscosity is taken to be $\eta/s = 0.16$ for the hydrodynamic
portion and to scale with the energy density below temperatures of 100 MeV.
Finally, a recent lattice equation of state \citep{Borsanyi:2010cj} was used for the hot region 
and the equation of state was forced to match an equilibrated hadron
resonance gas below T=150 MeV via the procedure discussed in Chapter 3.

The integration region was made up of fixed cells in configuration space
with grid densities of $dx = dy = d\eta_s = 0.15 $[fm], and a maximum proper 
time step of $d\tau = 0.05$ fm/c using the methods described in Section 4.1.5.  
Since the system is reflection symmetric
in every spatial dimension, only one octant was integrated with
boundary conditions that enforce symmetries across the boundaries.
Cells with temperatures below 30 MeV were culled and the hydrodynamic module was run 
until all cells in the integration region passed below T=150 MeV.
For comparison, the same conditions were also run with boost invariant
conditions.

\begin{figure}
\centerline{\includegraphics[width=0.7 \textwidth]{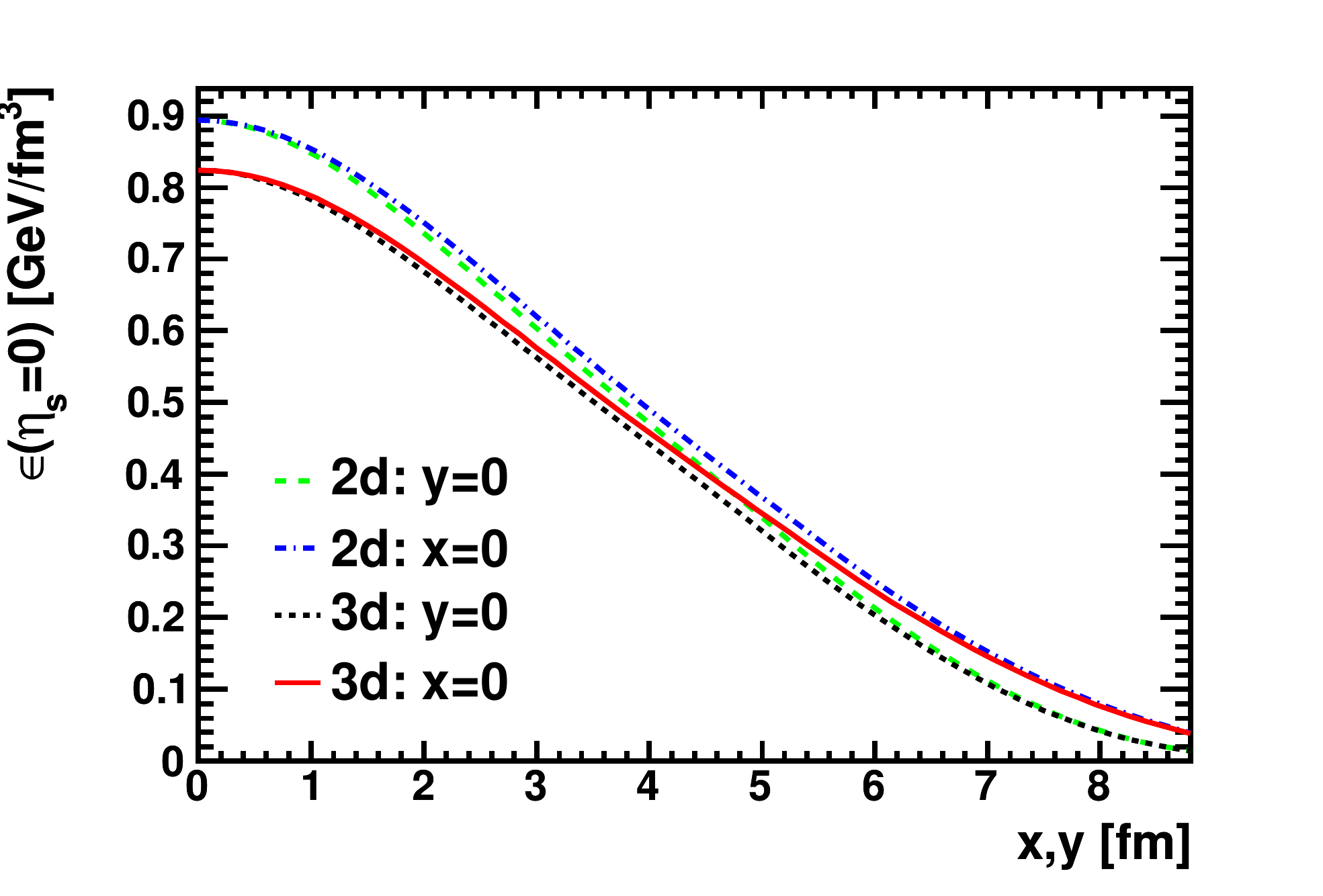}}
\caption{ 
Comparison of the energy density along the x- and y-axes at midrapidity
at $\tau = 5.52$ fm/c from a boost invariant and full 3d hydrodynamic 
treatment.
While the central energy density falls more rapidly when the longitudinal
expansion is included due to the increase in the longitudinal velocity gradient,
which can be seen by observing that both 3d lines (red full and black dotted) are
systematically lower than their 2d counterparts (blue dot-dashed and green dashed).
The difference diminishes in the tails of the distribution but is roughly 
5-10\% at the center of the fireball.
}
\label{fig:exey}
\end{figure}

\begin{figure}
\centerline{\includegraphics[width=0.7 \textwidth]{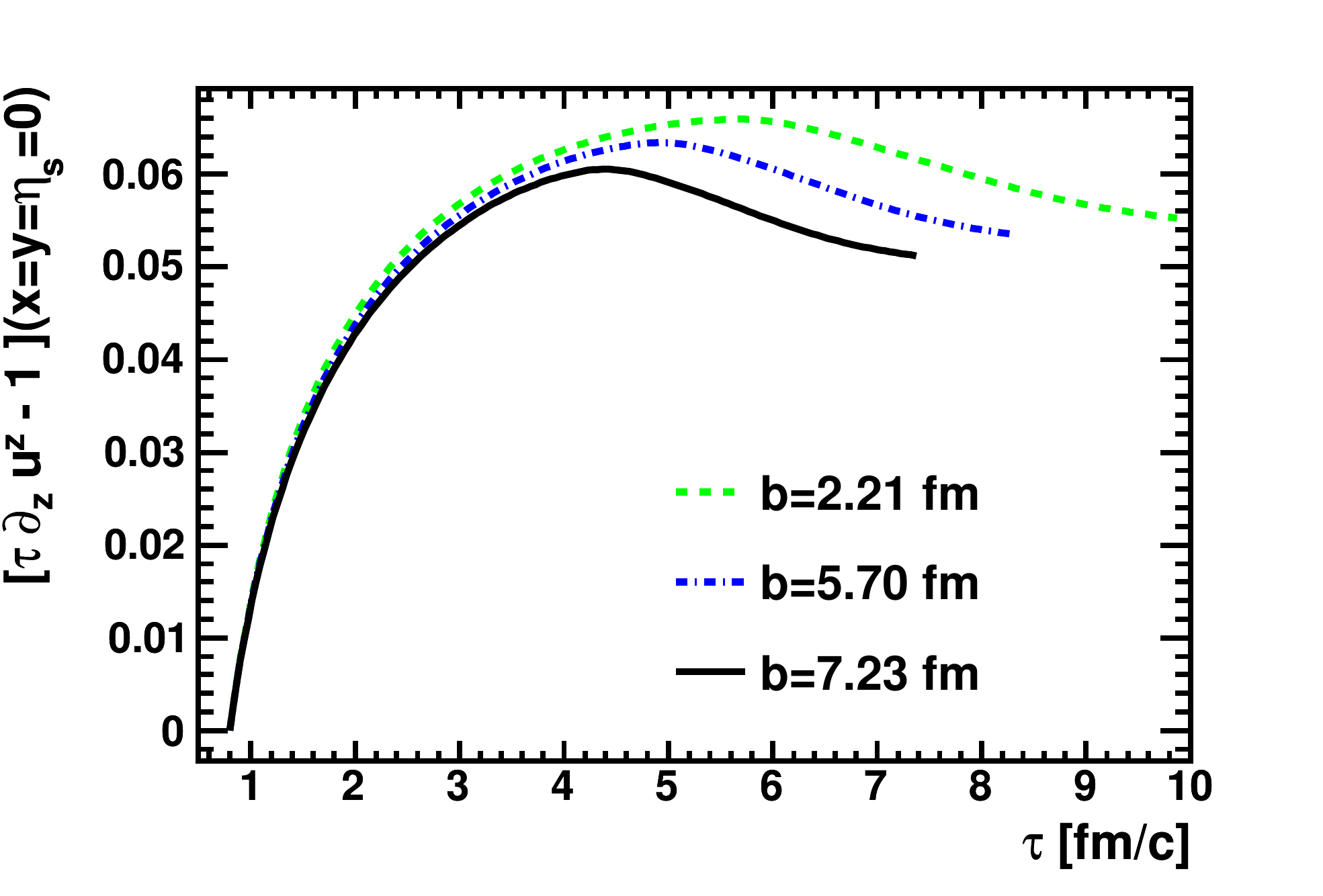}}
\caption{ 
The longitudinal velocity gradient multiplied by the proper time with
the trivial boost invariant portion ($1/\tau$) removed so that boost 
invariant solutions would be exactly zero.
The increase in the expansion rate is taken at the center of the fireball for 
impact parameters corresponding to centralities 0-5\%, 10-20\%, and
20-30\% respectively.
The behavior of the increase gradient in the
velocity gradient appears to be  independent of impact parameter
at early times and follows a fairly characteristic behavior as the system
matures.  This explains the faster decrease in the energy
density observed in Figure \ref{fig:exey}.
}
\label{fig:duzdz}
\end{figure}

The simplest way to look for changes due to the nontrivial longitudinal
expansion is to look at the values of the hydrodynamic fields 
along the symmetry axes of the system.  Figure \ref{fig:exey} shows
the energy density along the transverse axes at midrapidity at a 
time after the system has gone through a bit over half of its evolution.
We find that the energy density falls more rapidly in the center of the 
system in the 3d case, while the tails in the transverse direction are relatively unaffected.  
At the center of the fireball where the effect is the largest, the 
difference is less than 10\%.  We expect the difference to
continue to grow as the expansion progresses, and therefore
the source lifetime might be decreased by up to 20\% which 
we will investigate below.

This more rapid decrease of energy density comes from a
more rapid expansion in the longitudinal direction when
the non-trivial expansion is included due to the density gradient
in spatial rapidity.  The increased expansion
is demonstrated well by Figure \ref{fig:duzdz} which shows
the longitudinal velocity gradient at the center of the fireball
with the effects of the boost invariant expansion removed.
For a boost invariant expansion, this would be $\partial_z u^z = 1/\tau$.  
Therefore, in the boost invariant scenario, this measure would be exactly zero.
For the scenario with non-trivial longitudinal expansion,
this measure is positive because the expansion accelerates 
due to the presence of longitudinal density gradients.
The Gaussian shape produces linearly increasing longitudinal velocity 
as a function of proper time for small proper times near midrapidity 
that depends only on the longitudinal Gaussian radius.
This is demonstrated by the centrality independence of the longitudinal
expansion rate at early times.  
The longitudinal expansion is then slowed by viscosity, which tries to
equalize the much more rapid longitudinal expansion with the burgeoning
transverse expansion, and by the flattening of the density distribution
due to the non-trivial longitudinal expansion itself.
The increasing of the expansion rate lasts longer for central collisions but saturates
in the same manner around a $5\%$ increase in the expansion rate 
compared to the boost invariant case for the latter half of the evolution. 
Since the longitudinal expansion rate is increased by about the same amount
as the energy density at the center of the fireball is decreased
compared to the boost invariant result, it would seem that the more 
rapid decrease in energy density can be attributed entirely to the 
non-trivial expansion and a more complicated explanation need not be sought.

\begin{figure}
\centerline{\includegraphics[width=0.7 \textwidth]{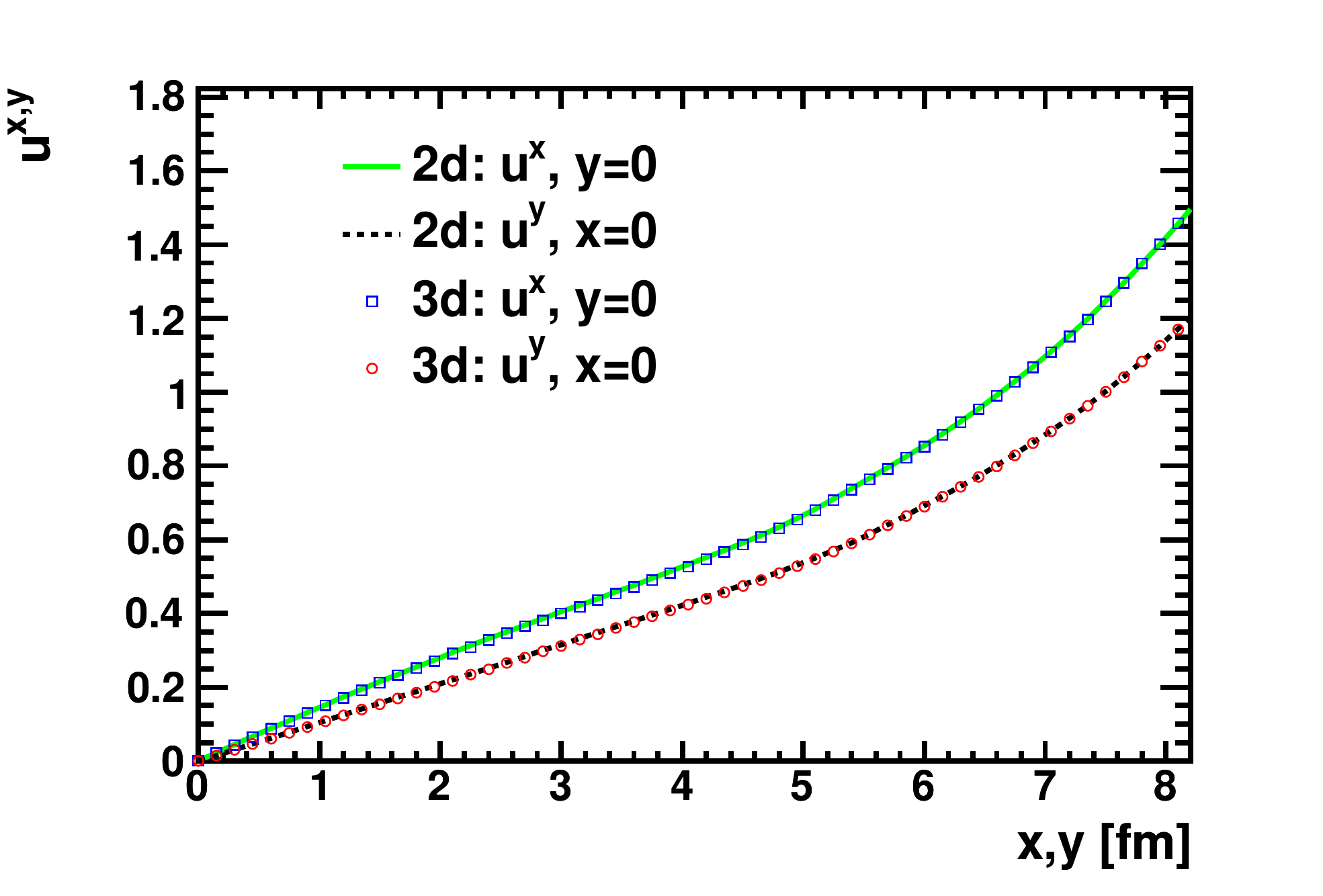}}
\caption{ 
The transverse velocity along the x- and y-axes at midrapidity
for hydrodynamic simulations with and without boost invariance
at $\tau = 5.52$ fm/c. The transverse collective velocity is altered by 
less than 1\% at midrapidity by the presence of a non-trivial 
boost invariant expansion.  This suggests that elliptic 
flow observables at midrapidity do not require a full treatment
of the longitudinal direction unless other factors on the basis
of the hydrodynamic evolution itself.
}
\label{fig:uxuy}
\end{figure}

If the energy density is affected in the center of the fireball at the 
10\% level, one might be concerned that the transverse collective velocity
might be affected in the same way.  Such a result would be 
important for flow observables and might propagate to the parameter
extraction based on these results.  Figure \ref{fig:uxuy} shows that 
this is not the case as the transverse collective velocity at the same time is 
affected by less than 1\%.  Apparently the transverse collective velocity is 
unaffected by the flatter distribution in the core, and any
other effects from the longitudinal expansion are unimportant 
to the description of the system at midrapidity.  This suggests
that (2+1)d hydrodynamic results are trustworthy for midrapidity
flow results, unless secondary influences like source lifetime 
or twisting provide unexpected corrections.

\begin{figure}
\centerline{\includegraphics[width=0.75 \textwidth]{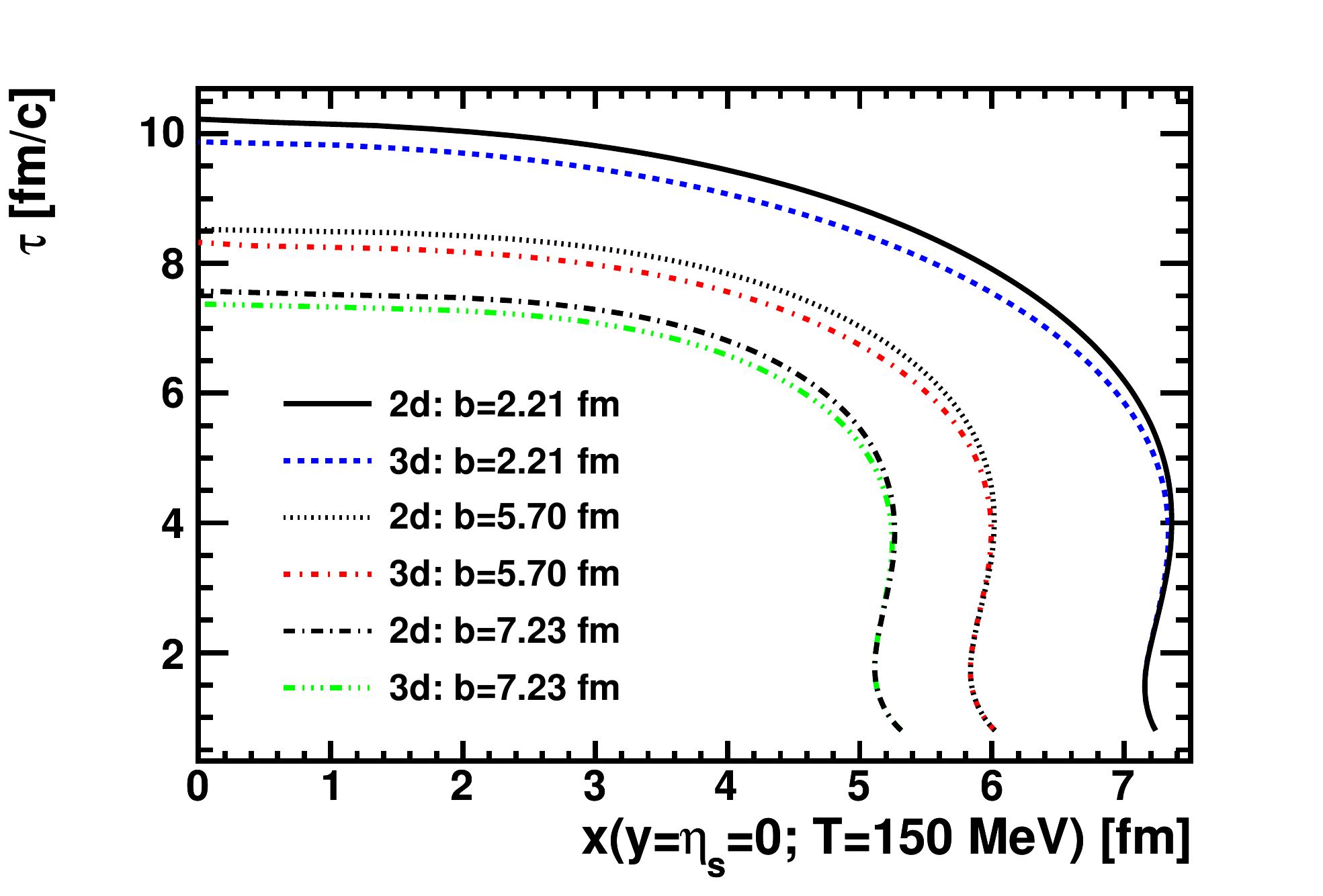}}
\caption{ 
The position of the freezeout surface along the x-axis as a function
of the proper time for collisions at three centralities.  In each case
the full integration shows the surface falling apart sooner but with
very similar structure to the boost invariant case.  This indicates
that boost invariant treatments might overestimate longitudinal
source lengths by 5\% but dramatic change to the results is unlikely.
}
\label{fig:fosx}
\end{figure}

Finally, we investigate the motion of the freezeout surface at 
midrapidity.  In the hybrid model that we will investigate later
in this chapter, the location of the freezeout surface and
the values of hydrodynamical fields there are the
output of the hydrodynamical module to the hadronic cascade.
Figure \ref{fig:fosx} shows exactly what one would have 
anticipated from Figure \ref{fig:exey} -- the
general structure of the freezeout surface is not drastically
altered.  Rather the system dissipates sooner but without
significantly altering the character of the motion.
Given that the collective velocity and the motion of the surface
are not significantly altered, it is difficult to imagine that 
theoretical conclusions regarding observables like the 
the anisotropic flow of the 
system will be significantly affected at midrapidity by
including the longitudinal dynamics more completely.  

However, we do expect the longitudinal source sizes to decrease 
if the system cools more rapidly at the 5-10\% level.  This is due 
to the relationship of the longitudinal source size to the duration
of the collision and the Bjorken expansion.  The Bjorken expansion
means that in the center of momentum frame, the longitudinal collective velocity
grows proportional to time.  Since HBT is sensitive to the ratio
of the collective velocity to the thermal velocity, there is a direct
relationship between the duration of the collision and the 
longitudinal source size measured by HBT.  Therefore more 
rapid decay of the thermal source will result in a smaller longitudinal
radius.  The overestimation of the longitudinal source size was a 
major issue in early hydrodynamic modeling \citep{Heinz:2001xi}, though
pre-equilibrium flow also plays a role in the resolution of the
so-called HBT puzzle \citep{Pratt:2008qv}.

There remains the possibility of more complicated model response 
to parameter changes but there is no evidence of this 
in these hydrodynamic results.  One
possible avenue for investigation would be to use midrapidity 
freezeout surfaces even from the full calculation and compare
to boost invariant hydrodynamic calculations.  Since there is little
evidence that there would be any effect from this, we instead 
abstain from drawing further conclusions until we have a more complete 
treatment of the dynamics away from midrapidity including 
the non-zero chemical potential and angular momentum of
the hot region.  Both of these effects are good candidates for 
future work with this code.  

\section{Boost Invariant Results}

We turn our attention to the boost invariant case and investigate
the predictions of the whole model where we can make direct
comparisons to the experimental data.
This entails producing many particles and looking at quantities
averaged over a large number of collisions.  While experiment observes 
particles at a very wide range of momenta, our goal is only to 
study the behavior of the many lower mass particles produced at low momentum
that come from the thermal source and to ignore the effects of hard processes
that produce the remaining high momentum and large mass particles.
Our interest is in determining  constraints on the structure of the quark matter described 
by the hydrodynamic phase.  For instance, it would be of great interest
to determine the shear viscosity of the quark matter and whether the 
equation of state calculated by the lattice is supported or constrained
by the experimental data.  In addition, improvements to the hydrodynamics
module that would allow the inclusion of known effects, such as the twisting
of the source, have the potential to spoil the agreement with 
experimental data.

The experimental data that we will consider here are the distribution
of low momentum, low mass particles produced in central and mid-central
collisions of gold nuclei at center of mass energy $\sqrt{s_{\text{NN}}} = 200$ GeV.  
We consider data only at midrapidity where 
effects of non-zero baryon number are expected to be 
small and the boost invariant code can make predictions.
Specifically, we look at the spectrum of charged pions, kaons, and protons 
produced in the 0-5\%, 10-20\%,  and 20-30\% most central collisions
which correspond to average impact parameters of 2.21, 5.70 and 7.37 fm 
respectively \citep{Adler:2003cb}.  Looking at multiple centralities allows one to test
the scaling of total particle multiplicity predicted by the 
various initial condition models. Anisotropic
flow in the most central collisions is dominated by fluctuations
that are not present in the initial conditions that we produce. 
Therefore, only multiplicity data is used the most central bin:
the pion multiplicity sets the energy scale, while kaon and proton
might be useful for constraining other parameters.
Incorrect eccentricity in the most central bin might be overcome by reorienting the reaction
plane in a Monte Carlo simulation to produce a realistic 
estimate of the initial eccentricity for a central event.
This results in more turbulent initial conditions with
steeper tails that can cause instability in the simulation.

\subsection{Initial State Uncertainty}

As mentioned in the previous chapter, there is significant uncertainty
in the structure of the initial state of the hydrodynamic phase.  
This uncertainty is not limited to the shape of the hot region but also 
the collective velocity and the six independent elements of the 
shear tensor.  While other groups have investigated effects of
varying the shape of the source and the shear tensor within 
reasonable theoretical bounds, we further include the effect
of including initial flow on spectrum and flow observables.

We first consider a set of model runs that were run at 
some interesting corners of the theoretically allowed parameter
space.  For this, the full model was run with 27 different parameter settings
coming from all possible combinations of
three settings for the three initial energy density models, 
three settings for the initial flow, and three settings for the
initial longitudinal pressure.  The models for the energy
density are a Glauber mixture with 85\% participant scaling
which we will refer to as the default Glauber model, the saturation model described 
in Equation \ref{eq:simpleSatIC}, and a smooth CGC model 
provided by Drescher et al. \citep{Drescher:2006pi} referred to in Chapter 6 as KLN
which predicts the entropy density instead of energy density
and must be converted through the equation of state.
The initial flow settings were the traditional setting of zero
initial flow, half of the initial flow predicted by 
Equation \ref{eq:earlyFlowResult}, and the full prediction of the same.
The initial value of the shear tensor is taken to be the set
$ \tilde{\pi}^{zz} = -\tilde{\pi}^{xx}/2 = -\tilde{\pi}^{yy}/2 = \{0,P/2, P\}$
respectively.

\begin{figure}
\centerline{\includegraphics[width=0.85 \textwidth]{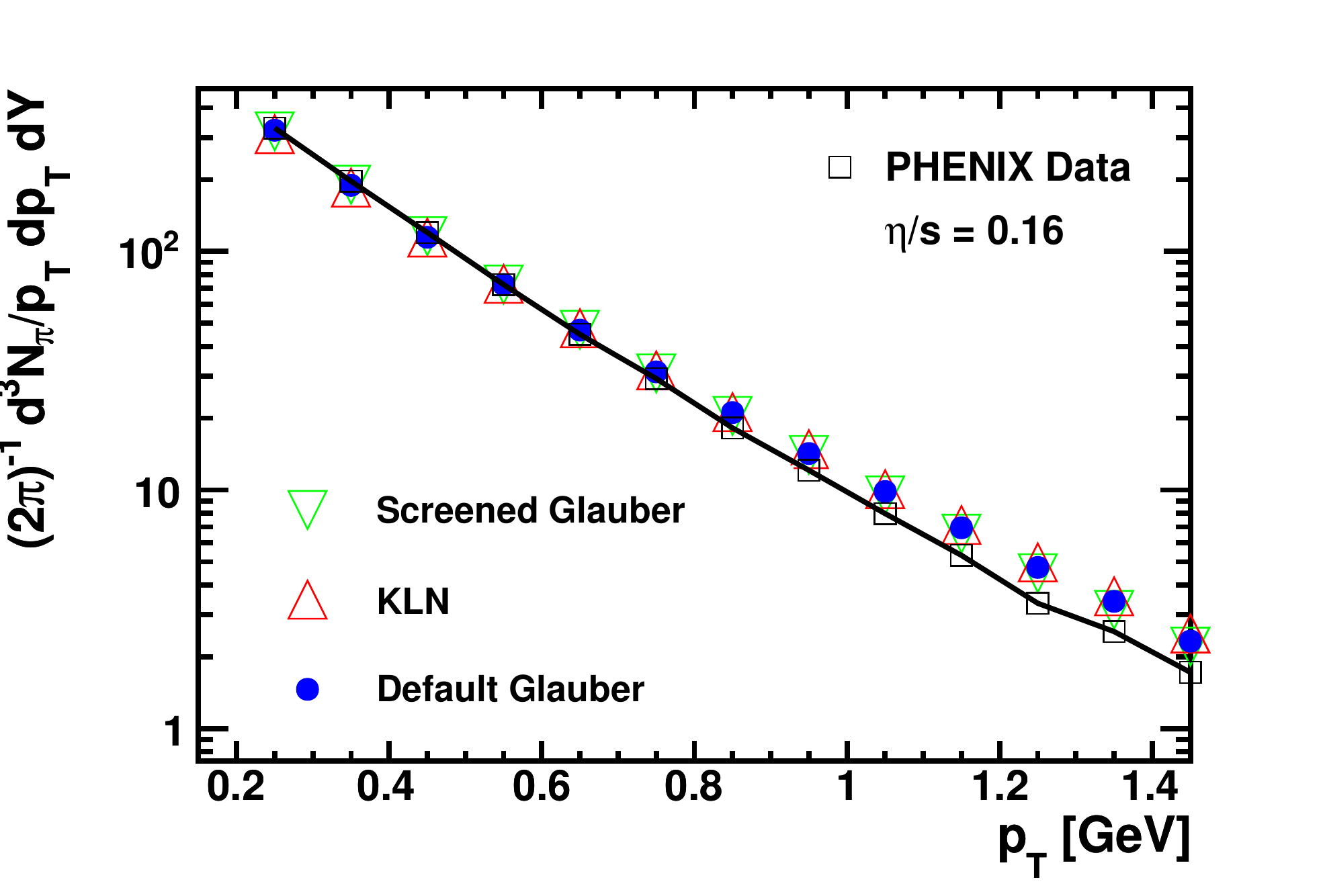}}
\caption{ 
Pion spectra at midrapidity for the most central (0-5\%) bin of
Au+Au collisions at $\sqrt{s} = 200$ GeV as a function of transverse
momentum.  The (black) line with squares is minimum bias data from the
PHENIX experiment \citep{Adler:2003cb} corrected for experimental acceptances and 
efficiencies.  The colored symbols are for three model runs with 
different initial conditions for the energy density where all three models
have been tuned to reproduce the total number of pions between $p_T = $ 
200 MeV and 1 GeV.  All of the models are easily capable of reproducing
the total multiplicity and there is not much variation in the slope meaning
that this observable is not likely to have significant resolving power.
}
\label{fig:specExample}
\end{figure}

The normalization for the total size and energy of the hot region
is a free parameter in all of the initial condition models that we 
consider.  For this study we wished to separate
effects due to the normalization from those due to the varied parameters. 
This requires running the hydrodynamic evolution
with a guess for the normalization, running the cascade module
from particle generation until collisions cease, and then
tallying particles within the various transverse momentum bins
in consideration.  This process was automated by a short Python script that 
was used to determine the normalization
for each set of initial conditions by comparing the number of 
low momentum pions produced to that measured by the PHENIX 
experiment between $p_T = $ 200 MeV and 1 GeV \citep{Adler:2003cb}.  
Figure \ref{fig:specExample}
shows the low momentum pion spectrum compared to model 
runs with each flavor of initial energy shape.  There is almost 
no variation between the model runs once they have been
tuned to reproduce the normalization, and all of the parameter
settings produce spectra that are too flat or overestimate the average
transverse momentum.  Some of this discrepancy can be 
attributed to the choice of maximum initial flow, but generally
we have found that models that reproduce the normalization 
somewhat overestimate the average transverse momentum.
This appearance of the effect is exaggerated in the figure
by the inclusion of higher momentum particles. Bose 
corrections in the cascade or increasing 
the freezeout temperature may help soften the spectra, 
but the overestimation of the transverse momentum is 5-10\% in these runs.
This is roughly the same as the level of variation between
the experiments \citep{Abelev:2008ab}.

\begin{figure}
\centerline{\includegraphics[width=0.6 \textwidth]{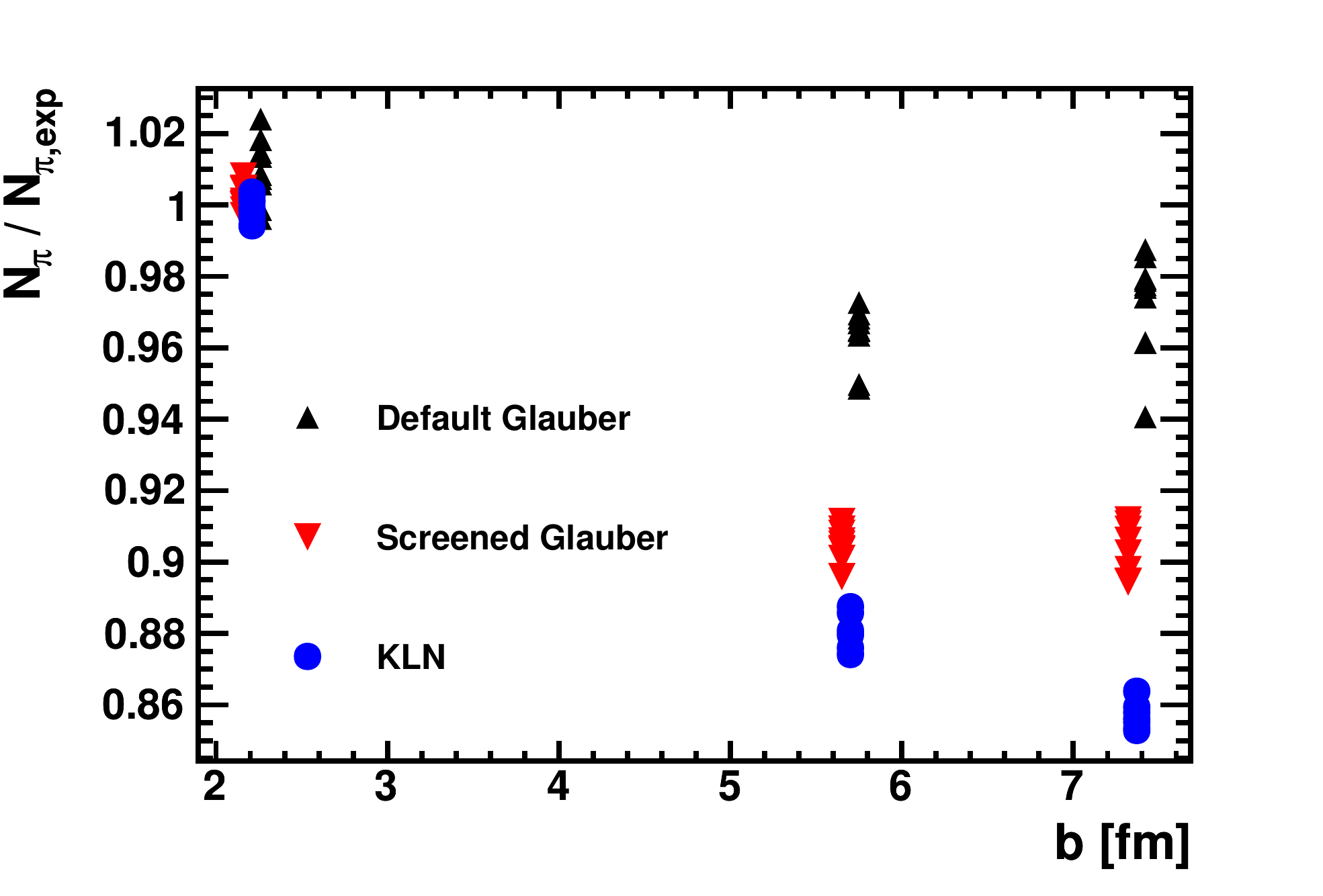}}
\caption{ 
Number of pions as a fraction of the number measured in the PHENIX
experiment in the low momentum window $p_T = $ 200 MeV and 1 GeV
displayed as a function of the impact parameters corresponding
to the 0-5\%, 10-20\%, and 20-30\% centrality bins.  The Glauber model
with 85\% participant scaling most accurately describes the experimental scaling with
all runs within 6\% of the correct multiplicity though a clear trend
to under predict the multiplicity at larger impact parameters.
Both the Glauber saturation and the fKLN models under-predict 
the number of pions by roughly 10\% in both of the mid-central bins.
Allowing the scattered gluon density to be proportional to the entropy
density instead of the energy density would help to counteract this issue.
}
\label{fig:NormSys}
\end{figure}

If we investigate the same quantity for larger impact parameters,
we unfortunately find that none scale correctly.  Figure \ref{fig:NormSys}
shows the pion multiplicity divided by the experimental pion multiplicity
as a function of impact parameter.  The default Glauber model 
most closely tracks with the data, though it is worth noting that the
fraction of participant and collisional scaling is tuned to reproduce this
data.  However, the other two models under-predict the experimental
scaling by 10-15\% in both the 10-20\% and the 20-30\% centrality bins
which is concerning.  This is an unexpected result based on a 
naive interpretation of Figure \ref{fig:ICNComp}, which showed that
the integrated density of the Glauber mixture and the saturating Glauber
have roughly the same scaling with impact parameter.
In principle, the deviation of the screening model and the KLN model 
might be addressed by allowing
the scattering densities to scale with the entropy density but it is our 
opinion that energy scaling is more justified.  
Notice also that the initial condition model has by far the largest effect 
on the multiplicity scaling as one would expect, although for the 
default Glauber model the variation within model runs 
due to initial flow and anisotropy can be as large as 5\%.
We note that it is possible that the multiplicity scaling errors 
observed in Figure \ref{fig:NormSys} may affect
the flow results, for instance by reducing the lifetime of the hot region,
but our assumption is that it will not affect the model response 
to other parameter changes.  Furthermore, later explorations 
of the larger parameter space will help illuminate this possibility.

\begin{figure}
\centerline{\includegraphics[width=0.8 \textwidth]{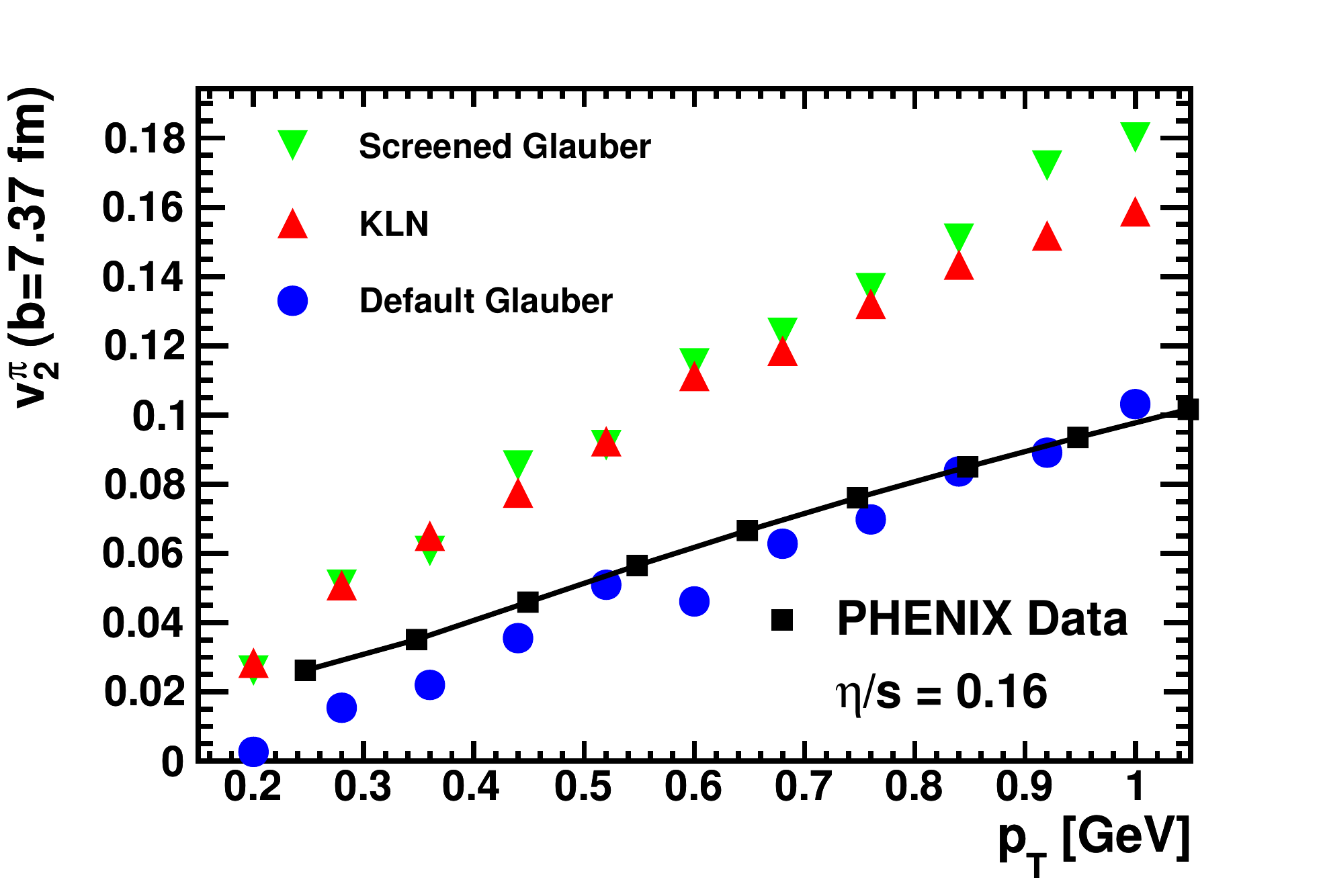}}
\caption{ 
Anisotropic flow for pions as measured by the PHENIX collaboration 
(black squares and line) for events in the 20-30\% centrality bin as compared to
model calculations with differing models for the initial energy eccentricity.
For this choice of the shear viscosity, both of the saturation models 
significantly overpredict the experimental data while the default
Glauber model describes the data.  
}
\label{fig:v2pt}
\end{figure}

We now turn to our main focus which is on the influence of modeling
choices in the initial state on the measured anisotropic flow ($v_2$), specifically
whether the initial velocity or the initial shear pressure in the system
strongly affect conclusions about the value of the shear viscosity
near the critical temperature.  As noted above, flow anisotropy 
in central collisions is expected to be dominated by fluctuations 
not present in our initial conditions and therefore we expect to
under-predict the data in central collisions.  Figure \ref{fig:v2pt} 
shows the anisotropic flow as a function of transverse momentum
for the same set of model runs shown in Figure \ref{fig:specExample}
with a shear viscosity near to what other groups have found, $\eta/s = 0.16$.
Statistical fluctuations are still present in the model output but the
 integrated $v_2$ was found to fluctuate by less than one
percent of the signal for a set of test runs in which the statistics were doubled.
The default Glauber comes the closest to describing the data, 
again a conclusion that we have in common with some studies 
\citep{Shen:2012vn}
though not all \citep{Luzum:2008cw},
but our model does show a different scaling with transverse
momentum than the data. This has been observed in other simulations
with significant shear viscosity and seems to come from the viscous corrections
to the phase space density at freezeout \citep{PhysRevC.77.064901,Dusling:2009df}.  
There is some ambiguity 
in this area, discussed in the previous chapter, and one would generally
expect the phase space distortion due to the viscous corrections
to the stress-energy tensor to be smaller for pions than 
for protons since the relaxation time for pions should be shorter once 
the species move independent of one another.  This is not included
in the freezeout algorithm presently but may explain this 
deviation.  Also, the saturation models that predict larger source 
eccentricities also produce much larger elliptic flow, and in this case
they over-predict the data by a factor of two.

\begin{figure}
\centerline{\includegraphics[width=0.8 \textwidth]{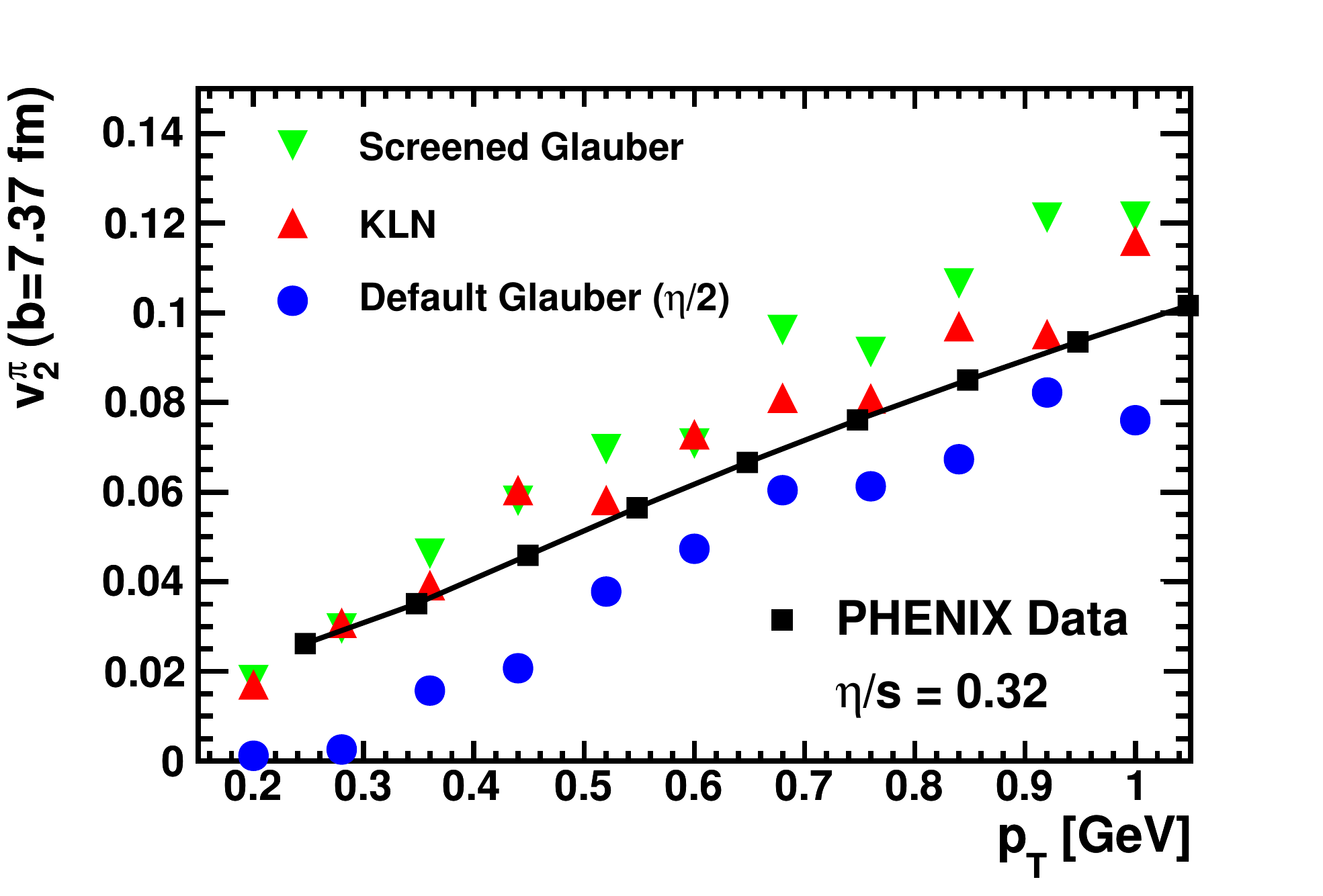}}
\caption{ 
Anisotropic flow for pions as measured by the PHENIX collaboration 
(black squares and line) for events in the 20-30\% centrality bin as compared to
model calculations with differing models for the initial energy eccentricity.
Compared to Figure \ref{fig:v2pt} the shear viscosity assumed in the 
fluid stage of the model is doubled.  This leads to the saturation models
giving a more accurate prediction while the default Glauber model
now significantly under predicts the data.  This suggests that the shear
viscosity will be difficult to determine independently from the initial
energy eccentricity without additional experimental data.
}
\label{fig:v2pt_DV}
\end{figure}

The same calculations were run with the shear viscosity doubled
and the analogous plot is Figure \ref{fig:v2pt_DV}.  The default
Glauber model now significantly under-predicts the experimental
data especially at lower momentum, while the saturation models
are in much closer agreement with the data.  Also note that the 
range in which the elliptic flow is consistent with zero now extends
up to 300 MeV for the default Glauber initial conditions, and 
also that the saturation models appear to develop this same feature
at low momentum.  Together with Figure \ref{fig:v2pt}, we find
that uncertainty in the initial energy density profile contributes 
at least a factor of two in the uncertainty of the shear viscosity
based solely on pion elliptic flow observables.  This means
that shear viscosity of at least 4-5 times the proposed conformal limit
\citep{Kovtun:2004de}
may be consistent with this data.

\begin{figure}
\centerline{\includegraphics[width=0.8 \textwidth]{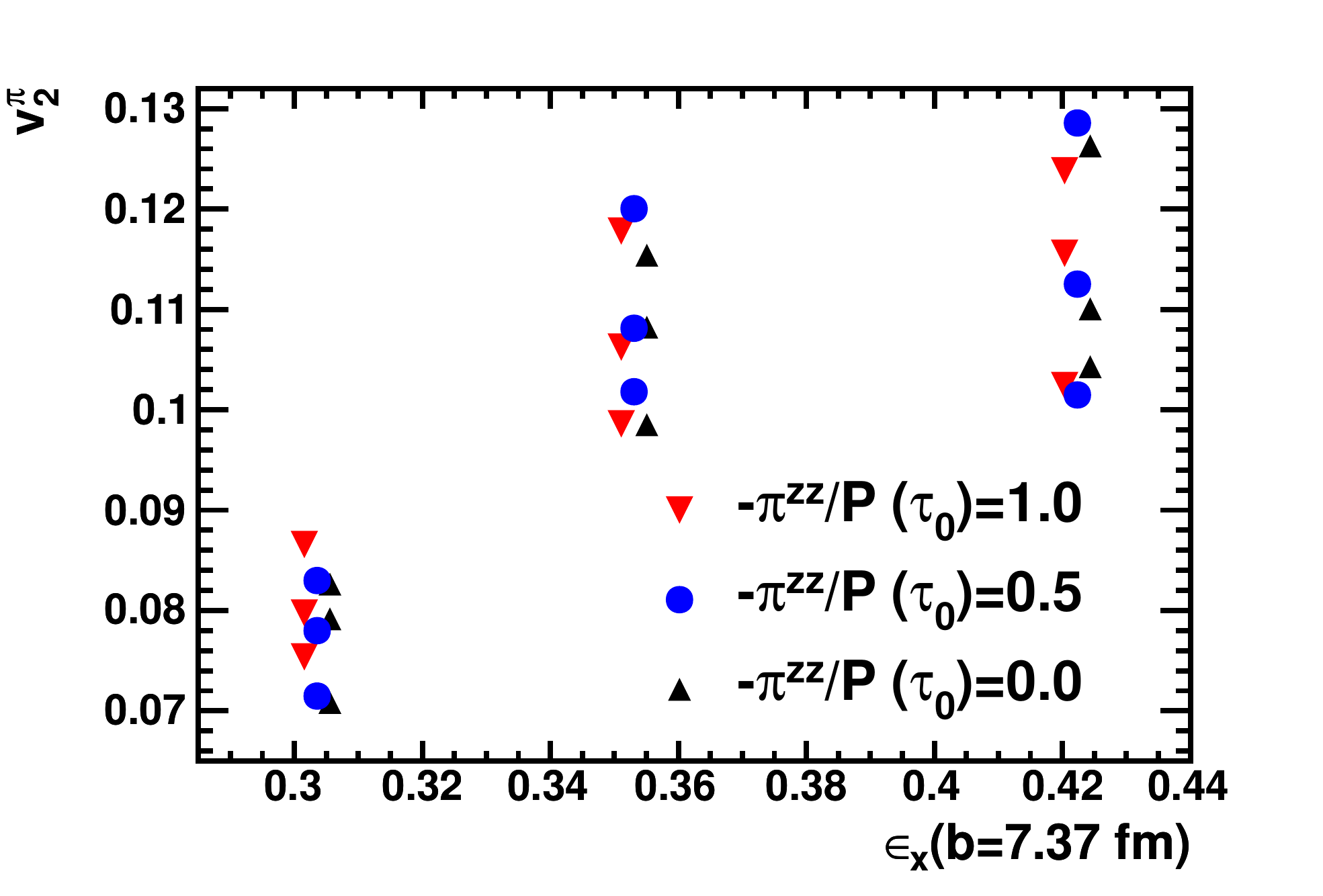}}
\caption{ 
The transverse momentum weighted, integrated anisotropic flow for the 20-30\%
centrality bin from the set of model runs described in the text.  The runs are separated
first by the source eccentricity due to the model choice for the initial
state along the x-axis, where the default Glauber model predicts the
smallest initial eccentricity followed by the KLN model and the screened
Glauber model. The model runs are then colored by the initial
anisotropy of the shear tensor with the largest initial anisotropy
as downward-pointing (red) triangles, no initial anisotropy 
as upward-pointing (black) trianges, and halfway in between
as (blue) circles.  If initial anisotropy were an important factor
in determining the elliptic flow, the symbols of one color would
be systematically above or below the others. However, ordering
appears random and not at all systematic and therefore
the initial value of the shear tensor appears to have no effect 
on the final anisotropic flow.  Also of interest is that the initial
eccentricity and elliptic flow do not scale with each other when
moving between initial energy density models:  the CGC model
and Glauber screening model have initial eccentricities that 
differ by 20\% but predict the same elliptic flow to within a few percent.
}
\label{fig:v2InitShear}
\end{figure}

As noted above, the energy density is only one of the ten
hydrodynamic variables that need to be set in the initial condition.
Six of the other variables are the viscous corrections to the stress energy
tensor.  Boost invariance and the absence of bulk viscosity constrain
four of these variables, and the further assumption that in the 
frame of the matter the transverse coordinates should be equivalent
means that there is only one shear correction to be set.  It is
convenient to set the longitudinal pressure correction.  This is
allowed to vary from zero to the pressure, $ 0 < \pi^{zz} < P$,
where zero would coincide with complete local equilibration and $P$
would coincide with a description in terms of free streaming particles.
This should cover the majority of the parameter space for this 
initialization for a fixed thermalization time, as corrections larger
than the pressure are unlikely to be well modeled by even
the second-order viscous equations of motion used.  Figure
\ref{fig:v2InitShear} shows all of the model runs performed 
with the smaller viscosity value colored in the figure according to the initial longitudinal
pressure correction.  The variance within the model output
appears to have no connection to the initial anisotropy, for instance, the 
output for runs with no initial longitudinal pressure are not
systematically above or below the other runs. The symbols
are frequently found in sets of three and in random order even
within those sets, meaning that the initial shear tensor
has no systematic influence on model output.  

That the initial value of the anisotropy is not an important parameter
for elliptic flow is not unexpected.
Regardless of the initial condition, the shear tensor will relax toward the Navier-Stokes
values on a time scale set by the relaxation time.  This means
that the initial value no longer plays a role for time scales
longer than the relaxation time.  For temperatures relevant to the hydrodynamic phase,
the relaxation time is less than 1.0 fm/c, for the shear viscosity to 
entropy density ratio used in these simulations, meaning that the initial value
only influences a small fraction of the evolution.

\begin{figure}
\centerline{\includegraphics[width=0.8 \textwidth]{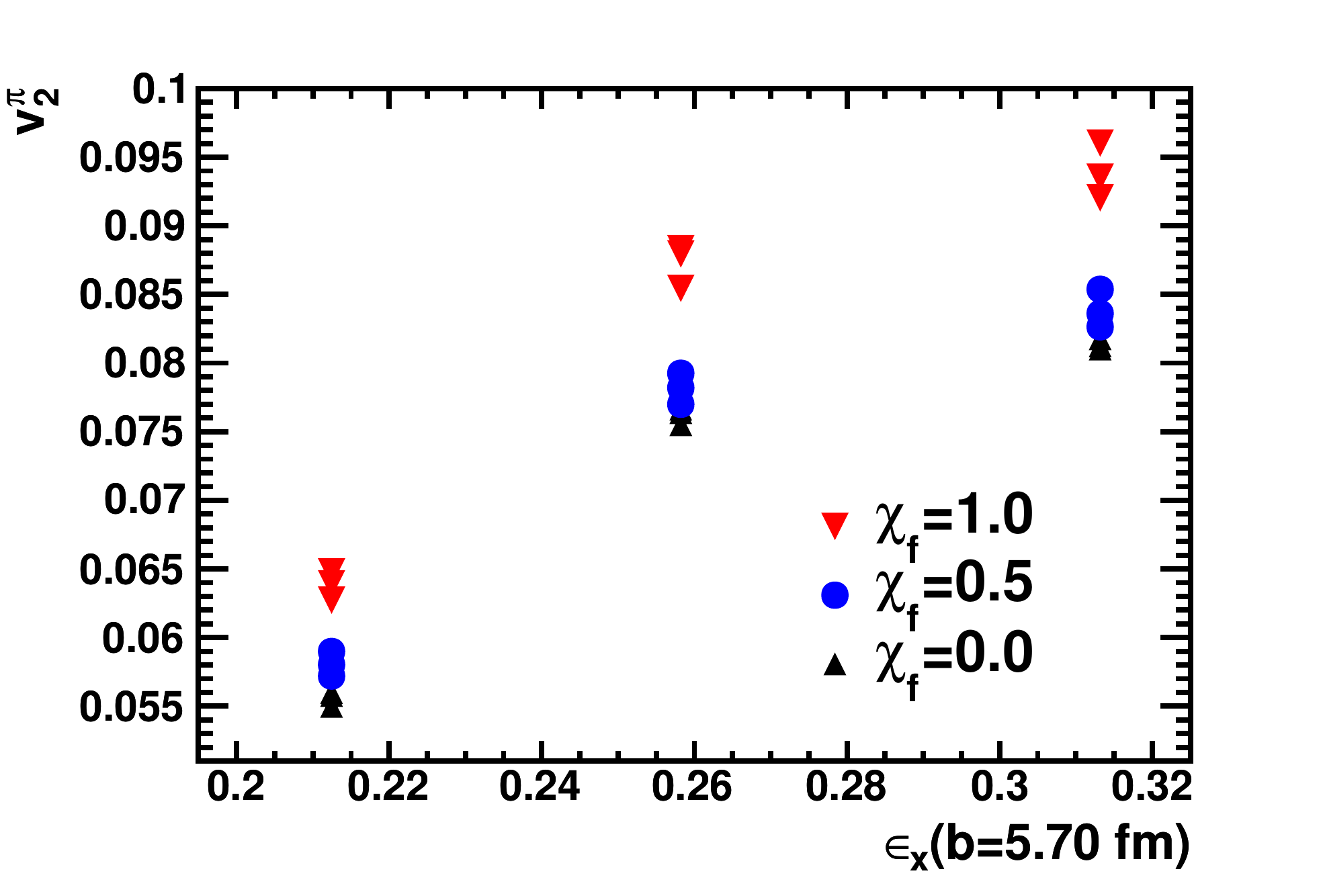}}
\caption{ 
The transverse momentum weighted integrated anisotropic flow for the 
10-20\% centrality bin from the set of model runs described 
in the text displayed in the same manner as Figure \ref{fig:v2InitShear}.
The model results are colored by the initial velocity as a fraction of
the results from Equation \ref{eq:earlyFlowResult} ($\chi_f$) where the (red)
downward triangles are the full result, the (blue) circles half, and the 
(black) upward facing triangles with no initial velocity.  The variation
of results within each initial density model is shown to be explained
by the changing of the initial flow as all of the runs are well-ordered 
by the strength of the initial flow.  The initial anisotropy, 
which are not differentiated from one another,
produce small, random changes in the observed elliptic flow.
}
\label{fig:v2InitFlow}
\end{figure}

In contrast to the initial anisotropy of the shear tensor, the
initial flow plays an important role in developing elliptic flow,
parameterized here in terms of $\chi_f$, which is the fraction
of the result in Equation \ref{eq:earlyFlowResult} used in the initial state.
This is evident in Figure \ref{fig:v2InitFlow} which shows the elliptic
flow produced in all model runs for the 10-20\% centrality bin, where
the key difference from Figure \ref{fig:v2InitShear} is that the symbols 
now indicate runs with the same initial flow.  The model runs are well-ordered
by the initial flow in every case with none of the zero initial flow output
even within 10\% of the full initial flow output.  Even more tellingly, 
this persists over all types of initial conditions.  This results in a combined
uncertainty between the initial profile and the initial flow of almost 100\%
in the model prediction of the elliptic flow in this variable.

\begin{figure}
\centerline{\includegraphics[width=0.9 \textwidth]{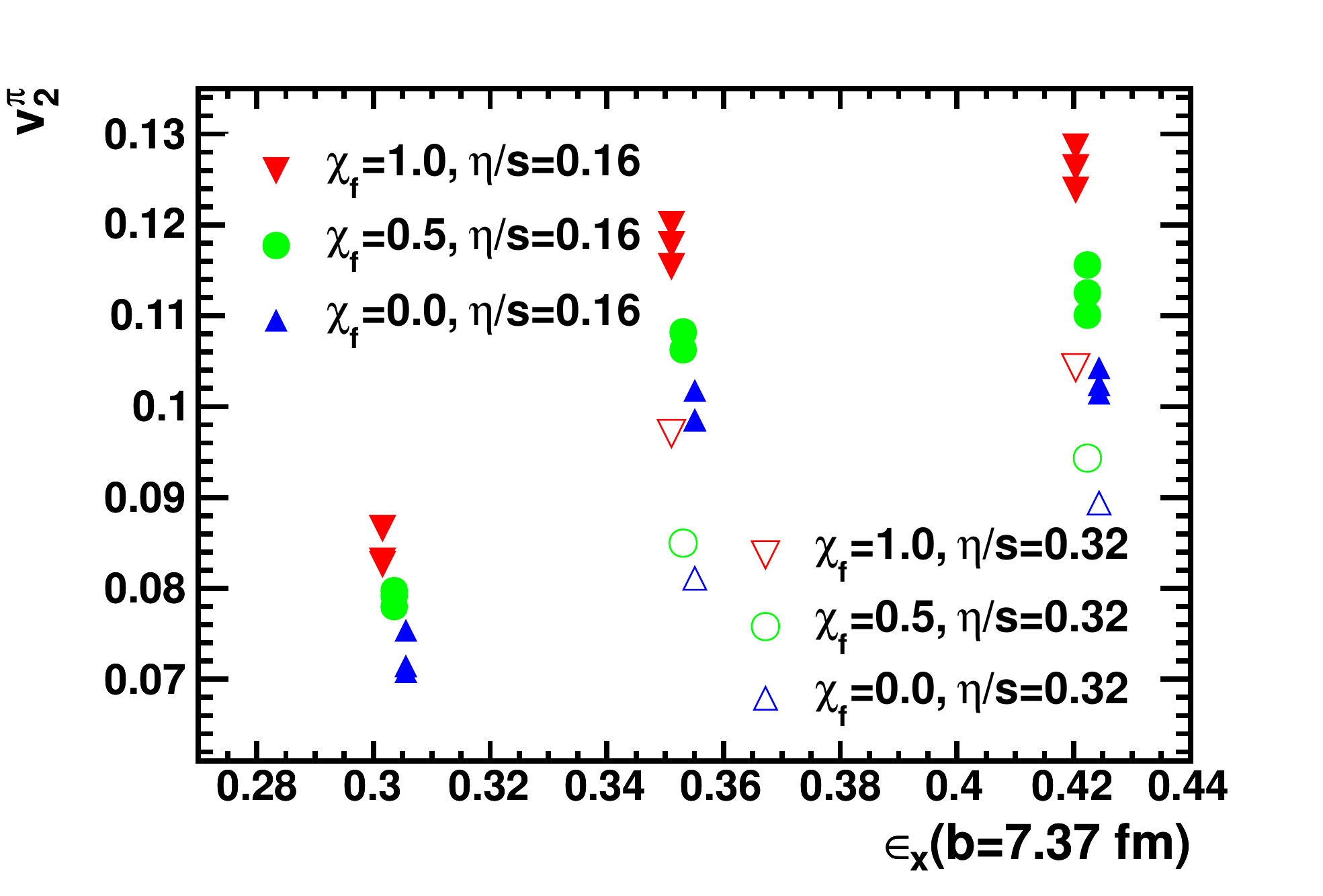}}
\caption{ 
The transverse momentum weighted integrated anisotropic flow for the 20-30\%
centrality bin from the set of model runs described in the text. The
solid marks are identical to those in Figure \ref{fig:v2InitShear} now
colored by initial flow ($\chi_f$) as described in Figure \ref{fig:v2InitFlow}
and arbitrarily offset for clarity.  Furthermore,
open symbols are now included for model runs with double the viscosity ($\eta/s=0.32$) for
the saturation models for initial conditions.   The models runs with double
the viscosity and the full amount of initial flow show the same amount of 
anisotropic flow as runs with the previous shear viscosity and no initial flow.
This indicates that uncertainty in the initial flow results in 100\% 
uncertainty in the value of the shear viscosity. 
}
\label{fig:v2InitFlow_DV}
\end{figure}

To see the effect of this uncertainty on the extraction of the shear viscosity 
near the phase transition, Figure \ref{fig:v2InitFlow_DV} again shows all
model runs distinguished by their initial flow.  We've returned to the
20-30\% centrality data in part to demonstrate that the effect is not unique
to mid-central collisions which is very clear from the lower viscosity data
presented.  Furthermore, we now include additional runs with twice the
original viscosity to confirm the findings from Figure \ref{fig:v2pt_DV}
which showed that initial shape contributed a factor of roughly two
to shear viscosity determination.  Figure \ref{fig:v2InitFlow_DV} confirms
that the uncertainty in shear viscosity extraction due to the initial shape
of the energy density is more than 100\%.  Runs with double the viscosity
produce more elliptic for the screening and saturation initial conditions 
that remain larger than that seen with the default Glauber output, meaning
that even more shear viscosity would have to be added to compensate
for changing the initial conditions from a default Glauber model to a
screening model.  This uncertainty appears to be additive with
the uncertainty contributed by uncertainty in the initial velocity profile
as the spread in the results is not decreased as the viscosity is increased.
Therefore even within a model that fixes the thermalization time, equation
of state, temperature dependence of the shear viscosity, bulk viscosity, etc.
one should expect a factor of three uncertainty in the extraction of the shear
viscosity. This conclusion is based only on integrated elliptic flow data
and it is possible that, for instance, the transverse momentum
dependence of the elliptic flow could provide some additional resolving 
power though, as mentioned before, such a conclusion may depend
more delicately upon details of the model.

\subsection{Multiple Parameter Extraction}

Efforts to understand the hydrodynamic parameter space 
is limited by the complexity of the hydrodynamic
model itself and the uncertainties that are inherent to the choices in 
constructing one.  This is underscored by the divisions between 
the several components of the model which are the initial condition
to hydrodynamics, the hydrodynamic evolution, the hadronic cascade,
and finally the analysis that calculates observables to be compared 
to experiment.  Each of these contain many decisions that likely 
affect theoretical conclusions about the matter created in heavy ion collisions.
For example, the initial condition to the hydrodynamic
phase is not well understood and should be constrained experimentally.
Therefore, one must investigate the sensitivity of any conclusion 
about the character of the matter to assumptions about 
the pre-equilibrium phase, that is, uncertainty about pre-equilibrium
dynamics must be fully propagated to any extraction of properties
of the the matter, like the shear viscosity or the equation of state.
So while there is a great deal of experimental data to be analyzed, 
there are also a large number of parameters to be considered.
Furthermore, the parameters do not necessarily have simple relationships
to observables and many parameters may be important to the 
calculation of a single observable.  The opposite is also true
as a single parameter may affect many observables, and these
effects may not be linear.

Model response throughout the parameter space was studied using the Markov Chain Monte Carlo 
(MCMC) method which takes random small steps through the parameter
space, where the model emulator was used instead of evaluating the model 
at each point. The Metropolis-Hastings method of performing
MCMC proceeds by always accepting steps to parameter settings that better
describe the data and accepting worse settings with probability
proportional to the ratio of the likelihoods, where the log-likelihood
is proportional to the sum of the squares of the deviation from 
the experimental data.  This means that 
after excluding a random walk near the first point chosen the
set of parameter space points visited in the MCMC trace is
proportional to the posterior likelihood -- that is, the likelihood
that a parameter setting is the true setting based on the model runs.
This is not the only method of exploring model response to parameter
variations, and other methods like gradient or Langevin search could
be used to provide an estimate of the most likely point in parameter space;
but they do not provide a direct estimate of the posterior distribution.

One difficulty of MCMC is that the model output must be known at 
every point in the parameter space.  Unless the model is extremely 
fast to evaluate, it is necessary to be able to
predict the output of the full model without evaluating it.   In the simplest
case, this could be done by interpolating linearly in each dimension
between the model runs nearest to the point of interest.  This has the
benefit of simplicity but may miss important features of the space.  
Instead, we chose to emulate the aggregated observables of our 
hybrid hydrodynamic-cascade model using a Gaussian process \citep{Rasmussen06gaussianprocesses}.
This method assumes that the local correlation structure in parameter
space is a multi-variate normal distribution and produces predictions
of the mean and variance, essentially a prediction and the uncertainty
in that prediction, at any point in the parameter space.  
To verify that the Gaussian process accurately infers
model outputs, the predictions of the Gaussian process
were tested against a withheld set of model runs which 
were all predicted to within an aggregate of three standard 
deviations over all model outputs.  A more complete discussion
of the Gaussian process model and its testing for this application
are available as a pre-print at the time of writing \citep{Novak:2013bqa}.

The Gaussian process requires model output from the region of parameter
space to be explored by the MCMC.  Since we are interested in applications
with a significant number of dimensions and the expected increase in 
required samplings scales exponentially with the number of dimensions,
it is important that the sampling be done efficiently.
The method used in this work which is Latin Hypercube sampling.
which divides each dimension of the space into $n$ bins and then
requires that exactly one sampling point to appear in each bin.
The result is that there is no redundancy if a dimension turns out to
be irrelevant and that many values of each parameter are present
in the sampling.  

\begin{table*}
\centering
\begin{tabular}{| c | C{4.3in} | c |}
\hline
Parameter & Description & Range\\
\hline
$(dE/dy)_{pp}$ & The initial energy per rapidity in the diffuse limit compared to measured value in $pp$ collision & 0.85--1.2\\
\hline
$\sigma_{\rm sat}$ & This controls how saturation sets in as function of areal density of the target or projectile. In the wounded nucleon model it is assumed to be the free nucleon-nucleon cross section of 42 mb & 30 mb--50 mb\\
\hline
$f_{wn}$ & Determines the relative weight of the wounded-nucleon and saturation formulas for the initial energy density described in (\ref{eq:wn},~\ref{eq:sat}) & 0--1\\
\hline
$F_{\rm flow}$ & Describes  the strength of the initial flow as a fraction of the amount described in \eqref{eq:earlyFlowResult} & 0.25--1.25\\
\hline
$\eta/s|_{T_c}$ & Viscosity to entropy ratio for $T=170$ MeV & 0 -- 0.5\\
\hline
$\alpha$ & Temperature dependence of $\eta/s$ for temperatures above freezeout, described in Equation \ref{eq:alphadef} & 0 - 5\\
\hline
\end{tabular}
\caption{\label{table:ICpars}
Summary of model parameters. Six model parameters were varied. The first four describe the initial state being fed into the hydrodynamic module, and the last two describe the viscosity and its energy dependence.}
\end{table*}

In our first effort, we explore six parameters simultaneously. 
We chose four related to the initial
condition and two related to the shear viscosity.  The 
normalization of the energy density is essentially a required
parameter to study plausible regions of parameter space
and we chose to couch this in terms of the relationship
to the total transverse energy produced in a minimum bias proton-proton
collision at the same center of mass energy per nucleon.
To test the importance of details of screening in the initial condition,
we chose two related parameters that set the extent to which
a screening model was used and the associated cross-section.
The energy density within the model was 
therefore given by the set of equations given in terms of the 
Glauber thickness functions ($T_{A,B}$) by
\begin{eqnarray}
\label{eq:wn}
\epsilon_{\rm wn}(x,y)&=&
(dE/dy)_{pp} \frac{\sigma_{\rm nn}}{2\sigma_{\rm sat}}
T_A(x,y)\left(1-\exp(-T_B(x,y)\sigma_{\rm sat})\right)\\
\nonumber
&&+ (dE/dy)_{pp} \frac{\sigma_{\rm nn}}{2\sigma_{\rm sat}}T_B(x,y)\left(1-\exp(-T_A(x,y)\sigma_{\rm sat})\right),\\
\label{eq:sat}
\epsilon_{\rm sat}(x,y)&=&(dE/dy)_{pp} \frac{\sigma_{\rm nn}}{\sigma_{\rm sat.}}
T_{\rm min}(x,y)\left(1-\exp(-T_{\rm max}(x,y)\sigma_{\rm sat})\right),\\
\nonumber
T_{\rm min}&=&\frac{2T_AT_B}{T_A+T_B},~~T_{\rm max}=(T_A+T_B)/2.
\end{eqnarray}
The parameter to be varied within the model is $\sigma_{\rm sat}$ as
$\sigma_{\rm nn} = 42$ mb is the inelastic nucleon-nucleon cross-section and
$(dE/dy)_{pp} = 2.613$ GeV is the transverse energy per unit rapidity
and is calculated directly from experimental data.  
The other two parameters are involved
in the final calculation of the energy density from
\begin{equation}
\epsilon(x,y)= (\kappa/\tau) \left[ f_{\rm wn}\epsilon_{\rm wn}(x,y) + (1-f_{\rm wn})\epsilon_{\rm sat}(x,y) \right]
\end{equation}
where $\kappa$ is unity if the total transverse energy is the sum over
the proton-proton collisions and increases in the total energy is larger,
and $f_{\rm wn}$ determines the fraction of the energy distribution
determined from the wounded nucleon scaling as opposed to screening scaling.
The fourth parameter is the last related to the initial state and it sets the initial
flow in the same way described in the previous subsection.
The final two parameters related to the shear viscosity ratio and were the
shear viscosity to entropy density ratio at the phase transition, 
$\left.\frac{\eta}{s}\right|_{T_c}$, with $T_c = 170$ MeV being the same as the
hydrodynamic freeze-out temperature, and the increase in the shear viscosity 
at higher temperatures that one expects from QCD arguments 
\citep{Csernai:2006zz,York:2008rr,Huot:2006ys}
moving away from the critical region,
\begin{equation}
\label{eq:alphadef}
\frac{\eta}{s}=\left.\frac{\eta}{s}\right|_{T_c}+\alpha\ln\left(\frac{T}{T_c}\right),
\end{equation}
where $\alpha$ is this final parameter to be varied.

Our statistical method samples this parameter space 
somewhat minimally -- $3^6 = 729$ points in the six dimensional space --
at points chosen by Latin hypercube sampling.
The produced particles were aggregated in the same way as the 
experimental data, for which we considered
\begin{itemize}
\item Low momentum pion yield at 0-5\% and 20-30\% centrality \citep{Adler:2003cb}.
\item Average transverse momentum for low momentum pions, kaons, and protons at each centrality \citep{Adler:2003cb}.
\item Three pion source radii from HBT at each centrality \citep{Abelev:2009tp}.
\item Pion elliptic flow reduced by 10\% to account for non-flow and with additional uncertainty 
to account for the lack of fluctuations in the initial conditions \citep{Adams:2004bi}.
\end{itemize}
A Gaussian process was trained to emulate the aggregated model predictions
and the space was sampling by MCMC as described above.

Among these, only HBT radii have not been discussed to this point.
HBT radii measure the extent of the outgoing phase space cloud
through two-particle momentum correlations.  Experiments can measure
the probability of detecting two particles at momenta $p_a$ and $p_b$
divided by the probability of finding particles of each momentum 
independently:
\begin{equation}
C^{ab} ( \vec{P}, \vec{q}) = \frac{ dN^{ab}/d^3 p_a d^3 p_b}{  dN^{a}/d^3 p_a \cdot dN^b/ d^3 p_b},
\end{equation}
where $\vec{P}$ is the summed pair momentum and $\vec{q}$ is their
relative momentum.  In the limit that the observation of particle
a is independent of the observation of particle b, $C^{ab} = 1$.
Instead, what is observed from heavy ion collisions is a rich
correlation structure at moderate momentum.  These correlations
can be explained by a thermal source of finite size, $S^{ab}$, and a knowledge of the 
two-particle wave-function, $\phi$, by the Koonin-Pratt equation,
\begin{equation}
C^{ab} ( \vec{P}, \vec{q})  = \int d^3 r ~ S^{ab}(\vec{P}, \vec{r}) | \phi(\vec{q}, \vec{r}) |^2 .
\end{equation}

Typically, the source function is taken to be a Gaussian and 
is parameterized in terms of the radius in the outward, sideward,
and longitudinal directions ($R_o$, $R_s$, $R_l$).  The separation of the longitudinal
radius is due to the expectation of differing dynamics in the 
direction.  The outward radius is measured along the 
direction of the transverse momentum, and the sideward
radius is orthogonal to the others.  The difference between
the outward radius and the sideward radius is then related
to the explosiveness of the collision environment.  Both
the longitudinal radius and the explosiveness of the 
collision environment were poorly described by 
early, ideal hydrodynamic simulations of heavy ion collisions \citep{Heinz:2002un},
which overestimated the longitudinal radius and explosiveness.

\begin{figure}
\centerline{\includegraphics[width=0.75 \textwidth]{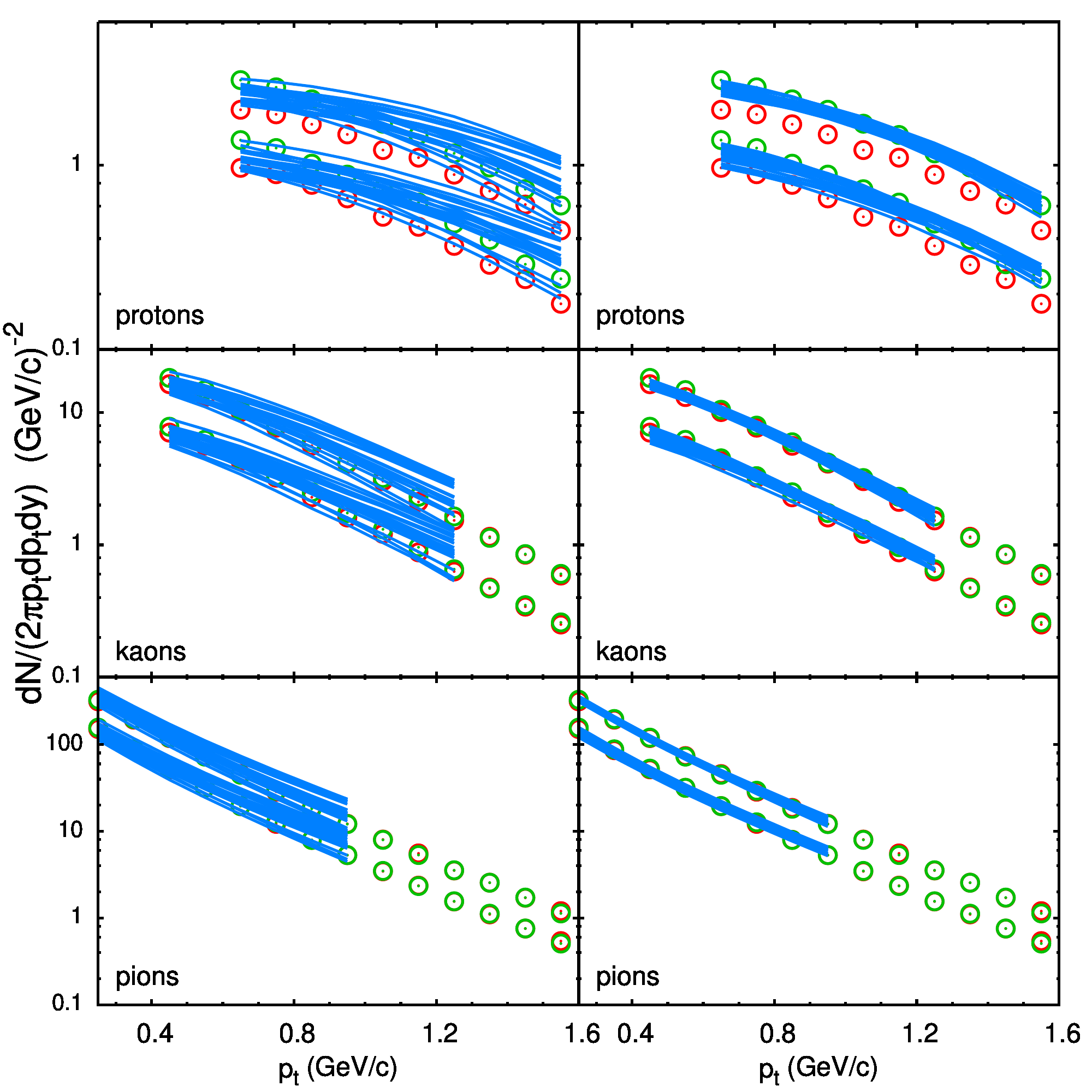}}
\caption{ 
Pion, kaon, and proton transverse momentum spectra for the 5\% 
most central collisions (above) and the 20-30\% centrality bin (below).  
The red symbols are the experimental data as published and the green
symbols are enhanced due to the absence of some chemical reactions
in the gas calculation.  The (blue) lines in the left panels show predictions
from random model sampled uniformly from the entire model space, whereas 
the right side shows model runs selected weighted by the posterior likelihood
distributions from the MCMC trace.  This demonstrates that all model
runs favored in the posterior distribution well describe the particle
spectrum data.
}
\label{fig:specMCMC}
\end{figure}

We compare twenty runs chosen randomly
from the parameter space to twenty runs chosen from the MCMC
trace.  This corresponds to a weighted draw from the prior and 
posterior distributions respectively. The produced particle spectrum for each species and 
centrality is shown in Figure \ref{fig:specMCMC}.  The randomly
chosen settings include many settings that can easily be excluded
by the experimental data.  Comparing this to those selected 
from the trace, we find all of the runs are in reasonable agreement
with the data and none could easily be excluded on these merits
given experimental uncertainties.  We expect that many parameter
settings will fit spectrum data but it is reassuring to find a region
of parameter space that fits this data at a high level.

\begin{figure}
\centerline{\includegraphics[width=0.85 \textwidth]{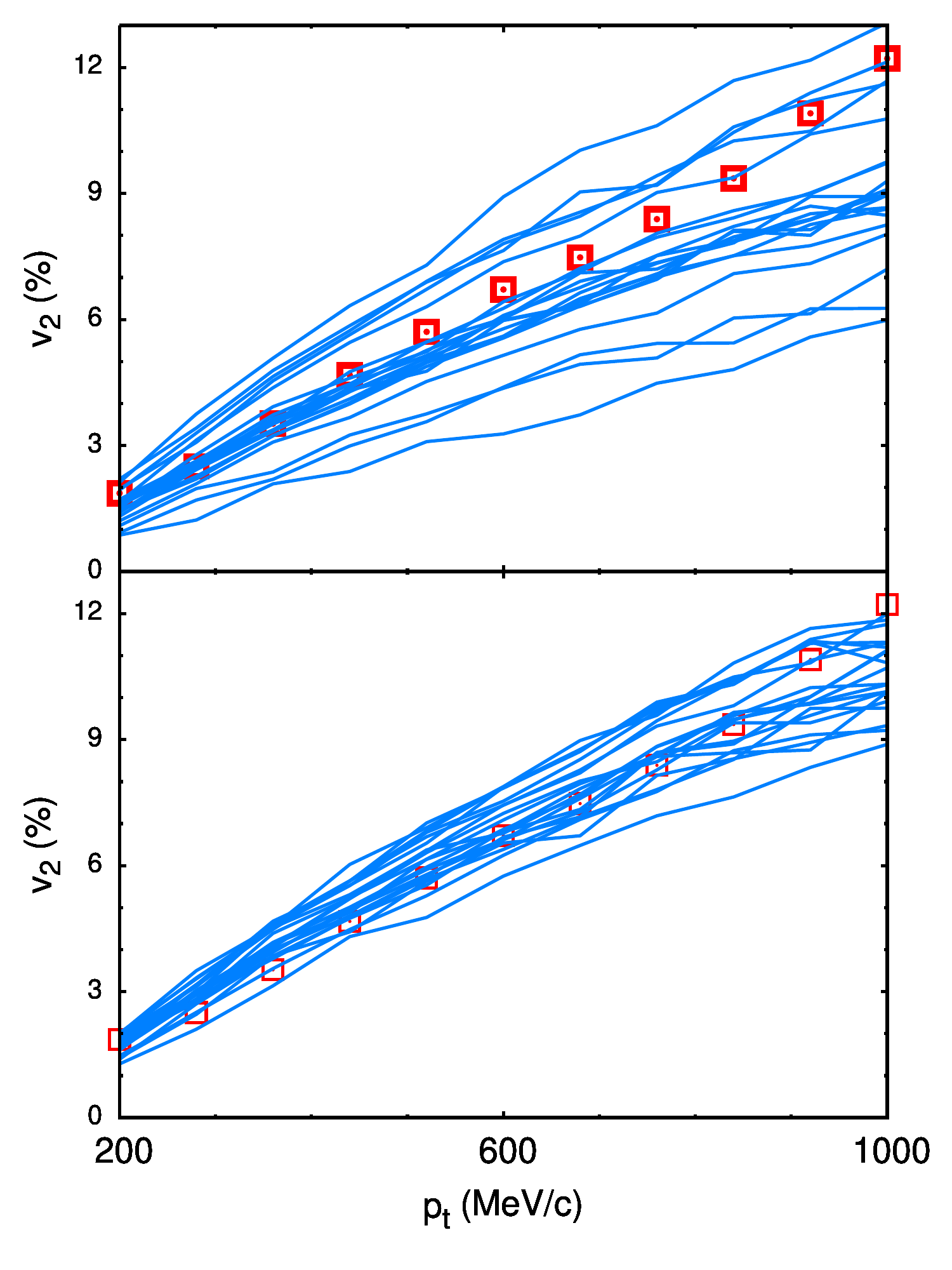}}
\caption{ 
The anisotropic flow as a function of transverse momentum for pions.  The
(red) squares are experimental data from the STAR collaboration.  The
twenty (blue) lines are model runs in each panel; in the top panel they are
chosen via uniform sampling of the parameter space while in the bottom 
panel they are taken randomly from the MCMC trace which samples more
likely regions more heavily.  The runs from the MCMC trace are considerably
closer to the experimental data but note that the posterior distribution contains
many settings that do not describe the anisotropic data with high precision.
}
\label{fig:v2_mcmc}
\end{figure}

For the pion elliptic flow output and data, Figure \ref{fig:v2_mcmc} shows 
the same model runs selected from the prior and posterior distributions.
Again the prior distribution contains many runs that deviate significantly
from the experimental data, whereas the posterior distribution is 
clustered around the data.  Here the deviation is larger than in the 
case of the spectrum which is at least in part due to the increased
uncertainty from the experiment which was 12\% \citep{Abelev:2008ae,Afanasiev:2009wq} 
as opposed to the 
multiplicities which had uncertainties of 6\% and the average transverse
momentum which had uncertainties of only 3\% \citep{Adler:2003cb,Abelev:2008ab}.  
This is larger than the experimental uncertainties since we do not include
initial condition fluctuations.  Also note that there
are some systematic differences in shape between the model runs 
and the experimental data, with the model runs over-predicting $v_2$
at lower momentum and under-predicting it at high momentum.  This
is in contrast to Figures \ref{fig:v2pt} and  \ref{fig:v2pt_DV} which
showed the opposite issue, though there are parameter differences
between those from the posterior distribution here and those taken
in that study including the freeze-out temperature which is higher 
in this case by 5 MeV.  

\begin{figure}
\centerline{\includegraphics[width=0.65\textwidth]{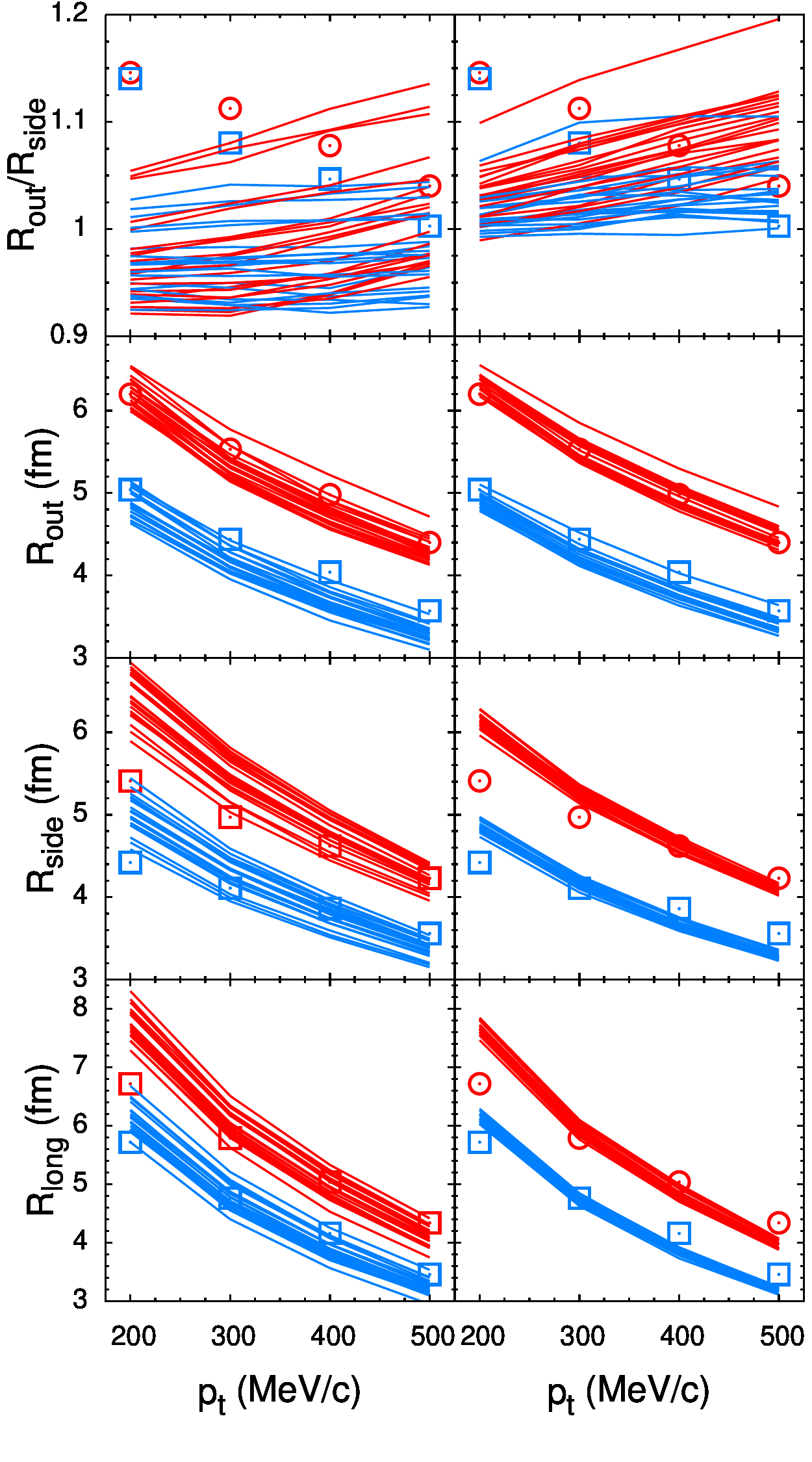}}
\caption{ 
Pion HBT radii as a function of transverse momentum for
central (red circles) and mid-central (blue squares).
In the left panel, the lines are from model runs with parameters
drawn from the flat prior distribution, whereas in the right panel,
the parameter settings are weighted by the posterior distribution.
The parameter settings from the posterior distribution exhibit
more explosiveness as evidenced by the increased $R_o/R_S$
ratio and more precise predictions in general that describe
the experimental data except for in the lowest momentum bin.
}
\label{fig:hbt_mcmc}
\end{figure}

The HBT radii were also reproduced well
by the runs pulled from the posterior distribution as 
evidenced in Figure \ref{fig:hbt_mcmc}.  The model
runs drawn from the prior distribution show more 
spread and tend to overestimate the sideward radius
and underestimate the outward radius. 
For runs from the posterior distribution, the outward
radius is described well at all momenta, while the 
other radii are overestimated in the lowest momentum
bin.  It is concerning that the momentum dependence
of the sideward radius appears to be too dramatic,
but the source of this discrepancy is not understood.

\begin{figure}
\centerline{\includegraphics[width=0.95 \textwidth]{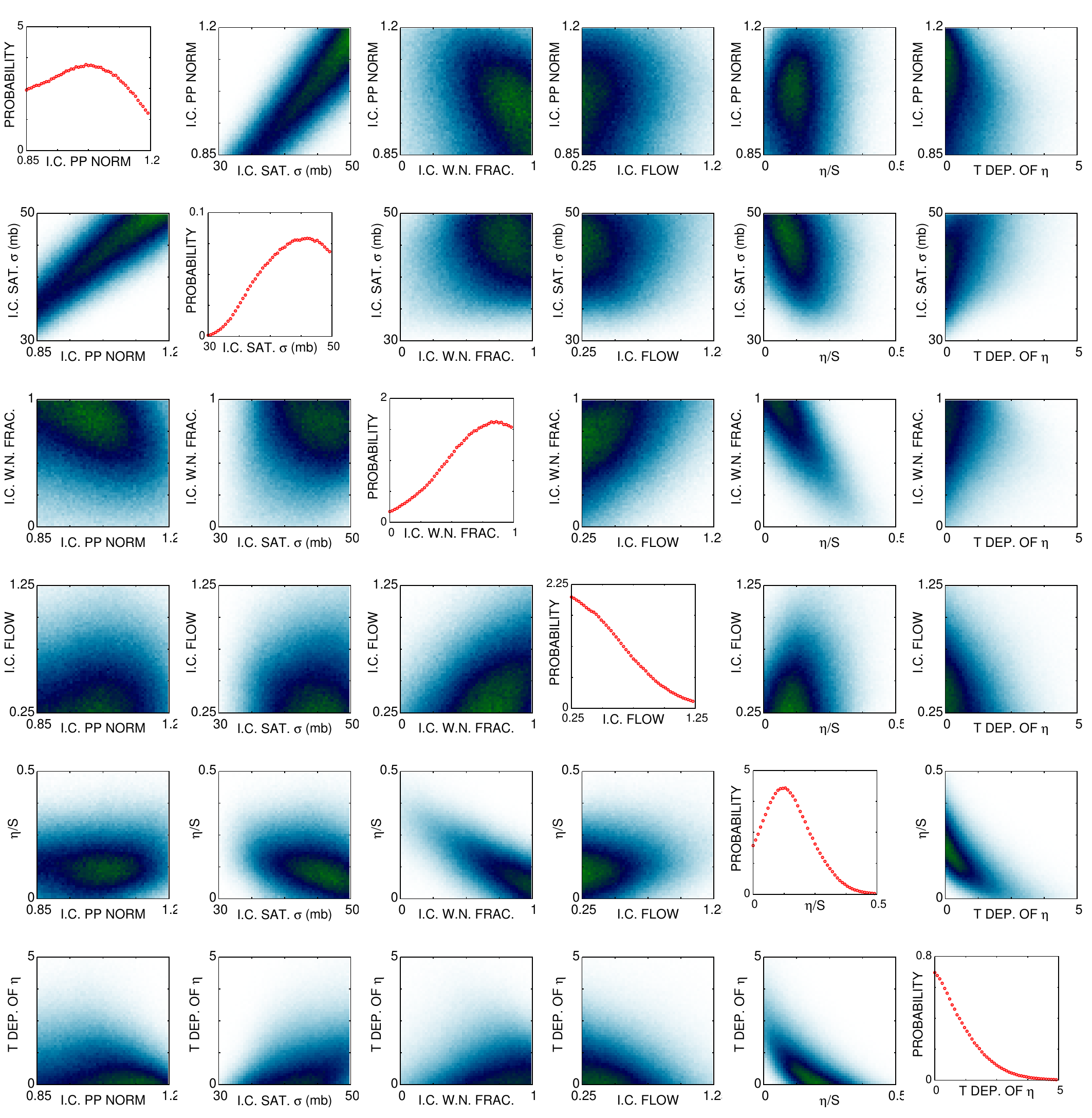}}
\caption{ 
Parameter space density plots for the Markov Chain Monte Carlo trace run 
over the Gaussian process emulation of the hybrid model.  On-diagonal plots
show the density projecting out all five remaining variables while
the off-diagonal plots show the two-dimensional density distribution
projecting out the four remaining variables.  The one-dimensional plots 
show that the most likely parameter sets feature
a small but non-zero value of the shear viscosity,
almost no increase to the shear viscosity at high temperatures,
small pre-equilibrium flow, and a density distribution more like wounded
nucleon than the screening model.  The two-dimensional plots show
how these conclusions depend on one another.  As expected, more eccentric
initial conditions from the screening model lead runs with a higher value of the 
shear viscosity to better agree with the data.  The sum of the shear
viscosity in the critical region and the slope of the shear viscosity above
the critical region appears well constrained; and likewise, the difference
between the saturation cross-section and the energy density normalization
appears well-constrained.
}
\label{fig:gpmcmc}
\end{figure}

Displaying the posterior distribution itself is a difficult
task as the posterior distribution is a six-dimensional density distribution.
Our method is to take all possible projections of the distribution in
one- and two-dimensions and this is shown for our space in Figure 
\ref{fig:gpmcmc}.  If one begins by looking at the one-dimensional 
projections, it appears that both parameters related to the shear
viscosity ($\eta/s$ and T\_DEP\_OF\_$ \eta $) are reduced to a fraction 
of their tested range, while the normalization (I.C.\_PP\_ NORM) and
the saturation cross-section (I.C.\_SAT.\_ $\sigma$ (mb)) are barely
constrained at all.  However, if one moves to the two-dimensional
plots we see that this is only because of our choice of variables and
that a linear combination of the normalization and the saturation
cross-section (roughly their difference) is constrained 
-- presumably by the multiplicities --
whereas the orthogonal combination (their sum) is only somewhat
constrained.  The physical explanation of this is not so clear but
perhaps a slow change in the centrality scaling of the 
multiplicity or the initial eccentricity due to changing
the saturation cross-section.

We find other expected parameter behaviors in the correlation
plots of Figure \ref{fig:gpmcmc}. For instance, the model better describes the data
when shear viscosity at the critical temperature is larger for models
containing saturation than for the wounded nucleon picture due to
larger eccentricities.  Also the optimal shear viscosity at the critical 
temperature is inversely related with the slope of the shear viscosity
at higher temperatures. This suggests that the model output is sensitive
to the average shear viscosity in the region near or above the critical temperature.
It is surprising then that initial conditions involving more saturation,
or more initial eccentricity, do not support a larger high temperature
slope but rather a lower one.   This implies a tension in the model 
output that is not immediately clear but warrants future investigation.
The conclusion that quark matter has small shear viscosity seems to be robust.
Just how small the shear viscosity is appears to be correlated with other parameters 
including the temperature dependence of the shear viscosity but also
the details of the nucleon screening model.  The lack of dependence on
the initial flow is in contrast to our results from the previous subsection.
Given that elliptic flow increases with increasing pre-equilibrium flow
and decreases with increasing shear viscosity, one would expect that 
elliptic flow would only be sensitive to the difference between these parameters.
This suggests that other experimental data is breaking this expected relationship
but the source of this is not well understood.
Another important future investigation might determine which experimental data
underpin this conclusion and attempt to identify other model parameters
that could compromise it.

Also of interest is that the posterior distribution dependence on normalization
is weak except when varying the saturation cross-section.
This suggests that future studies could avoid sampling in that parameter
by locating the normalization that reproduces experimental
multiplicities as done in the previous subsection.  This is attractive
because multiplicities require less than one hundred cascade events
to calculate accurately at the one percent level in contrast to the four thousand events
used to generate each point in parameter space sampled, in addition to 
reducing the dimensionality of the parameter space.

\section{Concluding Remarks}


The broad conclusion of this work is that adding the non-trivial aspects
of the longitudinal expansions to viscous hydrodynamic modeling
does not dramatically affect conclusions about the matter formed in 
heavy ion collisions.  We arrived at this conclusion by investigating
the matter in the hydrodynamic phase and noting that the 
flow velocity shows no change when fully including the longitudinal
expansion, as seen in Figure \ref{fig:uxuy}.
As the flow profile is the driver of elliptic flow, conclusions about 
the dynamics of the hot matter drawn from two dimensional simulations
are trustworthy. While Figure \ref{fig:fosx} shows a decrease in the predicted
source lifetime, the most dramatic effect of this is to decrease the longitudinal source size.
This proves a nice solution to an existing tension between theoretical
models and the experimental data.  These conclusions are based on
simulations done for collisions with beam energy of $\sqrt{s_{NN}} = 200$ GeV
and are expected to hold for larger beam energies, though they would need to be revisited
for smaller beam energies.


This does not imply that there is no reason to pursue further
three dimensional viscous models. Instead it means that 
it is not necessary to model midrapidity data at the highest RHIC energy, 
unless one pursues accuracy at better than the 5\% level.  Even
at small longitudinal rapidity, more experimental data become accessible
to an extended version of our model, 
including directed flow due to non-zero impact parameter
and the associated angular momentum of the source.  Attempts to include 
the effects of non-zero angular momentum in the initial condition \citep{Adil:2005qn}
within the current version of the model have demonstrated persistent
instability that have frustrated attempts to study its possible influence
at midrapidity.  Schematic changes to the integration of the basic
hydrodynamic equations may be necessary to undertake such a study.

These upgrades would make other effects available for study as well, 
for instance, a method that is stable against shocks would allow one 
to study the role of fluctuations.  In the model described here, hydrodynamic
modeling is a small fraction of the computational time compared to
the generation, manipulation and analysis of the particles within the 
resonance gas model.  Assuming that the described changes do not 
significantly alter this balance, one can imagine running around fifty
hydrodynamic events per centrality bin at a cost of less than a factor
of two in the run time.  This would allow investigation of fluctuation
data including average transverse momentum fluctuations and 
elliptic flow fluctuations both with and without boost invariance.

Other features of particle production away from midrapidity
include the effects of finite baryon number and the development
of elliptic flow away from midrapidity.  In this model, the 
finite baryon number from the protons and neutrons in the 
original nuclei is carried away.  While the nuclei recede 
rapidly, baryon current diffuses toward midrapidity and 
manifests itself as an imbalance between the number of protons 
and anti-protons at midrapidity.  
This baryon current would be a necessary feature of any model 
attempting to investigate effects away from midrapidity 
as it seems likely to affect other important observables like elliptic flow.
For instance, if the matter has more heavy baryons, then at the
same energy density the pressure is reduced, due to a decrease
in the mean velocity, and flow should
develop more slowly.  Since elliptic flow falls more rapidly 
than multiplicity as a function of pseudorapidity, this may be 
an interesting phenomenon and could provide consistency checks
on the initial conditions that underpin hydrodynamic models.


While there is much to be gained from continued study of three-dimensional
viscous hydrodynamic models, there is still quite a lot to be learned about
two-dimensional models as well.  While the six-dimensional parameter 
study shown in Figure \ref{fig:gpmcmc} demonstrates that we understand
some of the experimental constraints on model parameters, the full set 
of parameters is very large and many important parameters have yet to
be explored.  The equation of state calculated by lattice QCD seems 
to be a theoretical triumph especially given its agreement with the expectations
of the hadron resonance gas near the phase transition.  However, it
remains unknown whether or not experimental data gathered from 
heavy ion collisions support the conclusion that equilibrium dynamics 
of the quark-gluon plasma are well described by the equation of
state predicted by lattice QCD.  Direct experimental evidence for
lattice QCD would be a massive triumph and remains an important
goal of the field of relativistic heavy ion collisions.


\end{doublespace}

\bibliographystyle{model1-num-names}

\bibliography{PhD_Dissertation_Template}

\begin{thebibliography}{80}
\expandafter\ifx\csname natexlab\endcsname\relax\def\natexlab#1{#1}\fi
\providecommand{\url}[1]{\texttt{#1}}
\providecommand{\href}[2]{#2}
\providecommand{\path}[1]{#1}
\providecommand{\DOIprefix}{doi:}
\providecommand{\ArXivprefix}{arXiv:}
\providecommand{\URLprefix}{URL: }
\providecommand{\Pubmedprefix}{pmid:}
\providecommand{\doi}[1]{\href{http://dx.doi.org/#1}{\path{#1}}}
\providecommand{\Pubmed}[1]{\href{pmid:#1}{\path{#1}}}
\providecommand{\bibinfo}[2]{#2}
\ifx\xfnm\relax \def\xfnm[#1]{\unskip,\space#1}\fi
\bibitem[{Adams et~al.(2003)}]{Adams:2003im}
\bibinfo{author}{J.~Adams}, et~al. (\bibinfo{collaboration}{STAR
  Collaboration}),
\newblock \bibinfo{title}{{Evidence from d + Au measurements for final state
  suppression of high p(T) hadrons in Au+Au collisions at RHIC}},
\newblock \bibinfo{journal}{Phys.Rev.Lett.} \bibinfo{volume}{91}
  (\bibinfo{year}{2003}) \bibinfo{pages}{072304}.
\bibitem[{Abelev et~al.(2009)}]{Abelev:2008ab}
\bibinfo{author}{B.~Abelev}, et~al. (\bibinfo{collaboration}{STAR
  Collaboration}),
\newblock \bibinfo{title}{{Systematic Measurements of Identified Particle
  Spectra in $p p, d^+$ Au and Au+Au Collisions from STAR}},
\newblock \bibinfo{journal}{Phys.Rev.} \bibinfo{volume}{C79}
  (\bibinfo{year}{2009}) \bibinfo{pages}{034909}.
\bibitem[{Kolb(2002)}]{kolbThesis}
\bibinfo{author}{P.~F. Kolb}, \bibinfo{title}{Early Thermalization and
  Hydrodynamic Expansion in Nuclear Collisions at RHIC}, Ph.D. thesis,
  \bibinfo{year}{2002}.
\bibitem[{Adare et~al.(2007)}]{Adare:2006ti}
\bibinfo{author}{A.~Adare}, et~al. (\bibinfo{collaboration}{PHENIX
  Collaboration}),
\newblock \bibinfo{title}{{Scaling properties of azimuthal anisotropy in Au+Au
  and Cu+Cu collisions at s(NN) = 200-GeV}},
\newblock \bibinfo{journal}{Phys.Rev.Lett.} \bibinfo{volume}{98}
  (\bibinfo{year}{2007}) \bibinfo{pages}{162301}.
\bibitem[{Bearden et~al.(2005)}]{Bearden:2004yx}
\bibinfo{author}{I.~Bearden}, et~al. (\bibinfo{collaboration}{BRAHMS
  Collaboration}),
\newblock \bibinfo{title}{{Charged meson rapidity distributions in central
  Au+Au collisions at s(NN)**(1/2) = 200-GeV}},
\newblock \bibinfo{journal}{Phys.Rev.Lett.} \bibinfo{volume}{94}
  (\bibinfo{year}{2005}) \bibinfo{pages}{162301}.
\bibitem[{Borsanyi et~al.(2010)Borsanyi, Endrodi, Fodor, Jakovac, Katz
  et~al.}]{Borsanyi:2010cj}
\bibinfo{author}{S.~Borsanyi}, \bibinfo{author}{G.~Endrodi},
  \bibinfo{author}{Z.~Fodor}, \bibinfo{author}{A.~Jakovac},
  \bibinfo{author}{S.~D. Katz}, et~al.,
\newblock \bibinfo{title}{{The QCD equation of state with dynamical quarks}},
\newblock \bibinfo{journal}{JHEP} \bibinfo{volume}{1011} (\bibinfo{year}{2010})
  \bibinfo{pages}{077}.
\bibitem[{Huovinen and Petersen(2012)}]{Huovinen:2012is}
\bibinfo{author}{P.~Huovinen}, \bibinfo{author}{H.~Petersen},
\newblock \bibinfo{title}{{Particlization in hybrid models}}
  (\bibinfo{year}{2012}).
\bibitem[{Drescher et~al.(2006)Drescher, Dumitru, Hayashigaki, and
  Nara}]{Drescher:2006pi}
\bibinfo{author}{H.-J. Drescher}, \bibinfo{author}{A.~Dumitru},
  \bibinfo{author}{A.~Hayashigaki}, \bibinfo{author}{Y.~Nara},
\newblock \bibinfo{title}{{The Eccentricity in heavy-ion collisions from color
  glass condensate initial conditions}},
\newblock \bibinfo{journal}{Phys.Rev.} \bibinfo{volume}{C74}
  (\bibinfo{year}{2006}) \bibinfo{pages}{044905}.
\bibitem[{Adler et~al.(2004)}]{Adler:2003cb}
\bibinfo{author}{S.~Adler}, et~al. (\bibinfo{collaboration}{PHENIX
  Collaboration}),
\newblock \bibinfo{title}{{Identified charged particle spectra and yields in
  Au+Au collisions at S(NN)**1/2 = 200-GeV}},
\newblock \bibinfo{journal}{Phys.Rev.} \bibinfo{volume}{C69}
  (\bibinfo{year}{2004}) \bibinfo{pages}{034909}.
\bibitem[{Bass et~al.(1999)Bass, Bleicher, Cassing, Dumitru, Drescher
  et~al.}]{Bass:1999zq}
\bibinfo{author}{S.~Bass}, \bibinfo{author}{M.~Bleicher},
  \bibinfo{author}{W.~Cassing}, \bibinfo{author}{A.~Dumitru},
  \bibinfo{author}{H.~Drescher}, et~al.,
\newblock \bibinfo{title}{{Last call for RHIC predictions}},
\newblock \bibinfo{journal}{Nucl.Phys.} \bibinfo{volume}{A661}
  (\bibinfo{year}{1999}) \bibinfo{pages}{205--260}.
\bibitem[{Adamczyk et~al.(2012)}]{Adamczyk:2012ku}
\bibinfo{author}{L.~Adamczyk}, et~al. (\bibinfo{collaboration}{STAR
  collaboration}),
\newblock \bibinfo{title}{{Inclusive charged hadron elliptic flow in Au + Au
  collisions at $\sqrt{s_{NN}}$ = 7.7 - 39 GeV}},
\newblock \bibinfo{journal}{Phys.Rev.} \bibinfo{volume}{C86}
  (\bibinfo{year}{2012}) \bibinfo{pages}{054908}.
\bibitem[{Adcox et~al.(2002)}]{Adcox:2001jp}
\bibinfo{author}{K.~Adcox}, et~al. (\bibinfo{collaboration}{PHENIX
  Collaboration}),
\newblock \bibinfo{title}{{Suppression of hadrons with large transverse
  momentum in central Au+Au collisions at $\sqrt{s_{NN}}$ = 130-GeV}},
\newblock \bibinfo{journal}{Phys.Rev.Lett.} \bibinfo{volume}{88}
  (\bibinfo{year}{2002}) \bibinfo{pages}{022301}.
\bibitem[{Adams et~al.(2005)}]{Adams:2005ph}
\bibinfo{author}{J.~Adams}, et~al. (\bibinfo{collaboration}{STAR
  Collaboration}),
\newblock \bibinfo{title}{{Distributions of charged hadrons associated with
  high transverse momentum particles in pp and Au + Au collisions at
  s(NN)**(1/2) = 200-GeV}},
\newblock \bibinfo{journal}{Phys.Rev.Lett.} \bibinfo{volume}{95}
  (\bibinfo{year}{2005}) \bibinfo{pages}{152301}.
\bibitem[{Adler et~al.(2003)}]{Adler:2002tq}
\bibinfo{author}{C.~Adler}, et~al. (\bibinfo{collaboration}{STAR
  Collaboration}),
\newblock \bibinfo{title}{{Disappearance of back-to-back high $p_{T}$ hadron
  correlations in central Au+Au collisions at $\sqrt{s_{NN}}$ = 200-GeV}},
\newblock \bibinfo{journal}{Phys.Rev.Lett.} \bibinfo{volume}{90}
  (\bibinfo{year}{2003}) \bibinfo{pages}{082302}.
\bibitem[{Ackermann et~al.(2001)}]{Ackermann:2000tr}
\bibinfo{author}{K.~Ackermann}, et~al. (\bibinfo{collaboration}{STAR
  Collaboration}),
\newblock \bibinfo{title}{{Elliptic flow in Au + Au collisions at
  (S(NN))**(1/2) = 130 GeV}},
\newblock \bibinfo{journal}{Phys.Rev.Lett.} \bibinfo{volume}{86}
  (\bibinfo{year}{2001}) \bibinfo{pages}{402--407}.
\bibitem[{Adler et~al.(2003)}]{Adler:2003kt}
\bibinfo{author}{S.~Adler}, et~al. (\bibinfo{collaboration}{PHENIX
  Collaboration}),
\newblock \bibinfo{title}{{Elliptic flow of identified hadrons in Au+Au
  collisions at s(NN)**(1/2) = 200-GeV}},
\newblock \bibinfo{journal}{Phys.Rev.Lett.} \bibinfo{volume}{91}
  (\bibinfo{year}{2003}) \bibinfo{pages}{182301}.
\bibitem[{Abelev et~al.(2008)}]{Abelev:2008ae}
\bibinfo{author}{B.~Abelev}, et~al. (\bibinfo{collaboration}{STAR
  Collaboration}),
\newblock \bibinfo{title}{{Centrality dependence of charged hadron and strange
  hadron elliptic flow from s(NN)**(1/2) = 200-GeV Au + Au collisions}},
\newblock \bibinfo{journal}{Phys.Rev.} \bibinfo{volume}{C77}
  (\bibinfo{year}{2008}) \bibinfo{pages}{054901}.
\bibitem[{Afanasiev et~al.(2009)}]{Afanasiev:2009wq}
\bibinfo{author}{S.~Afanasiev}, et~al. (\bibinfo{collaboration}{PHENIX
  Collaboration}),
\newblock \bibinfo{title}{{Systematic Studies of Elliptic Flow Measurements in
  Au+Au Collisions at s**(1/2) = 200-GeV}},
\newblock \bibinfo{journal}{Phys.Rev.} \bibinfo{volume}{C80}
  (\bibinfo{year}{2009}) \bibinfo{pages}{024909}.
\bibitem[{Armesto et~al.(2012)Armesto, Cole, Gale, Horowitz, Jacobs
  et~al.}]{Armesto:2011ht}
\bibinfo{author}{N.~Armesto}, \bibinfo{author}{B.~Cole},
  \bibinfo{author}{C.~Gale}, \bibinfo{author}{W.~A. Horowitz},
  \bibinfo{author}{P.~Jacobs}, et~al.,
\newblock \bibinfo{title}{{Comparison of Jet Quenching Formalisms for a
  Quark-Gluon Plasma 'Brick'}},
\newblock \bibinfo{journal}{Phys.Rev.} \bibinfo{volume}{C86}
  (\bibinfo{year}{2012}) \bibinfo{pages}{064904}.
\bibitem[{Kolb and Heinz(2003)}]{Kolb:2003dz}
\bibinfo{author}{P.~F. Kolb}, \bibinfo{author}{U.~W. Heinz},
\newblock \bibinfo{title}{{Hydrodynamic description of ultrarelativistic heavy
  ion collisions}}  (\bibinfo{year}{2003}).
\bibitem[{Adams et~al.(2005)}]{Adams:2005dq}
\bibinfo{author}{J.~Adams}, et~al. (\bibinfo{collaboration}{STAR
  Collaboration}),
\newblock \bibinfo{title}{{Experimental and theoretical challenges in the
  search for the quark gluon plasma: The STAR Collaboration's critical
  assessment of the evidence from RHIC collisions}},
\newblock \bibinfo{journal}{Nucl.Phys.} \bibinfo{volume}{A757}
  (\bibinfo{year}{2005}) \bibinfo{pages}{102--183}.
\bibitem[{Heinz and Kolb(2002)}]{Heinz:2002un}
\bibinfo{author}{U.~W. Heinz}, \bibinfo{author}{P.~F. Kolb},
\newblock \bibinfo{title}{{Two RHIC puzzles: Early thermalization and the HBT
  problem}}  (\bibinfo{year}{2002}).
\bibitem[{Pratt(2009)}]{Pratt:2008qv}
\bibinfo{author}{S.~Pratt},
\newblock \bibinfo{title}{{Resolving the HBT Puzzle in Relativistic Heavy Ion
  Collision}},
\newblock \bibinfo{journal}{Phys.Rev.Lett.} \bibinfo{volume}{102}
  (\bibinfo{year}{2009}) \bibinfo{pages}{232301}.
\bibitem[{Schmidt(2006)}]{Schmidt:2006us}
\bibinfo{author}{C.~Schmidt},
\newblock \bibinfo{title}{{Lattice QCD at finite density}},
\newblock \bibinfo{journal}{PoS} \bibinfo{volume}{LAT2006}
  (\bibinfo{year}{2006}) \bibinfo{pages}{021}.
\bibitem[{Huovinen and Petreczky(2011)}]{Huovinen:2011xc}
\bibinfo{author}{P.~Huovinen}, \bibinfo{author}{P.~Petreczky},
\newblock \bibinfo{title}{{Equation of state at finite baryon density based on
  lattice QCD}},
\newblock \bibinfo{journal}{J.Phys.} \bibinfo{volume}{G38}
  (\bibinfo{year}{2011}) \bibinfo{pages}{124103}.
\bibitem[{Song and Heinz(2008)}]{Song:2007ux}
\bibinfo{author}{H.~Song}, \bibinfo{author}{U.~W. Heinz},
\newblock \bibinfo{title}{{Causal viscous hydrodynamics in 2+1 dimensions for
  relativistic heavy-ion collisions}},
\newblock \bibinfo{journal}{Phys.Rev.} \bibinfo{volume}{C77}
  (\bibinfo{year}{2008}) \bibinfo{pages}{064901}.
\bibitem[{Luzum and Romatschke(2008)}]{Luzum:2008cw}
\bibinfo{author}{M.~Luzum}, \bibinfo{author}{P.~Romatschke},
\newblock \bibinfo{title}{{Conformal Relativistic Viscous Hydrodynamics:
  Applications to RHIC results at s(NN)**(1/2) = 200-GeV}},
\newblock \bibinfo{journal}{Phys.Rev.} \bibinfo{volume}{C78}
  (\bibinfo{year}{2008}) \bibinfo{pages}{034915}.
\bibitem[{Dusling and Teaney(2008)}]{Dusling:2007gi}
\bibinfo{author}{K.~Dusling}, \bibinfo{author}{D.~Teaney},
\newblock \bibinfo{title}{{Simulating elliptic flow with viscous
  hydrodynamics}},
\newblock \bibinfo{journal}{Phys.Rev.} \bibinfo{volume}{C77}
  (\bibinfo{year}{2008}) \bibinfo{pages}{034905}.
\bibitem[{Bjorken(1983)}]{Bjorken:1982qr}
\bibinfo{author}{J.~Bjorken},
\newblock \bibinfo{title}{{Highly Relativistic Nucleus-Nucleus Collisions: The
  Central Rapidity Region}},
\newblock \bibinfo{journal}{Phys.Rev.} \bibinfo{volume}{D27}
  (\bibinfo{year}{1983}) \bibinfo{pages}{140--151}.
\bibitem[{Bazavov et~al.(2009)Bazavov, Bhattacharya, Cheng, Christ, DeTar
  et~al.}]{Bazavov:2009zn}
\bibinfo{author}{A.~Bazavov}, \bibinfo{author}{T.~Bhattacharya},
  \bibinfo{author}{M.~Cheng}, \bibinfo{author}{N.~Christ},
  \bibinfo{author}{C.~DeTar}, et~al.,
\newblock \bibinfo{title}{{Equation of state and QCD transition at finite
  temperature}},
\newblock \bibinfo{journal}{Phys.Rev.} \bibinfo{volume}{D80}
  (\bibinfo{year}{2009}) \bibinfo{pages}{014504}.
\bibitem[{Huang(1987)}]{huangStatMech}
\bibinfo{author}{K.~Huang}, \bibinfo{title}{Statistical Mechanics},
  \bibinfo{edition}{2} ed., \bibinfo{publisher}{Wiley}, \bibinfo{year}{1987}.
\bibitem[{Romatschke(2010)}]{Romatschke:2009im}
\bibinfo{author}{P.~Romatschke},
\newblock \bibinfo{title}{{New Developments in Relativistic Viscous
  Hydrodynamics}},
\newblock \bibinfo{journal}{Int.J.Mod.Phys.} \bibinfo{volume}{E19}
  (\bibinfo{year}{2010}) \bibinfo{pages}{1--53}.
\bibitem[{Betz et~al.(2009)Betz, Henkel, and Rischke}]{Betz:2008me}
\bibinfo{author}{B.~Betz}, \bibinfo{author}{D.~Henkel},
  \bibinfo{author}{D.~Rischke},
\newblock \bibinfo{title}{{From kinetic theory to dissipative fluid dynamics}},
\newblock \bibinfo{journal}{Prog.Part.Nucl.Phys.} \bibinfo{volume}{62}
  (\bibinfo{year}{2009}) \bibinfo{pages}{556--561}.
\bibitem[{Huovinen and Molnar(2009)}]{Huovinen:2008te}
\bibinfo{author}{P.~Huovinen}, \bibinfo{author}{D.~Molnar},
\newblock \bibinfo{title}{{The Applicability of causal dissipative
  hydrodynamics to relativistic heavy ion collisions}},
\newblock \bibinfo{journal}{Phys.Rev.} \bibinfo{volume}{C79}
  (\bibinfo{year}{2009}) \bibinfo{pages}{014906}.
\bibitem[{Pratt(2008)}]{Pratt:2007gj}
\bibinfo{author}{S.~Pratt},
\newblock \bibinfo{title}{{Formulating viscous hydrodynamics for large velocity
  gradients}},
\newblock \bibinfo{journal}{Phys.Rev.} \bibinfo{volume}{C77}
  (\bibinfo{year}{2008}) \bibinfo{pages}{024910}.
\bibitem[{Muronga(2002)}]{Muronga:2001zk}
\bibinfo{author}{A.~Muronga},
\newblock \bibinfo{title}{{Second order dissipative fluid dynamics for
  ultrarelativistic nuclear collisions}},
\newblock \bibinfo{journal}{Phys.Rev.Lett.} \bibinfo{volume}{88}
  (\bibinfo{year}{2002}) \bibinfo{pages}{062302}.
\bibitem[{Baier et~al.(2008)Baier, Romatschke, Son, Starinets, and
  Stephanov}]{Baier:2007ix}
\bibinfo{author}{R.~Baier}, \bibinfo{author}{P.~Romatschke},
  \bibinfo{author}{D.~T. Son}, \bibinfo{author}{A.~O. Starinets},
  \bibinfo{author}{M.~A. Stephanov},
\newblock \bibinfo{title}{{Relativistic viscous hydrodynamics, conformal
  invariance, and holography}},
\newblock \bibinfo{journal}{JHEP} \bibinfo{volume}{0804} (\bibinfo{year}{2008})
  \bibinfo{pages}{100}.
\bibitem[{Muronga(2007{\natexlab{a}})}]{Muronga:2006zw}
\bibinfo{author}{A.~Muronga},
\newblock \bibinfo{title}{{Relativistic Dynamics of Non-ideal Fluids: Viscous
  and heat-conducting fluids. I. General Aspects and 3+1 Formulation for
  Nuclear Collisions}},
\newblock \bibinfo{journal}{Phys.Rev.} \bibinfo{volume}{C76}
  (\bibinfo{year}{2007}{\natexlab{a}}) \bibinfo{pages}{014909}.
\bibitem[{Muronga(2007{\natexlab{b}})}]{Muronga:2006zx}
\bibinfo{author}{A.~Muronga},
\newblock \bibinfo{title}{{Relativistic Dynamics of Non-ideal Fluids: Viscous
  and heat-conducting fluids. II. Transport properties and microscopic
  description of relativistic nuclear matter}},
\newblock \bibinfo{journal}{Phys.Rev.} \bibinfo{volume}{C76}
  (\bibinfo{year}{2007}{\natexlab{b}}) \bibinfo{pages}{014910}.
\bibitem[{Grad(1949)}]{ref:grad}
\bibinfo{author}{H.~Grad},
\newblock \bibinfo{journal}{Commun. Pure App. Math} \bibinfo{volume}{2}
  (\bibinfo{year}{1949}) \bibinfo{pages}{381}.
\bibitem[{Baier et~al.(2006)Baier, Romatschke, and Wiedemann}]{Baier:2006um}
\bibinfo{author}{R.~Baier}, \bibinfo{author}{P.~Romatschke},
  \bibinfo{author}{U.~A. Wiedemann},
\newblock \bibinfo{title}{{Dissipative hydrodynamics and heavy ion
  collisions}},
\newblock \bibinfo{journal}{Phys.Rev.} \bibinfo{volume}{C73}
  (\bibinfo{year}{2006}) \bibinfo{pages}{064903}.
\bibitem[{Huovinen and Petreczky(2010)}]{Huovinen:2009yb}
\bibinfo{author}{P.~Huovinen}, \bibinfo{author}{P.~Petreczky},
\newblock \bibinfo{title}{{QCD Equation of State and Hadron Resonance Gas}},
\newblock \bibinfo{journal}{Nucl.Phys.} \bibinfo{volume}{A837}
  (\bibinfo{year}{2010}) \bibinfo{pages}{26--53}.
\bibitem[{Press et~al.(1992)Press, Teukolsky, Vetterling, and
  Flannery}]{Press:1992:NRC:148286}
\bibinfo{author}{W.~H. Press}, \bibinfo{author}{S.~A. Teukolsky},
  \bibinfo{author}{W.~T. Vetterling}, \bibinfo{author}{B.~P. Flannery},
  \bibinfo{title}{Numerical recipes in C (2nd ed.): the art of scientific
  computing}, \bibinfo{publisher}{Cambridge University Press},
  \bibinfo{address}{New York, NY, USA}, \bibinfo{year}{1992}.
\bibitem[{Paech and Pratt(2006)}]{Paech:2006st}
\bibinfo{author}{K.~Paech}, \bibinfo{author}{S.~Pratt},
\newblock \bibinfo{title}{{Origins of bulk viscosity in relativistic heavy ion
  collisions}},
\newblock \bibinfo{journal}{Phys.Rev.} \bibinfo{volume}{C74}
  (\bibinfo{year}{2006}) \bibinfo{pages}{014901}.
\bibitem[{Weinberg(1995)}]{citeulike:712984}
\bibinfo{author}{S.~Weinberg}, \bibinfo{title}{{The Quantum Theory of Fields
  (Volume 1)}}, \bibinfo{edition}{1} ed., \bibinfo{publisher}{Cambridge
  University Press}, \bibinfo{year}{1995}. \URLprefix
  \url{http://www.worldcat.org/isbn/0521550017}.
\bibitem[{Prakash et~al.(1993)Prakash, Prakash, Venugopalan, and
  Welke}]{Prakash:1993bt}
\bibinfo{author}{M.~Prakash}, \bibinfo{author}{M.~Prakash},
  \bibinfo{author}{R.~Venugopalan}, \bibinfo{author}{G.~Welke},
\newblock \bibinfo{title}{{Nonequilibrium properties of hadronic mixtures}},
\newblock \bibinfo{journal}{Phys.Rept.} \bibinfo{volume}{227}
  (\bibinfo{year}{1993}) \bibinfo{pages}{321--366}.
\bibitem[{Song et~al.(2011)Song, Bass, and Heinz}]{Song:2010aq}
\bibinfo{author}{H.~Song}, \bibinfo{author}{S.~A. Bass},
  \bibinfo{author}{U.~Heinz},
\newblock \bibinfo{title}{{Viscous QCD matter in a hybrid
  hydrodynamic+Boltzmann approach}},
\newblock \bibinfo{journal}{Phys.Rev.} \bibinfo{volume}{C83}
  (\bibinfo{year}{2011}) \bibinfo{pages}{024912}.
\bibitem[{Niemi et~al.(2012)Niemi, Denicol, Huovinen, Molnar, and
  Rischke}]{Niemi:2012ry}
\bibinfo{author}{H.~Niemi}, \bibinfo{author}{G.~Denicol},
  \bibinfo{author}{P.~Huovinen}, \bibinfo{author}{E.~Molnar},
  \bibinfo{author}{D.~Rischke},
\newblock \bibinfo{title}{{Influence of a temperature-dependent shear viscosity
  on the azimuthal asymmetries of transverse momentum spectra in
  ultrarelativistic heavy-ion collisions}},
\newblock \bibinfo{journal}{Phys.Rev.} \bibinfo{volume}{C86}
  (\bibinfo{year}{2012}) \bibinfo{pages}{014909}.
\bibitem[{Csernai et~al.(2006)Csernai, Kapusta, and McLerran}]{Csernai:2006zz}
\bibinfo{author}{L.~P. Csernai}, \bibinfo{author}{J.~Kapusta},
  \bibinfo{author}{L.~D. McLerran},
\newblock \bibinfo{title}{{On the Strongly-Interacting Low-Viscosity Matter
  Created in Relativistic Nuclear Collisions}},
\newblock \bibinfo{journal}{Phys.Rev.Lett.} \bibinfo{volume}{97}
  (\bibinfo{year}{2006}) \bibinfo{pages}{152303}.
\bibitem[{Huot et~al.(2007)Huot, Jeon, and Moore}]{Huot:2006ys}
\bibinfo{author}{S.~C. Huot}, \bibinfo{author}{S.~Jeon}, \bibinfo{author}{G.~D.
  Moore},
\newblock \bibinfo{title}{{Shear viscosity in weakly coupled N = 4 super
  Yang-Mills theory compared to QCD}},
\newblock \bibinfo{journal}{Phys.Rev.Lett.} \bibinfo{volume}{98}
  (\bibinfo{year}{2007}) \bibinfo{pages}{172303}.
\bibitem[{Boris and Book(1973)}]{shasta}
\bibinfo{author}{J.~Boris}, \bibinfo{author}{D.~Book},
\newblock \bibinfo{journal}{J. Comput. Phys.} \bibinfo{volume}{11}
  (\bibinfo{year}{1973}) \bibinfo{pages}{38}.
\bibitem[{Kurganov and Tadmor(2000)}]{ktHydro}
\bibinfo{author}{A.~Kurganov}, \bibinfo{author}{E.~Tadmor},
\newblock \bibinfo{journal}{Journal of Computational Physics}
  \bibinfo{volume}{160} (\bibinfo{year}{2000}) \bibinfo{pages}{214}.
\bibitem[{Pratt and Torrieri(2010)}]{Pratt:2010jt}
\bibinfo{author}{S.~Pratt}, \bibinfo{author}{G.~Torrieri},
\newblock \bibinfo{title}{{Coupling Relativistic Viscous Hydrodynamics to
  Boltzmann Descriptions}},
\newblock \bibinfo{journal}{Phys.Rev.} \bibinfo{volume}{C82}
  (\bibinfo{year}{2010}) \bibinfo{pages}{044901}.
\bibitem[{Holopainen and Huovinen(2012)}]{Holopainen:2012id}
\bibinfo{author}{H.~Holopainen}, \bibinfo{author}{P.~Huovinen},
\newblock \bibinfo{title}{{Dynamical Freeze-out in Event-by-Event
  Hydrodynamics}},
\newblock \bibinfo{journal}{J.Phys.Conf.Ser.} \bibinfo{volume}{389}
  (\bibinfo{year}{2012}) \bibinfo{pages}{012018}.
\bibitem[{Holopainen et~al.(2011)Holopainen, Niemi, and
  Eskola}]{Holopainen:2010gz}
\bibinfo{author}{H.~Holopainen}, \bibinfo{author}{H.~Niemi},
  \bibinfo{author}{K.~J. Eskola},
\newblock \bibinfo{title}{{Event-by-event hydrodynamics and elliptic flow from
  fluctuating initial state}},
\newblock \bibinfo{journal}{Phys.Rev.} \bibinfo{volume}{C83}
  (\bibinfo{year}{2011}) \bibinfo{pages}{034901}.
\bibitem[{Cooper and Frye(1974)}]{CooperFrye}
\bibinfo{author}{F.~Cooper}, \bibinfo{author}{G.~Frye},
\newblock \bibinfo{journal}{Phys. Rev. D} \bibinfo{volume}{10}
  (\bibinfo{year}{1974}) \bibinfo{pages}{186}.
\bibitem[{William E.~Lorensen(1987)}]{marchingCubes}
\bibinfo{author}{H.~E.~C. William E.~Lorensen},
\newblock \bibinfo{title}{Marching cubes: A high resolution 3d surface
  construction algorithm},
\newblock \bibinfo{journal}{Computer Graphics} \bibinfo{volume}{21}
  (\bibinfo{year}{1987}) \bibinfo{pages}{4}.
\bibitem[{Miller et~al.(2007)Miller, Reygers, Sanders, and
  Steinberg}]{Miller:2007ri}
\bibinfo{author}{M.~L. Miller}, \bibinfo{author}{K.~Reygers},
  \bibinfo{author}{S.~J. Sanders}, \bibinfo{author}{P.~Steinberg},
\newblock \bibinfo{title}{{Glauber modeling in high energy nuclear
  collisions}},
\newblock \bibinfo{journal}{Ann.Rev.Nucl.Part.Sci.} \bibinfo{volume}{57}
  (\bibinfo{year}{2007}) \bibinfo{pages}{205--243}.
\bibitem[{Kolb et~al.(2001)Kolb, Heinz, Huovinen, Eskola, and
  Tuominen}]{Kolb:2001qz}
\bibinfo{author}{P.~Kolb}, \bibinfo{author}{U.~W. Heinz},
  \bibinfo{author}{P.~Huovinen}, \bibinfo{author}{K.~Eskola},
  \bibinfo{author}{K.~Tuominen},
\newblock \bibinfo{title}{{Centrality dependence of multiplicity, transverse
  energy, and elliptic flow from hydrodynamics}},
\newblock \bibinfo{journal}{Nucl.Phys.} \bibinfo{volume}{A696}
  (\bibinfo{year}{2001}) \bibinfo{pages}{197--215}.
\bibitem[{Kharzeev et~al.(2004)Kharzeev, Levin, and Nardi}]{Kharzeev:2002ei}
\bibinfo{author}{D.~Kharzeev}, \bibinfo{author}{E.~Levin},
  \bibinfo{author}{M.~Nardi},
\newblock \bibinfo{title}{{QCD saturation and deuteron nucleus collisions}},
\newblock \bibinfo{journal}{Nucl.Phys.} \bibinfo{volume}{A730}
  (\bibinfo{year}{2004}) \bibinfo{pages}{448--459}.
\bibitem[{McLerran and Venugopalan(1994)}]{McLerran:1993ni}
\bibinfo{author}{L.~D. McLerran}, \bibinfo{author}{R.~Venugopalan},
\newblock \bibinfo{title}{{Computing quark and gluon distribution functions for
  very large nuclei}},
\newblock \bibinfo{journal}{Phys.Rev.} \bibinfo{volume}{D49}
  (\bibinfo{year}{1994}) \bibinfo{pages}{2233--2241}.
\bibitem[{Lappi and Venugopalan(2006)}]{Lappi:2006xc}
\bibinfo{author}{T.~Lappi}, \bibinfo{author}{R.~Venugopalan},
\newblock \bibinfo{title}{{Universality of the saturation scale and the initial
  eccentricity in heavy ion collisions}},
\newblock \bibinfo{journal}{Phys.Rev.} \bibinfo{volume}{C74}
  (\bibinfo{year}{2006}) \bibinfo{pages}{054905}.
\bibitem[{Iancu and Venugopalan(2003)}]{Iancu:2003xm}
\bibinfo{author}{E.~Iancu}, \bibinfo{author}{R.~Venugopalan},
\newblock \bibinfo{title}{{The Color glass condensate and high-energy
  scattering in QCD}}  (\bibinfo{year}{2003}).
\bibitem[{Drescher and Nara(2007)}]{Drescher:2006ca}
\bibinfo{author}{H.-J. Drescher}, \bibinfo{author}{Y.~Nara},
\newblock \bibinfo{title}{{Effects of fluctuations on the initial eccentricity
  from the Color Glass Condensate in heavy ion collisions}},
\newblock \bibinfo{journal}{Phys.Rev.} \bibinfo{volume}{C75}
  (\bibinfo{year}{2007}) \bibinfo{pages}{034905}.
\bibitem[{Kolb et~al.(2000)Kolb, Sollfrank, and Heinz}]{Kolb:2000sd}
\bibinfo{author}{P.~F. Kolb}, \bibinfo{author}{J.~Sollfrank},
  \bibinfo{author}{U.~W. Heinz},
\newblock \bibinfo{title}{{Anisotropic transverse flow and the quark hadron
  phase transition}},
\newblock \bibinfo{journal}{Phys.Rev.} \bibinfo{volume}{C62}
  (\bibinfo{year}{2000}) \bibinfo{pages}{054909}.
\bibitem[{Hirano and Tsuda(2003)}]{Hirano:2002hv}
\bibinfo{author}{T.~Hirano}, \bibinfo{author}{K.~Tsuda},
\newblock \bibinfo{title}{{Collective flow and HBT radii from a full 3-D
  hydrodynamic model with early chemical freezeout}},
\newblock \bibinfo{journal}{Nucl.Phys.} \bibinfo{volume}{A715}
  (\bibinfo{year}{2003}) \bibinfo{pages}{821--824}.
\bibitem[{Schenke et~al.(2010)Schenke, Jeon, and Gale}]{Schenke:2010nt}
\bibinfo{author}{B.~Schenke}, \bibinfo{author}{S.~Jeon},
  \bibinfo{author}{C.~Gale},
\newblock \bibinfo{title}{{(3+1)D hydrodynamic simulation of relativistic
  heavy-ion collisions}},
\newblock \bibinfo{journal}{Phys.Rev.} \bibinfo{volume}{C82}
  (\bibinfo{year}{2010}) \bibinfo{pages}{014903}.
\bibitem[{Vredevoogd and Pratt(2009)}]{Vredevoogd:2008id}
\bibinfo{author}{J.~Vredevoogd}, \bibinfo{author}{S.~Pratt},
\newblock \bibinfo{title}{{Universal Flow in the First Stage of Relativistic
  Heavy Ion Collisions}},
\newblock \bibinfo{journal}{Phys.Rev.} \bibinfo{volume}{C79}
  (\bibinfo{year}{2009}) \bibinfo{pages}{044915}.
\bibitem[{Teaney et~al.(2001)Teaney, Lauret, and Shuryak}]{Teaney:2001av}
\bibinfo{author}{D.~Teaney}, \bibinfo{author}{J.~Lauret},
  \bibinfo{author}{E.~Shuryak},
\newblock \bibinfo{title}{A hydrodynamic description of heavy ion collisions at
  the sps and rhic}  (\bibinfo{year}{2001}).
\bibitem[{Heinz and Kolb(2002)}]{Heinz:2001xi}
\bibinfo{author}{U.~W. Heinz}, \bibinfo{author}{P.~F. Kolb},
\newblock \bibinfo{title}{{Early thermalization at RHIC}},
\newblock \bibinfo{journal}{Nucl.Phys.} \bibinfo{volume}{A702}
  (\bibinfo{year}{2002}) \bibinfo{pages}{269--280}.
\bibitem[{Shen and Heinz(2012)}]{Shen:2012vn}
\bibinfo{author}{C.~Shen}, \bibinfo{author}{U.~Heinz},
\newblock \bibinfo{title}{{Collision Energy Dependence of Viscous Hydrodynamic
  Flow in Relativistic Heavy-Ion Collisions}},
\newblock \bibinfo{journal}{Phys.Rev.} \bibinfo{volume}{C85}
  (\bibinfo{year}{2012}) \bibinfo{pages}{054902}.
\bibitem[{Song and Heinz(2008)}]{PhysRevC.77.064901}
\bibinfo{author}{H.~Song}, \bibinfo{author}{U.~Heinz},
\newblock \bibinfo{title}{Causal viscous hydrodynamics in 2 + 1 dimensions for
  relativistic heavy-ion collisions},
\newblock \bibinfo{journal}{Phys. Rev. C} \bibinfo{volume}{77}
  (\bibinfo{year}{2008}) \bibinfo{pages}{064901}.
\bibitem[{Dusling et~al.(2010)Dusling, Moore, and Teaney}]{Dusling:2009df}
\bibinfo{author}{K.~Dusling}, \bibinfo{author}{G.~D. Moore},
  \bibinfo{author}{D.~Teaney},
\newblock \bibinfo{title}{{Radiative energy loss and v(2) spectra for viscous
  hydrodynamics}},
\newblock \bibinfo{journal}{Phys.Rev.} \bibinfo{volume}{C81}
  (\bibinfo{year}{2010}) \bibinfo{pages}{034907}.
\bibitem[{Kovtun et~al.(2005)Kovtun, Son, and Starinets}]{Kovtun:2004de}
\bibinfo{author}{P.~Kovtun}, \bibinfo{author}{D.~Son},
  \bibinfo{author}{A.~Starinets},
\newblock \bibinfo{title}{{Viscosity in strongly interacting quantum field
  theories from black hole physics}},
\newblock \bibinfo{journal}{Phys.Rev.Lett.} \bibinfo{volume}{94}
  (\bibinfo{year}{2005}) \bibinfo{pages}{111601}.
\bibitem[{Rasmussen(2006)}]{Rasmussen06gaussianprocesses}
\bibinfo{author}{C.~E. Rasmussen},
\newblock \bibinfo{title}{Gaussian processes for machine learning},
\newblock \bibinfo{publisher}{MIT Press}, \bibinfo{year}{2006}.
\bibitem[{Novak et~al.(2013)Novak, Novak, Pratt, Coleman-Smith, and
  Wolpert}]{Novak:2013bqa}
\bibinfo{author}{J.~Novak}, \bibinfo{author}{K.~Novak},
  \bibinfo{author}{S.~Pratt}, \bibinfo{author}{C.~Coleman-Smith},
  \bibinfo{author}{R.~Wolpert},
\newblock \bibinfo{title}{{Determining Fundamental Properties of Matter Created
  in Ultrarelativistic Heavy-Ion Collisions}}  (\bibinfo{year}{2013}).
\bibitem[{York and Moore(2009)}]{York:2008rr}
\bibinfo{author}{M.~A. York}, \bibinfo{author}{G.~D. Moore},
\newblock \bibinfo{title}{{Second order hydrodynamic coefficients from kinetic
  theory}},
\newblock \bibinfo{journal}{Phys.Rev.} \bibinfo{volume}{D79}
  (\bibinfo{year}{2009}) \bibinfo{pages}{054011}.
\bibitem[{Abelev et~al.(2009)}]{Abelev:2009tp}
\bibinfo{author}{B.~Abelev}, et~al. (\bibinfo{collaboration}{STAR
  Collaboration}),
\newblock \bibinfo{title}{{Pion Interferometry in Au+Au and Cu+Cu Collisions at
  RHIC}},
\newblock \bibinfo{journal}{Phys.Rev.} \bibinfo{volume}{C80}
  (\bibinfo{year}{2009}) \bibinfo{pages}{024905}.
\bibitem[{Adams et~al.(2005)}]{Adams:2004bi}
\bibinfo{author}{J.~Adams}, et~al. (\bibinfo{collaboration}{STAR
  Collaboration}),
\newblock \bibinfo{title}{{Azimuthal anisotropy in Au+Au collisions at
  s(NN)**(1/2) = 200-GeV}},
\newblock \bibinfo{journal}{Phys.Rev.} \bibinfo{volume}{C72}
  (\bibinfo{year}{2005}) \bibinfo{pages}{014904}.
\bibitem[{Adil and Gyulassy(2005)}]{Adil:2005qn}
\bibinfo{author}{A.~Adil}, \bibinfo{author}{M.~Gyulassy},
\newblock \bibinfo{title}{{3D jet tomography of twisted strongly coupled quark
  gluon plasmas}},
\newblock \bibinfo{journal}{Phys.Rev.} \bibinfo{volume}{C72}
  (\bibinfo{year}{2005}) \bibinfo{pages}{034907}.

\end{thebibliography}

\end{document}